\documentclass[reprint, showpacs, superscriptaddress, english, twocolumn, notitlepage, longbibliography, nofootinbib]{revtex4-1}
\usepackage[T1]{fontenc}
\newcommand\thefontsize{\expandafter\string\the\font}

\usepackage[utf8]{inputenc}
\pdfoutput=1
\usepackage{times}
\usepackage{color}
\usepackage{array}
\usepackage{verbatim}
\usepackage{multirow}
\usepackage{amsmath}
\usepackage{amssymb}
\usepackage{dsfont}
\usepackage{graphicx}
\usepackage{esint}
\usepackage[unicode=true,
 bookmarks=true,bookmarksnumbered=true,bookmarksopen=false,
 breaklinks=true,pdfborder={0 0 0},backref=false,colorlinks=true]
 {hyperref}
\usepackage{enumitem}
\usepackage{titlesec}
\usepackage{cleveref}
\usepackage[normalem]{ulem}

\hypersetup{
    linkcolor=red,          
    citecolor=blue,        
    filecolor=magenta,      
    urlcolor=magenta           
}

\crefname{appendix}{Sec.}{Secs.}
\crefname{equation}{Eq.}{Eqs.}
\crefname{figure}{Fig.}{Figs.}
\crefname{table}{Table}{Tables}
\crefname{section}{Sec.}{Secs.}

\renewcommand{\paragraph}[1]{\vspace{0.2cm}{\bf \textit{#1}}}
\def\ie{i.e.,\ }
\def\eg{e.g.,\ }
\def\etc{etc.\ }

\definecolor{Gray}{gray}{0.85}
\newcolumntype{a}{>{\columncolor{Gray}}c}

\usepackage{simpler-wick}
\allowdisplaybreaks[1] %

\newcommand{\td}{\widetilde}
\newcommand{\ovl}{\overline}

\def\pare#1{\left( #1 \right)}
\def\brak#1{\left[#1\right]}
\def\brace#1{\left\{#1\right\}}
\def\bra#1{\langle #1 |}
\def\ket#1{| #1 \rangle}
\def\Bra#1{\left\langle #1 \right|}
\def\Ket#1{\left| #1 \right\rangle}
\def\inn#1{\langle #1 \rangle}
\def\Inn#1{\left\langle #1 \right\rangle}
\def\abs#1{\left| #1 \right|}
\def\Im{\mathrm{Im}}
\def\Re{\mathrm{Re}}

\def\mF{\mathcal{F}}
\def\ii{\mathrm{i}}
\def\pp{\mathbf{p}}
\def\Tr{\mathrm{Tr}}
\def\kk{\mathbf{k}}
\def\GG{\mathbf{G}}
\def\mA{\mathcal{A}}
\def\mD{\mathcal{D}}
\def\RR{\mathbf{R}}
\def\tt{\mathbf{t}}

\begin{document}
\title{Intrinsic (Axion) Statistical Topological Insulator}

\author{Xi Chen}
\thanks{These authors contribute equally to this work.}
\affiliation{International Center for Quantum Materials, School of Physics, Peking University, Beijing 100871, China}

\author{Fa-Jie Wang}
\thanks{These authors contribute equally to this work.}
\affiliation{International Center for Quantum Materials, School of Physics, Peking University, Beijing 100871, China}

\author{Zhen Bi}
\affiliation{Department of Physics, The Pennsylvania State University, University Park, Pennsylvania 16802, USA}

\author{Zhi-Da Song}
\email{songzd@pku.edu.cn}
\affiliation{International Center for Quantum Materials, School of Physics, Peking University, Beijing 100871, China}
\affiliation{Hefei National Laboratory, Hefei 230088, China}
\affiliation{Collaborative Innovation Center of Quantum Matter, Beijing 100871, China}

\date{\today}

\begin{abstract}
Ensembles that respect symmetries on average exhibit richer topological states than those in pure states with exact symmetries, leading to the concept of average symmetry-protected topological states (ASPTs).
The free-fermion counterpart of ASPT is the so-called statistical topological insulator (STI) in disordered ensembles. 
In this work, we demonstrate the existence of an {\it intrinsic} STI, which has no clean counterpart.
Using a real space construction (topological crystal), we find an axion STI characterized by the average axion angle $\bar{\theta}=\pi$, protected by an average $C_4T$ symmetry with $(C_4T)^4=1$. 
While the exact $C_{4}T$ symmetry reverses the sign of $\theta$ angle, and hence seems to protect a $\mathbb{Z}_2$ classification of $\theta\!=\!0,\pi$, we prove that the $\theta\!=\!\pi$ state cannot be realized in the clean limit if $(C_{4}T)^4 \!=\! 1$.
Therefore, the axion STI lacks band insulator correspondence and is thus {\it intrinsic}.
To illustrate this state, we construct a lattice model and numerically explore its phase diagram, identifying an axion STI phase separated from both band insulators and trivial Anderson insulators by a metallic phase, revealing the intrinsic nature of the STI.
We also argue that the intrinsic STI is {\it robust} against electron-electron interactions. 
Our work thus provides the first intrinsic crystalline ASPT and its lattice realization. 
\end{abstract}

\maketitle

\textbf{\textit{Introduction.}}
Topological insulators (TIs) \cite{TI1, TI2, fu_topological_2007, TI3, TI4}, including topological crystalline insulators (TCIs) \cite{TCI1, hsieh2012topological, TCI2, TCI3, TCI4, TCI5}, are band insulators that are not adiabatically connected to atomic limits without breaking the protecting symmetry \cite{soluyanov_wannier_2011,TCI8, TCI6, TCI7}. 
Their classification in clean systems has been well established in various schemes \cite{AZ_CLASSIFICATION2, AZ_CLASSIFICATION1, AZ_CLASSIFICATION3, TCI_CLASSIFICATION_1, TCI_CLASSIFICATION_3, TCI_CLASSIFICATION_4, song_topological_2017, TCS_0, TCI9, TCS_1, TCI11, TCI12, PhysRevB.105.045112}. 
For TIs protected by local symmetries, disorder, as an inevitable effect in realistic materials, does not change the classifications, provided the disorder respects the protecting symmetries \cite{AZ_CLASSIFICATION1, AZ_CLASSIFICATION3, TAI}.
Surprisingly, Refs.~\cite{fu_topology_2012, ringel_strong_2012, mong_quantum_2012} showed that some TIs can remain robust even when disorder breaks the protecting symmetries, as long as symmetries are respected {\it on average}.
Such TIs were later conceptually clarified in Ref.~\cite{STI} and termed as statistical TIs (STIs). 
An STI is defined as an ensemble of disordered insulators whose boundaries are prevented from localization by the average symmetry.
While STIs may exhibit unique boundary properties \cite{chaou2024disordered}, all the known examples of STIs can be adiabatically connected to the clean limit with exact symmetries. 
We hence call them {\it extrinsic} STIs. 

A closely related concept to STI is the average symmetry-protected topological (ASPT) state \cite{ASPT_PRX} - the many-body counterpart of STI -  that has attracted attention recently. Notably, {\it intrinsic} ASPT states \cite{ASPT_INTRINSIC}, which cannot be adiabatically connected to clean, gapped symmetric systems, have been identified. 
Inspired by this, we introduce and explore the concept of {\it intrinsic} STIs, referring to TIs that are only realizable in strongly disordered systems.

The 3D $\mathbb{Z}_2$ TI protected by time-reversal symmetry ($T$) exhibits a topological magneto-electric effect described by the Lagrangian term $\mathcal{L}_{\theta} \!=\! \frac{\theta e^2}{4 \pi^2 \hbar} \mathbf{E} \cdot \mathbf{B}$, with the axion angle $\theta\!=\!\pi$ \cite{TRS_AXION,wang_equivalent_2010}. 
If $T$ is broken, other $\theta$-odd symmetries, such as inversion \cite{wang_equivalent_2010,TCI4,AXION4,khalaf_higher-order_2018,kooi_inversion-symmetry_2018}, roto-reflections \cite{AXION2, fang_bulk_2012, varjas_bulk_2015, miert_higher-order_2018}, translation followed by $T$ \cite{TCI2,zhang_topological_2015}, and rotations ($C_{n=2,3,4,6}$) followed by $T$ \cite{zhang_topological_2015, schindler_higher-order_2018, varnava_surfaces_2018, wieder2018axion, ezawa_magnetic_2018, ahn_symmetry_2019, li_pfaffian_2020,PhysRevB.106.195144}, will also quantize $\theta$ to 0 or $\pi$. 
It is therefore generally believed that these symmetries can protect TCIs with $\theta\!=\!\pi$, known as axion TIs \cite{nenno2020axion, 10.1063/5.0038804, PhysRevB.101.155130, Leung_2020, PhysRevB.94.085102}. 
However, counterintuitively, we prove that a $C_{4}T$ symmetry, which satisfies $(C_{4}T)^{4}\!=\!1$, {\it cannot} protect axion TIs in the clean limit.
(Note that such a $C_4T$ can protect fragile topological insulators  \cite{TCI1,zhang_topological_2015,PhysRevB.102.115117}, which have no stable surface state.) 
Surprisingly, we show that an axion STI characterized by the {\it average} axion angle $\bar{\theta}=\pi$ can arise in Anderson insulators respecting  $C_4T$ symmetry on average, even if $(C_4T)^4=1$.
To illustrate this state, we construct a disordered lattice model and examine its phase diagram using the finite-size scaling of quasi-1D localization length \cite{chalker,transfer_matrix,one_parameter_scaling}.
An STI phase is identified through delocalized surface states and a half-quantized magneto-electric polarization in the bulk.
The STI phase is separated from both clean insulators and trivial Anderson insulators by an unavoidable metal phase, demonstrating its intrinsic nature. 
Our discovery marks the first example of intrinsic STI, highlighting a new type of topological phases.

\begin{figure}[t]
	\centering
    \includegraphics[width=1\linewidth]{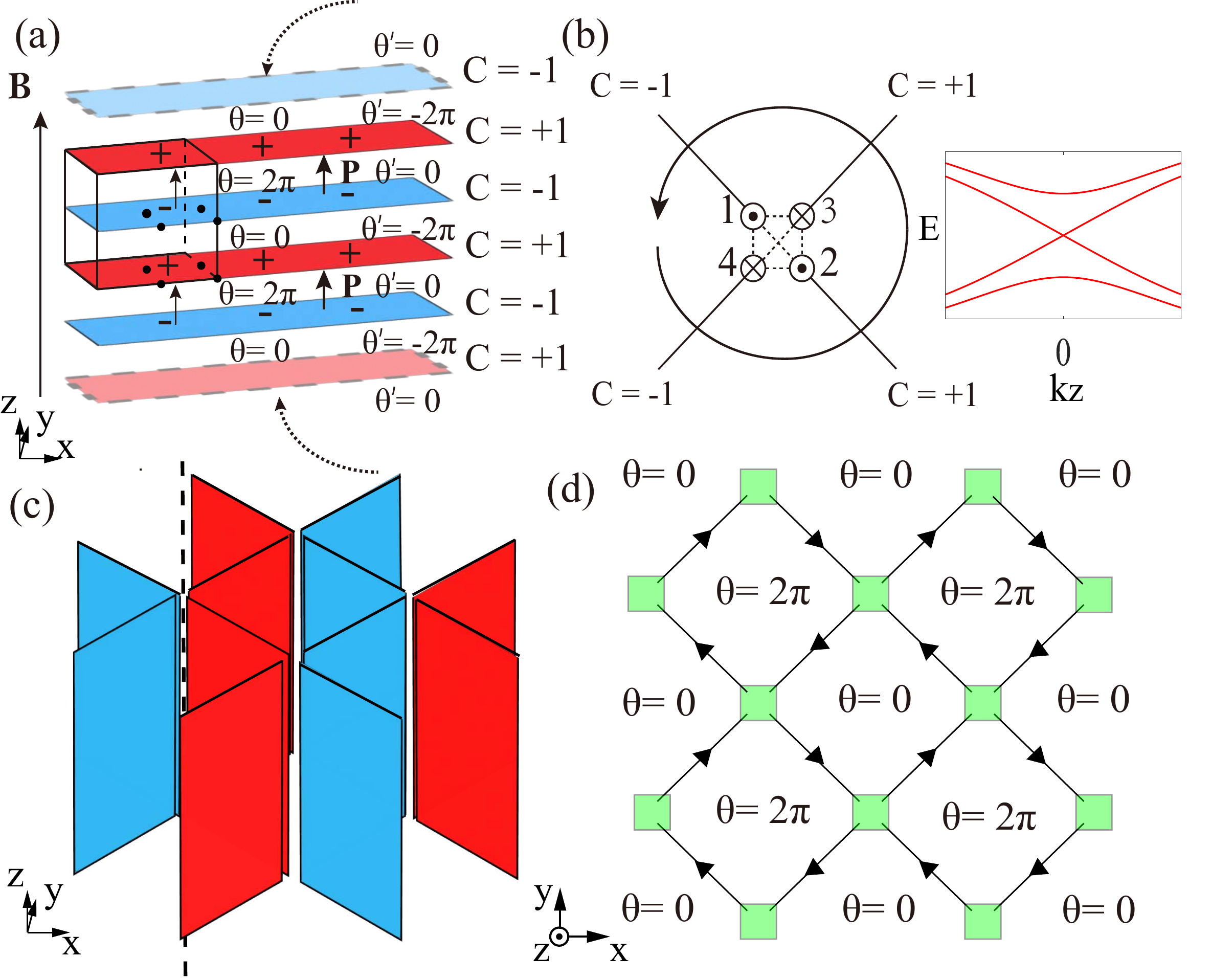}
    \caption[]{ 
    (a) The simplest topological crystal for inversion-protected axion TI. Inversion centers are shown by the black dots. Red (blue) planes show Chern insulators with $C = 1 (-1)$ with normal $+z$. $\theta$ and $\theta'$ show local axion angles before and after adding the pair of dashed layers.
    (b) A closed loop in an insulating topological crystal must enclose edge modes with zero net chirality, ensuring that $\theta$ is single-valued.
    For a $C_4T$-symmetric hinge in (c), if $(C_4T)^4=1$, the $(C_4T)^2 = -1$ subspace hosts Kramer helical states, while the $(C_4T)^2 = 1$ subspace can be locally gapped, as shown by the 1D band structure on the right.
    (c) A $C_4T$-symmetric topological crystal, with $C_4T$ axes occupied by intersecting hinges of Chern layers. Red (blue) planes are Chern insulators with $C = 1 (-1)$ with normals $(\mathbf{e}_x \pm \mathbf{e}_y)/\sqrt{2}$. 
    (d) The local axion angle of each 3D block embedded between Chern insulators in (c). The arrows show chiral modes on the $z$ surface, separating 2D blocks with $C = \pm 1/2$.
    }
\label{fig:extrinsic-STI}
\end{figure}

\textbf{\textit{Extrinsic axion STI.}}
To demonstrate the importance of the average axion angle, we first consider an extrinsic axion STI protected by an average inversion symmetry \cite{song_delocalization_2021,li_critical_2021}. 
By adding trivial local states, clean TCIs can be adiabatically deformed to ``topological crystals'' \cite{song_topological_2017, TCS_0, TCS_1}, which consists of building blocks made up of lower-dimensional TIs, where the electron correlation length is much smaller than the lattice constant. 
The simplest topological crystal for an inversion-protected axion TI is illustrated in \cref{fig:extrinsic-STI}(a), where all the planes at integer (half-integer) $z$-coordinates are decorated with insulators with Chern number $C\!=\!1$ ($-1$). 
The cubic with unit length represents a unit cell. 
Each unit cell has eight inversion centers at $x,y,z=0,\frac12$, which are all occupied by Chern insulators.

We now examine the spatial dependence of the local axion angle $\theta$ in the topological crystal by analyzing the magneto-electric response \(\boldsymbol{\mathcal{P}} = \frac{\theta}{2\pi} \frac{e}{\Phi_0} \mathbf{B}\) \cite{TRS_AXION, wang_equivalent_2010}, where $\Phi_0=h/e$ is the flux quantum. 
Suppose the 3D block above (below) a Chern insulator has local axion angle $\theta_1$ ($\theta_2$), with its volume given by $V_1=S\cdot L_1$ ($V_2=S\cdot L_2$), where $S$ represents the area of the Chern insulator. 
Then, under a  perpendicular magnetic field $\bold{B} = B \bold{e_z}$, the electric charge accumulated on the Chern insulator is given by
\begin{equation}
\Delta Q = -\frac{\mathcal{P}_1 V_1}{L_1} + \frac{\mathcal{P}_2 V_2}{L_2} = \frac{\theta_2 - \theta_1}{2\pi} \frac{e}{\Phi_0} B S \ .
\end{equation}
On the other hand, there must be $\Delta Q = C \frac{e}{\Phi_0} B S$ according to the Streda formula of quantum Hall states \cite{streda}, where \(C\) is the Chern number. 
The two expressions of $\Delta Q$ together imply \(\theta_1 - \theta_2 = -2\pi C\), meaning that a variation of \(\pm 2\pi\) in \(\theta\) is accompanied by crossing a Chern insulator with \(C = \mp 1\). Within each 3D block embedded between Chern insulators, \(\theta\) must remain constant, as there is no electric accumulation inside the region.

Since \(\theta = 0\) in the vacuum, the value of \(\theta\) in each 3D block in \cref{fig:extrinsic-STI}(a) is uniquely determined to be either 0 or \(2\pi\).
While regions with $\theta=0$ and $\theta=2\pi$ are topologically equivalent in the bulk, their difference manifests by $C=\pm 1$ on their common boundaries.
The average axion angle \(\bar{\theta} = \pi\), provided that the distances between adjacent Chern insulators are equal, as required by the inversion symmetry.
As shown in \cref{fig:extrinsic-STI}(a), by adding a pair of Chern insulators with $C=\pm 1$ (represented by the dashed red and blue planes) on opposite boundaries of the topological crystal, a global change of $-2\pi$ in the local axion angle can be achieved ($\theta \to \theta'$), resulting in $\bar{\theta}= \pi \to \bar{\theta'}=-\pi$.
Since this process on the boundary does not affect the bulk topology, it demonstrates that \(\bar{\theta} \mod 2\pi\) is a $\mathbb{Z}_2$ topological invariant. 

We further demonstrate that  \(\theta\) is always single-valued in insulating topological crystals, ensuring that \(\bar{\theta}\) is well-defined.
In a general topological crystal, including the \(C_4 T\)-symmetric state discussed below, a closed loop in space may cross multiple Chern insulators (\cref{fig:extrinsic-STI}(b)).
For the system to be insulating, edge states of the Chern insulators must be locally gapped at the hinges where the Chern insulators intersect. 
This condition enforces the net chirality (of the edge modes) enclosed by the loop to be zero, meaning that \(\theta\) undergoes an equal number of changes of \(\pm 2\pi\) around the loop.
A \(2n\pi\) winding of \(\theta\) corresponds to \(n\) unpaired chiral modes enclosed by the loop, which should not appear in an insulating TCI.

Topological crystals for axion TIs remain robust under weak quenched disorder respecting an average $\theta$-odd symmetry. 
Intra-Chern-layer disorder below the quantum Hall transition threshold does not disrupt the layers’ topology, so $\theta$ stays well-defined and single-valued. 
Inter-Chern-layer disorder induces local fluctuations of $\theta$, but the average $\theta$-odd symmetry ensures that  $\bar{\theta}$ remains to be $\pi$ \cite{song_delocalization_2021}.

Another extrinsic example closely related to the following sections is that $C_4T$, as a $\theta$-odd symmetry, can protect a higher-order TI characterized by $\bar\theta=\pi$ provided $(C_4T)^4=-1$ \cite{schindler_higher-order_2018, li_pfaffian_2020}. 
This state corresponds to the topological crystal shown in \cref{fig:extrinsic-STI}(c),~(d), comprising four types of Chern layer blocks with normal vectors $(\mathbf{e}_x + \mathbf{e}_y)/\sqrt{2}$, $(-\mathbf{e}_x + \mathbf{e}_y)/\sqrt{2}$, $(-\mathbf{e}_x - \mathbf{e}_y)/\sqrt{2}$, $(\mathbf{e}_x - \mathbf{e}_y)/\sqrt{2}$ respectively, each with a Chern number of one along its normal.
These blocks are connected via an array of $C_4T$ axes, at each of which four Chern layers intersect and cyclically transform under $C_4T$ symmetry. 
An axis hosts two pairs of chiral (anti-chiral) edge modes propagating along $\mathbf{e}_z$ ($-\mathbf{e_z}$) (\cref{fig:extrinsic-STI}(b)).
Assuming sufficiently large Chern gaps, the bulk low-energy physics of the topological crystal arises from these edge modes. 
Since the net chirality is zero, one may expect a symmetric gap to be allowed at each axis. This expectation holds true if $(C_4T)^4 = -1$, as demonstrated in Refs.~\cite{peng_topological_2022,sup}, leading to a gapped topological crystal that realizes the clean axion TI.

\textbf{\textit{Intrinsic axion STI.}}
For $(C_4T)^4=1$, a symmetric gap is forbidden in the clean limit, ruling out the clean axion TI. 
Consider the four edge modes at a $C_4T$ axis, labeled as 1, 2, 3, 4 (\cref{fig:extrinsic-STI}(b)).
Without loss of generality, we assume 1, 2 and 3, 4 are chiral and anti-chiral, respectively. 
\(C_4 T\) transforms these modes as $1\to4\to2\to3\to1$. 
Since \((C_4 T)^4 = 1\), the 4D Hilbert space splits into two 2D subspaces with \(C_2 = (C_4 T)^2\) eigenvalues $\pm1$. 
The $C_2=-1$ subspace is spanned by \(\frac{1}{\sqrt{2}} (|1\rangle - |2\rangle)\) and \(\frac{1}{\sqrt{2}} (|3\rangle - |4\rangle)\), which have opposite chirality and form a Dirac point at $k_z=0$. 
Since $(C_4T)^2 = -1$ in this subspace, the Kramers' degeneracy protects the Dirac point from being gapped. 
In contrast, the $C_2=1$ subspace, with $(C_4T)^2 = 1$, lacks Kramers' degeneracy and can be gapped.
Thus, the absence of clean axion TI arises from Kramers' degeneracy in the $C_2=-1$ subspace. 
In Ref.~\cite{sup}, we further show there is no other nontrivial topological crystal construction satisfying \((C_4 T)^4 = 1\), so that such symmetry does not protect any clean (stable) TI.
In addition, we also mathematically prove $\theta=0 \mod 2\pi$ for $C_4T$-symmetric band insulators with $(C_4T)^4 = 1$ in the momentum space formalism \cite{sup}.

To realize an {\it intrinsic} STI, we introduce Gaussian disorder with variance $W$ on the hinges where Chern layers intersect, breaking the exact $C_4T$ symmetry while preserving it on average. 
Since the bulk's low-energy physics is quasi-1D, a small $W$ can localize the system without affecting the topology of the Chern layers, as the disorder is confined to the hinges. 
The local axion angle $\theta$ of each 3D block between Chern layers can be determined (\cref{fig:extrinsic-STI}(d)), following the method used for \cref{fig:extrinsic-STI}(a).
The average value of $\theta$ over all 3D blocks is $\pi$. 
In the clean limit, $\theta$ is ill-defined around metallic helical modes, but with disorder, $\theta$ becomes well-defined and single-valued everywhere as the helical modes localize.
Therefore, the disordered topological crystal behaves as an Anderson insulator with $\bar{\theta} = \pi$.
We now consider the effect of disorders away from the hinges.
First, intra-Chern-layer disorder cannot change the layers' topology without inducing quantum Hall transitions and closing the bulk mobility gap.
Second, inter-Chern-layer disorder can only induce local fluctuations of $\theta$, while $\bar{\theta}$ is quantized to $\pi$ by the average $C_4T$ symmetry.
Therefore, the axion STI is robust against all types of average $C_4T$ symmetric disorder that does not close the bulk mobility gap.

\textbf{\textit{Bulk-boundary correspondence.}}
The nontrivial topology of the disordered topological crystal is also manifested through delocalized surface states.
Specifically, consider the surface in the $z$ direction. Chiral edge modes emerge at the boundaries of Chern layers, forming a network structure. As shown in \cref{fig:extrinsic-STI}(d), they resemble Chalker's model that describes the quantum Hall transition \cite{chalker}.
To further deepen the analogy with the quantum Hall transition, we assign effective Chern numbers ($C$'s) to the 2D blocks on the surface. 
First, the $C$'s of two adjacent surface 2D blocks, which are the surfaces of adjacent 3D bulk blocks with $\theta = 0$, $2\pi$, should differ by $\pm 1$ since they are separated by a chiral mode.
Second, the $C_4T$ symmetry, which exchanges the two types of 2D blocks, imposes that their $C$'s  must be opposite.
Thus, we assign effective Chern numbers $C=\pm1/2$ to the two types of surface 2D blocks.
The chiral modes can then be interpreted as domain walls between regions with $C =1/2$ and $-1/2$.
In the clean limit, the chiral modes connect to the bulk helical modes at the $C_4T$ axes represented by green squares. 
In the presence of disorders that localize bulk states, the green squares represent scattering nodes with random strengths and phase shifts.
The average $C_4T$ symmetry ensures that regions with $C =1/2$, $-1/2$ occupy equal areas. This symmetry condition pins the surface state at the critical point of the quantum Hall transition, where chiral modes percolate between $C = \pm 1/2$ regions and remain delocalized, manifesting the nontrivial bulk topology.

\begin{figure}[t!]
	\centering
    \includegraphics[width=1\linewidth]{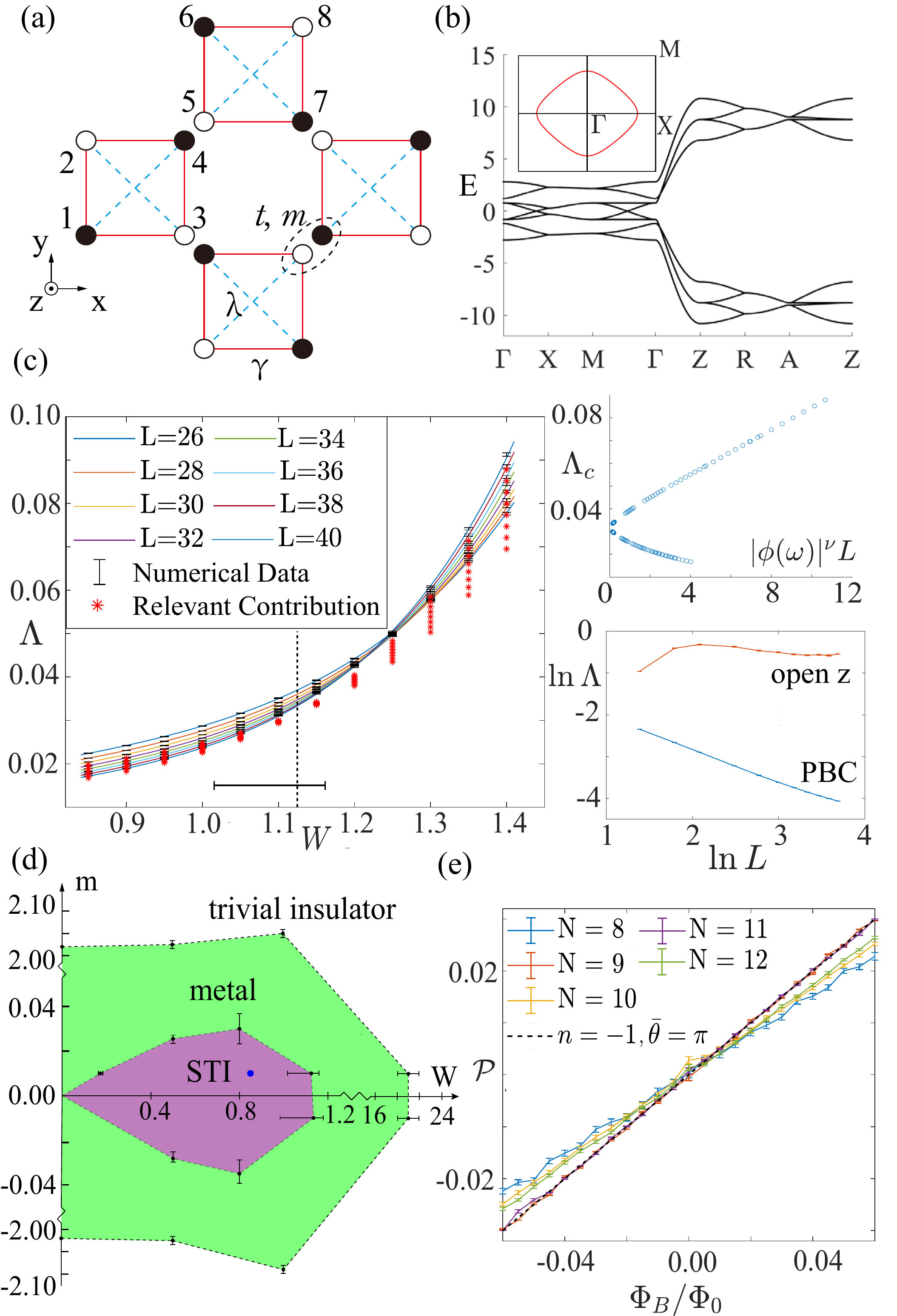}
    \caption[]{ 
    (a) The lattice model. Red (blue) lines represent $\gamma$ ($\lambda$). Black (white) circles represent chiral and anti-chiral modes.
    (b) Band structure of the lattice model with $t=2, \gamma = 1, \lambda = 0.01$ and $m = 0.8$. Inset shows the nodal line on the $k_z = 0$ plane.
    (c) Numerical normalized quasi-1D localization length $\Lambda$ with $t = 2, \gamma = 1, \lambda = 0.01, m=0.01$ and the Fermi level $E_F = -0.01$. Left: Numerical data and polynomial fitting of $\Lambda$ near $W_c \approx 1.125$. Red stars show the calculated relevant contribution to $\Lambda$ by setting $u_2(\Tilde{r}) = 0$. Upper right: Collapse of the relevant contribution to a one-parameter scaling function. Lower right: Comparison of $\Lambda$ with $W=0.85$ between periodic boundary condition in $y,z$ directions and open $z$ boundary condition. 
    (d) Phase diagram of the disordered lattice model with $t = 2, \gamma = 1, \lambda = 0.01$ and $E_F = -0.01$. The axion STI phase (purple) is completely surrounded by a gapless phase (green), which is in turn surrounded by the trivial insulator phase (white).
    (e) The topological magneto-electric response with $m=0.01, W = 0.85$. Other parameters are the same as (d). 
    }
\label{fig_data}
\end{figure}

\textbf{\textit{A lattice model.}}
Inspired by the topological crystal, we now construct a lattice model for the intrinsic STI. 
A pair of chiral modes on opposite edges of a Chern insulator can be regularized to a wire lattice: $2t k_z\sigma_z \to 2t \sin{k_z}\sigma_z + 2t(1-\cos{k_z})\sigma_x + m \sigma_x$, where $\sigma_z = \pm 1$ represents the chiral and anti-chiral modes around $k_z=0$, and the mass term $m\sigma_x$ mimics a coupling between the two edges. 
If the Chern layer reaches thermodynamic limit within one unit cell, there must be $m=0$.
Here we use $m$ as a tuning parameter.
A finite $m$ drives the low energy physics away from the quasi-1D limit, as discussed below.
Applying this regularization to the Chern layers in \cref{fig:extrinsic-STI}(c), we obtain the tight-binding model in \cref{fig_data}(a), where each pair of nearest neighbors in the $xy$ plane form a wire that simulates a Chern layer.
The red ($\gamma$) and dashed blue ($\lambda$) bonds represent couplings between the chiral and anti-chiral modes around the $C_4T$ axes. 
The Hamiltonian is 
\begin{widetext}
\begin{align}
H =\sum_{k_z, i} \zeta_i 2t\sin(k_z) c_{k_z, i}^{\dagger} c_{k_z,i}
+ \sum_{k_z, \langle i, j\rangle} (m+2t(1-\cos k_z)) c_{k_z, i}^{\dagger} c_{k_z, j} 
+ \sum_{k_z, \langle\!\langle i, j\rangle\!\rangle} \gamma c_{k_z, i}^{\dagger} c_{k_z,j}
+\sum_{k_z, \langle\!\langle\!\langle i, j\rangle\!\rangle\!\rangle} \lambda c_{k_z,i}^{\dagger} c_{k_z,j}
\label{eq:lattice_model_main}
\end{align}
\end{widetext}
where $i,j$ label sites in $xy$ plane, $\zeta_i = 1$, $-1$ for black (chiral) and white (anti-chiral) sites respectively, and $\langle \cdot \rangle, \langle\!\langle\cdot \rangle\!\rangle, \langle\!\langle\!\langle \cdot \rangle\!\rangle\!\rangle$ represent first, second, and third nearest neighbor pairs. 
The model respects a $C_{4}T$ symmetry satisfying $(C_4T)^4=1$, which transforms the eight orbitals as $1\to 3\to 4\to 2\to 1$, $5\to 7\to 8 \to 6\to 5$.
Even all the parameters are real, the time-reversal symmetry is inherently broken as chiral and anti-chiral modes from the same wire are coupled to different $C_4T$ centers. 
We identify all crystalline symmetries of \cref{eq:lattice_model_main}, including a $PT$ symmetry that squares to 1 in Ref.~\cite{sup}, where we also show that breaking the symmetries except $C_4T$ does not affect the topology of the STI. 
Below we focus on \cref{eq:lattice_model_main} at half filling. 
For $m = 0$, the model's low energy physics consists of decoupled 1D helical modes, as explained in the topological crystal argument, and the entire $k_z = 0$ plane lies on the Fermi surface. 
For $0<|m|<2\gamma$, the model is a nodal line semi-metal protected by $PT$ (\cref{fig_data}(b)). 
For $|m| > 2\gamma$, the system becomes a trivial band insulator.

Disorder is introduced by adding random imaginary components, $\ii \gamma'$, to the hopping $\gamma$ along the red bonds. 
The random variables $\gamma'$ follow a Gaussian distribution with a variant $W$.
$\gamma'$s on different bonds are assumed uncorrelated. 
The $C_4T$ symmetry is preserved on average since it maps one disorder configuration to another with equal probability, leaving the disorder ensemble invariant \cite{STI, ASPT_PRX}.
(We also numerically verify that the choice of disorder term does not affect the existence of an STI phase \cite{sup}.)

\textbf{\textit{Phase diagram.}}
The disordered lattice model is studied using the transfer matrix method in quasi-1D geometry \cite{chalker,transfer_matrix,one_parameter_scaling}, where the longitudinal size $M$ (along the $x$ direction) is much larger than the transverse size $L \times L$ (along the $y$ and $z$ directions).
To study the localization of bulk states, we impose periodic boundary conditions in both transverse directions $y$ and $z$. The localization length $\xi_{\rm 1D}(L)$ along the longitudinal direction, which depends on $L$, is obtained from the Lyapunov exponents of the transfer matrix.
The normalized quasi-1D localization length $\Lambda(L) = \xi_{\rm 1D}(L)/L$ characterizes the (de)localization behavior: Metallic, critical, and localized states correspond to divergent, finite, and zero $\lim_{L \to \infty} \Lambda(L)$, respectively.

On the insulating side of a metal-insulator phase transition, the 3D localization length $\xi_{\rm 3D}$ diverges as ${|r - r_c|}^{-\nu}$, where $r$ is the tuning parameter, $r_c$ is the critical value, and $\nu>0$ is a universal exponent. 
$\Lambda(r, L)$ follows the one-parameter scaling law with the scaling variable $L/\xi_{\rm 3D}$ for sufficiently large $L$ \cite{one_parameter_scaling}. 
For small $L$, an irrelevant correction to $\Lambda$ due to the finite-size effect must be considered, and $\Lambda$ takes the following form \cite{scaling_irr}:
\begin{equation}
    \Lambda(r, L) = F(u_1(\Tilde{r})L^{1/\nu}, u_2(\Tilde{r})L^{y})
\label{eq:scaling}
\end{equation}
where $\Tilde{r} = (r - r_c)/r_c$, $y<0$, and $F, u_1, u_2$ are undetermined functions that can be Taylor expanded near the phase transition point. 
The phase diagram in \cref{fig_data}(d) is obtained by fitting \cref{eq:scaling} to numerically computed $\Lambda$. 
For example, in the insulator-metal transition shown in \cref{fig_data}(c), we use $W$ as the tuning parameter, with all other parameters fixed.
The black error bars in \cref{fig_data}(c) show numerically computed $\Lambda$, while the red stars show the relevant contribution $F(u_1(\Tilde{r})L^{1/\nu}, 0)$, calculated using the fitted scaling function.
From \cref{fig_data}(c), we obtain $W_c = 1.125$ $[1.015, 1.162]$ and $\nu = 1.40$ $[1.24, 1.56]$, which is consistent with previously reported value $\nu = 1.443$ $[1.437, 1.449]$ for 3D Anderson transitions in the unitary class \cite{slevin2016estimate}.
Other points on the phase boundary are determined similarly by choosing $r=W$ or $m$. 
Further numerical details are provided in Ref.~\cite{sup}.

When $m=0$, the bulk's low energy states are quasi-1D and can be localized by arbitrarily small disorder strength $W$.
This corresponds to a localization transition occurring at $W_{c1}(0) = 0$. As will be discussed in the next paragraph, the resulting insulating phase is identified as the intrinsic STI.
For sufficiently large $W$, the system evolves into a trivial Anderson insulator. This implies a delocalized phase transition within the disorder range $[W_{c2}(0), W_{c3}(0)]$ that separates the STI from the trivial Anderson insulator, where $W_{c2}(0)$ and $W_{c3}(0)$ are estimated as $1.12$ and $20.0$, respectively.
When $m$ is small but nonzero, the bulk becomes a nodal-line semi-metal \cite{burkov_topological_2011}.
As $W$ increases, the system undergoes a localization transition at a finite $W_{c1}(m)$ \cite{shindou_critical_2020,jia2023phase,goncalves_disorder_nodal_2020}, entering the STI phase, followed by successive insulator-metal-insulator transitions at $W_{c2}(m)$ and $W_{c3}(m)$, eventually reaching the trivial Anderson insulator phase.
Notably, $W_{c1}(m)$ and $W_{c2}(m)$ merge at $m \approx \pm 0.03$, enclosing the intrinsic STI phase, which is separated from both the trivial Anderson insulator and clean insulators.

\textbf{\textit{Topological surface state.}}
To demonstrate the delocalized surface state of the STI, we recalculate the normalized localization length $\Lambda'(L)$ under an open boundary condition in the $z$-direction. 
This setup allows us to observe surface-specific phenomena, which are otherwise hidden under periodic boundary conditions.
As illustrated in \cref{fig:extrinsic-STI}(d), the surface state is expected to exhibit quantum Hall criticality.
Using the parameters corresponding to the blue point within the STI phase in \cref{fig_data}(d), we find that $\Lambda'(L)$ indeed exhibits critical behavior - it remains constant as $L$ approaches infinity.
This behavior is in sharp contrast to $\Lambda(L)$ with periodic boundary conditions, which decays with $L$ and reveals bulk localization, as illustrated in \cref{fig_data}(c).

\textbf{\textit{Topological magneto-electric effect.}}
To further illustrate the nontrivial topology of the intrinsic STI, we directly compute the magneto-electric response of the disordered lattice model in a cubic geometry with $N^3$ sites.
Periodic boundary condition is imposed along the $z$ direction to avoid delocalized surface states, while open boundary conditions are applied along  $x$ and $y$ directions.
A magnetic field $B$ is introduced along the $x$ direction, and the resulting electric polarization is measured as $\mathcal{P} = \langle \hat{X} \rangle / N^3$, where $\langle \hat{X} \rangle$ is the expectation value of the $x$-coordinate for occupied states.
Theoretically, $\mathcal{P}$ is given by $-(n+ P_3) {\Phi_B}/{\Phi_0}$,
where $\Phi_{B}$ is the flux through a unit cell, $n$ is a surface-dependent {\it integer}, and $P_3=\bar\theta/(2\pi)$ represents the magneto-electric polarization.  
The disorder-averaged $\mathcal{P}$ (over 100 configurations) is plotted as a function of $B$ in \cref{fig_data}(e). The slope $d\mathcal{P}/(d\Phi_B/\Phi_0)$ converges to $\frac{1}{2}$ as $N$ increases, confirming the topological magneto-electric effect characterized by $P_3 = \frac{1}{2}$.

\textbf{\textit{Discussions.}} 
We argue that the STI becomes a well-defined intrinsic fermionic ASPT in the presence of electron-electron interactions (see Ref.~\cite{sup} for details). Within the topological crystal framework, an intrinsic ASPT is characterized by an obstruction to anomaly-free SPT decoration, where the system cannot be symmetrically gapped even with interactions in the absence of disorder. On the other hand, introducing average symmetric disorder enables localization, leading to a short-range entangled bulk, while the boundary remains delocalized.
In our case, the decorated Chern insulators around a $C_4T$ axis generate a helical hinge mode exhibiting a $U(1)$ and $C_4T$ mixed anomaly.
The interacting clean 1D helical hinge can be mapped to the edge of a 2D TI with exact time-reversal symmetry, where edge states are known to be delocalized before spontaneous symmetry breaking \cite{wu2006helical,xu2006stability,chou2018gapless}.
Thus, the obstruction in the clean limit persists to the interacting case. 
After breaking $C_4T$ symmetry down to an average symmetry through disorder, the anomaly constraint is lifted, allowing these modes to become localized and rendering the system’s ground state short-range entangled.
As depicted by \cref{fig:extrinsic-STI}(d), without interaction, the surface perpendicular to the $z$ direction is similar to a delocalized critical Chalker's network, manifesting the nontrivial topology of the STI.
The Chalker's network belongs to the same universality class as the quantum Hall transition, which is known to be stable against short-ranged interactions \cite{huckestein1999integer,wang2002electron,pruisken2008non,kumar2022interaction}.
Therefore, the delocalized surface states are also stable against interactions, illustrating the interaction stability of the bulk topology through bulk-boundary correspondence.
We hence conclude that the intrinsic STI becomes an intrinsic crystalline ASPT with interaction present. 

In Ref.~\cite{sup}, we also discuss experimental signatures distinguishing the STI from conventional 3D TIs, such as a statistical version of the Witten effect \cite{witten1979dyons, PhysRevB.82.035105, PhysRevB.108.155104}. We also generalize the discussion to a $C_2T$ symmetry squaring to $-1$. 

\textbf{\textit{Acknowledgments.}}
We are grateful to Yang Qi and Rui-Xing Zhang for helpful discussions. 
Z.-D. S., X. C., and F.-J. W. were supported by National Natural Science Foundation of China (General Program No.~12274005), National Key Research and
Development Program of China (No.~2021YFA1401900), and Innovation Program for Quantum Science and Technology (No.~2021ZD0302403). Z. B. acknowledges support from NSF under award number DMR-2339319.

\clearpage
\onecolumngrid 
\appendix{\bf{Appendix}}

\crefname{appendix}{Sec.}{Secs.}
\crefname{equation}{Eq.}{Eqs.}
\crefname{figure}{Fig.}{Figs.}
\crefname{table}{Table}{Tables}
\crefname{section}{Sec.}{Secs.}

\renewcommand{\paragraph}[1]{\vspace{0.2cm}{\bf \textit{#1}}}
\renewcommand\thesection{\Roman{section}}
\renewcommand\thesubsection{\Alph{subsection}}
\def\ie{i.e.,\ }
\def\eg{e.g.,\ }
\def\etc{etc.\ }

\definecolor{Gray}{gray}{0.85}
\newcolumntype{a}{>{\columncolor{Gray}}c}

\allowdisplaybreaks[1] %

\def\pare#1{\left( #1 \right)}
\def\brak#1{\left[#1\right]}
\def\brace#1{\left\{#1\right\}}
\def\bra#1{\langle #1 |}
\def\ket#1{| #1 \rangle}
\def\Bra#1{\left\langle #1 \right|}
\def\Ket#1{\left| #1 \right\rangle}
\def\inn#1{\langle #1 \rangle}
\def\Inn#1{\left\langle #1 \right\rangle}
\def\abs#1{\left| #1 \right|}
\def\Im{\mathrm{Im}}
\def\Re{\mathrm{Re}}

\def\mF{\mathcal{F}}
\def\ii{\mathrm{i}}
\def\pp{\mathbf{p}}
\def\Tr{\mathrm{Tr}}
\def\kk{\mathbf{k}}
\def\GG{\mathbf{G}}
\def\mA{\mathcal{A}}
\def\mD{\mathcal{D}}
\def\RR{\mathbf{R}}
\def\tt{\mathbf{t}}

\renewcommand{\thefigure}{S\arabic{figure}}
\renewcommand{\theequation}{S\arabic{equation}}

\tableofcontents

\clearpage

\section{Absence of clean axion insulator 
\texorpdfstring{($(C_4T)^4\!=\!1$)}{}: Momentum space proof \label{sec:momentum_proof}}

In this section, we prove that a $C_4T$-symmetric band insulator must have a  vanishing magneto-electric polarization $P_3$ if $(C_4T)^4=1$.

\subsection{\texorpdfstring{$P_3$}{P3} and the winding number of sewing matrix}

In a band insulator, the magneto-electric polarization $P_3 = \frac{\theta}{2\pi}$ can be calculated as a Chern-Simons integral \cite{fu_topological_2007,TRS_AXION}
\begin{equation} \label{eq:P3-def}
P_3 = \frac1{16\pi^2} \int d^3 \kk \ \epsilon_{ijk} 
    \Tr\brak{ \pare{ \mF^{ij}(\kk) - \frac23 \ii \mA^{i}(\kk) \mA^{j}(\kk) } \mA^{k}(\kk)  } \quad \mod 1\ . 
\end{equation}
where the indices $i,j,k=x,y,z$ are implicitly summed when repeated, $\epsilon_{ijk}$ is the Levi-Civita symbol. 
The non-Abelian Berry connection and curvature are defined as
\begin{equation}
    \mA^i_{mn}(\kk) = - \ii \inn{u_{m}(\kk) | \partial_i | u_n(\kk)},\qquad 
    \mF^{ij} = \partial_i \mA^{j} - \partial_j \mA^i + \ii [\mA^{i},\mA^{j}]\ ,
\end{equation}
respectively, where $\partial_i\equiv \partial_{k_i}$. One should not confuse the minus sign in $\mA^i$. 
The band indices $m,n$ are limited to the occupied bands. 
Upon a gauge transformation within occupied bands $\ket{u_{n}(\kk)} \to \ket{u_{l}(\kk)} U_{ln} (\kk)$, where repeated band indices are implicitly summed, the Berry connection and curvature transform as 
\begin{equation}
    \mA^i \to U \mA^i U^\dagger - \ii U \partial_i U^\dagger,\qquad 
    \mF^{ij} \to U \mF^{ij} U^\dagger \ ,
\end{equation}
respectively. 
Substituting the gauge transformation into \cref{eq:P3-def}, we obtain the change of the integral
\begin{align}\label{eq:DeltaP3}
\Delta P_3 =& \frac1{16\pi^2} \int d^3\kk \ \epsilon_{ijk} \Tr \bigg[ 
    \frac23 \pare { U \partial_i U^\dagger } \pare { U \partial_j U^\dagger } \pare { U \partial_k U^\dagger }  
    + \ii \mF^{ij} (U^\dagger \partial_k U )
    + \frac23 \mA^{i} \mA^{j} (U^\dagger \partial_k U) 
    + \frac23 (U^\dagger \partial_i U)  \mA^{j} \mA^{k} \nonumber\\
&   + \frac23 \mA^{i} (U^\dagger \partial_j U)   \mA^{k} 
    + \frac{2\ii}3 (U^\dagger \partial_i U) (U^\dagger \partial_j U) \mA^k 
    + \frac{2\ii}3 \mA^i (U^\dagger \partial_j U) (U^\dagger \partial_k U) 
    + \frac{2\ii}3  (U^\dagger \partial_i U) \mA^j (U^\dagger \partial_k U) 
    \bigg] \ .
\end{align}
Using the cyclic condition of trace, we simplify it to 
\begin{align} 
\Delta P_3 =& \frac1{16\pi^2} \int d^3\kk \ \epsilon_{ijk} \Tr \bigg[ 
    \frac23 \pare { U \partial_i U^\dagger } \pare { U \partial_j U^\dagger } \pare { U \partial_k U^\dagger } 
    + \ii \mF^{ij} (U^\dagger \partial_k U )
    + 2 \mA^{i} \mA^{j} (U^\dagger \partial_k U) 
    - 2\ii  \mA^j (\partial_k U^\dagger) (\partial_i U) 
    \bigg] \ .
\end{align}
Expanding $\mF^{ij}$ in terms of the Berry connection, we find that the last three terms in the above equation sum to a full derivative term 
\begin{align}
 & \frac1{16\pi^2} \int d^3\kk \ \epsilon_{ijk} \Tr \bigg[
    2 \ii (\partial_i \mA^j) (U^\dagger \partial_k U )
    - 2 \mA^{i} \mA^j (U^\dagger \partial_k U) 
    + 2 \mA^{i} \mA^j (U^\dagger \partial_k U) 
    + 2\ii \mA^j (\partial_i U^\dagger) (\partial_k U ) \bigg] \nonumber\\
=&  \frac1{16\pi^2} \int d^3\kk \ \epsilon_{ijk} \cdot \partial_i \cdot 
    \Tr [2\ii \  \mA^j (U^\dagger \partial_k U) ] = 0\ . 
\end{align}
Then 
\begin{equation}
   \Delta P_3 = \frac1{24\pi^2} \int d^3\kk \ \epsilon_{ijk} \Tr [ 
    \pare { U \partial_i U^\dagger } \pare { U \partial_j U^\dagger } \pare { U \partial_k U^\dagger } ] 
\end{equation}
has the form of a 3D winding number, which is an integer. 
Therefore, $P_3$ defined in \cref{eq:P3-def} is a gauge invariant quantity after modulo 1. 

Now let us see how $P_3$ is quantized by a $C_4 T$ symmetry.
We do not specify $(C_4T)^4=1$ or $-1$ in this subsection. 

The following derivation parallels that in Refs.~\cite{TRS_AXION,fang_bulk_2012,schindler_higher-order_2018}. 
We define the $C_4 T$ sewing matrix as 
\begin{equation}
    B_{mn}(\kk) = \inn{ u_{m}( C_4T \cdot \kk) | C_{4}T | u_n(\kk)}, 
\end{equation}
where $C_4 T \cdot \kk = (k_y, -k_x, -k_z)$. 
It is direct to verify that $B_{mn}(\kk)$ is unitary and periodic over the Brillouin zone. 
It immediately follows
\begin{equation} \label{eq:C4T-sewing}
C_{4}T \ket{ u_n(\kk)} = \ket{ u_{m}(C_4T\cdot \kk) } \cdot B_{m n}(\kk),\qquad 
\ket{ u_{m}(C_4T\cdot \kk)  } = B_{mn}^*(\kk) \cdot C_{4}T \cdot \ket{ u_n(\kk)} \ ,
\end{equation}
where implicit summations over repeated band indices are limited to occupied bands. 
We write $C_4 T \cdot \kk$ as $\kk'$ for simplicity.  
The coordinate transformation has the form $k'_i = R_{ij} k_j$, with $R\cdot R^T = 1$, $\det R=-1$. 
The derivatives with respect to $k$ ($\partial$) and $k'$ ($\partial'$) are related by $\partial_i = R_{ji} \partial_j'$, $\partial_i' = R_{ij} \partial_j$. 
The non-Abelian Berry connection defined in the new coordinate is
{\small
\begin{align} \label{eq:A-tilde}
\td\mA^i_{mn} (\kk') \equiv & - \ii \inn{ u_{m}(\kk') | \partial_i' | u_n(\kk')  }
= - \ii \inn{ u_{m}(\kk') | R_{ij} \partial_j  | u_n(\kk')  } 
= - \ii R_{ij} B_{mm'}(\kk)  \inn{ C_4 T \cdot u_{m'}(\kk) | \partial_j \Big[ | C_{4}T \cdot u_{n'}(\kk)  } B_{nn'}^*(\kk) \Big] \nonumber\\
=& - \ii   R_{ij}  B_{mm'}(\kk)\inn{ u_{m'}^*(\kk) | \partial_j  | u_{n'}^*(\kk)  } B_{nn'}^*(\kk) - \ii R_{ij} B_{ml}(\kk) \partial_j B_{nl}^*(\kk) \nonumber\\
=& - R_{ij} [ B(\kk) \mA^{j*}(\kk)  B^{\dagger}(\kk) ]_{mn} 
    - \ii R_{ij} [ B(\kk) \partial_j  B^{\dagger}(\kk) ]_{mn}\ . 
\end{align}}
After a few steps of algebra, one can also derive 
\begin{equation} \label{eq:F-tilde}
\td\mF^{ij} (\kk') \equiv  
\partial_i' \td \mA^{j}(\kk') - \partial_j' \td \mA^i(\kk') + \ii [\td \mA^{i}(\kk'), \td \mA^{j}(\kk')] = - R_{i i'} R_{jj'} B(\kk) \mF^{i'j'*} (\kk) B^\dagger(\kk)\ . 
\end{equation}
Suppose 
\begin{align}
q + P_3  = &\frac1{16\pi^2} \int d^3 \kk \ \epsilon_{ijk} 
    \Tr\brak{ \pare{ \mF^{ij}(\kk) - \frac23 \ii \mA^{i}(\kk) \mA^{j}(\kk) } \mA^{k}(\kk)  }  \nonumber\\
 =&  \frac1{16\pi^2} \int d^3 \kk' \epsilon_{ijk} 
    \Tr\brak{ \pare{ \td\mF^{ij}(\kk') - \frac23 \ii \td\mA^{i}(\kk') \td\mA^{j}(\kk') } \td\mA^{k}(\kk')  } \ . 
\end{align}
The two rows are the same quantity expressed in two coordinate systems, hence they equal to each other. 
Here $q$ is the gauge-dependent integer part of the integral, and $P_3$ ranges from 0 to 1.  
Inserting \cref{eq:A-tilde,eq:F-tilde} to the second expression, we obtain 
{\small
\begin{align}
q + P_3 \!=\!  \frac1{16\pi^2} \int d^3 \kk \epsilon_{ijk} R_{ii'}R_{jj'} R_{kk'}
    \Tr\brak{ \pare{ B \mF^{i'j'*} B^\dagger \!+\! \frac{2\ii}3 
    (B \mA^{i'*} B^\dagger \!+\! \ii B\partial_{i'} B^\dagger) 
    (B\mA^{j'*}B^\dagger \!+\! \ii B\partial_{j'} B^\dagger) } 
    (B\mA^{k'*}B^\dagger \!+\! \ii B\partial_{k'}B^\dagger)  } . 
\end{align}}
Since $\epsilon_{ijk} R_{ii'}R_{jj'} R_{kk'} = \det R \cdot \epsilon_{i'j'k'}$ and $\det R=-1$, the above equation becomes 
{\small
\begin{align}
q + P_3 =&  - \frac1{16\pi^2} \int d^3 \kk \ \epsilon_{ijk} 
    \Tr\brak{ \pare{ B \mF^{i j *} B^\dagger + \frac{2\ii}3 (B \mA^{i*} B^\dagger + \ii B\partial_{i} B^\dagger) (B\mA^{j*}B^\dagger + \ii B\partial_{j} B^\dagger) } (B\mA^{k*}B^\dagger + \ii B\partial_{k}B^\dagger)  } \nonumber\\
=& - \frac1{16\pi^2} \int d^3 \kk \ \epsilon_{ijk}  \Tr\brak{ \pare{ \mF^{ij}(\kk) - \frac23 \ii \mA^{i}(\kk) \mA^{j}(\kk) } \mA^{k}(\kk)  }^*
  - \frac{1}{24\pi^2} \int d^3 \kk \ \epsilon_{ijk}  
  \Tr\brak{ (B\partial_i B^\dagger)  (B\partial_j B^\dagger)   (B\partial_k B^\dagger)  }\nonumber\\
& - \frac1{16\pi^2} \int d^3 \kk \ \epsilon_{ijk} \Tr \bigg[ 
  - \ii \mF^{ij*} (B^\dagger \partial_k B) 
  + \frac{2}3 \mA^{i*} \mA^{j*} (B^\dagger \partial_k B) 
  + \frac{2}3 \mA^{i*} (B^\dagger \partial_j B)  \mA^{k*} 
  + \frac{2}3  (B^\dagger \partial_i B )\mA^{j*} \mA^{k*}  \nonumber\\
&\qquad \qquad 
    - \frac{2\ii}3 (B^\dagger \partial_i B) (B^\dagger \partial_j B) \mA^{k*} 
    - \frac{2\ii}3 (B^\dagger \partial_i B) \mA^{j*} (B^\dagger \partial_k B) 
    - \frac{2\ii}3 \mA^{i*} (B^\dagger \partial_j B)  (B^\dagger \partial_k B)  \bigg] \ .
\end{align}}
The first term in the second row equals to $-q - P_3$.
The complex conjugations of the last seven terms (in the third and fourth rows) share the same form as the last seven terms in \cref{eq:DeltaP3}, which have been shown as a full derivative term. 
Then, there must be 
\begin{equation} \label{eq:P3-winding}
2P_3 = - \frac{1}{24\pi^2} \int d^3 \kk \ \epsilon_{ijk}  
  \Tr\brak{ (B\partial_i B^\dagger)  (B\partial_j B^\dagger)   (B\partial_k B^\dagger)  } \quad \mod 2\ . 
\end{equation}
Since the integral on the right hand side is a 3D winding number (an integer), $P_3$ is either 0 or $\frac12$ mod 1. 
Therefore, in the presence of a $C_4 T$ symmetry, regardless of $(C_4T)^4=1$ or $-1$, $P_3$ can only take value 0 or $\frac12$. 

We have assumed a smooth $B(\kk)$ to validate the differentials in \cref{eq:P3-winding}. 
This can be fulfilled when the system does not exhibit 3D Chern numbers: For topological states protected by symmetries (translation excluded), one can always choose a symmetry-breaking Wannier gauge \cite{soluyanov_wannier_2011} where the states are smooth over the Brillouin zone.

\subsection{Trivial \texorpdfstring{$P_3$}{P3}}

\begin{figure}[th]
\centering
\includegraphics[width=0.25\linewidth]{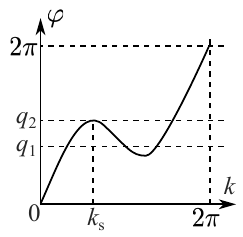}
\caption{A Map $f$ from $\mathrm{S}^1$ to $\mathrm{S}^1$. 
    To calculate the parity of $\mathrm{deg}[f]$, one can choose a regular value, {\it e.g.}, $q_1$, in the target manifold  and count its preimages. The parity of the number of preimages gives $\mathrm{deg}[f]$ mod 2. 
    One should not use a critical value, {\it e.g.}, $q_2$, for the counting. 
}
\label{fig:mapping}
\end{figure}

In this subsection we prove $P_3=0$ if $(C_4T)^4=1$ by showing that the winding number on the right hand side of \cref{eq:P3-winding} is always even. 
First, we block-diagonalize the $C_4T$ sewing matrix $B$ into two-by-two and one-by-one blocks. 
A one-by-one block always contributes zero to the winding number.
Each two-by-two block $B^r$ ($r=1,2\cdots$) realizes a mapping from the 3D torus $\mathrm{T}^3$ to $\mathrm{SU(2)}\sim \mathrm{S}^3$ and contributes a factor $\mathrm{deg}[B^r]$ -  degree of the mapping- to the winding number \cite{wang_equivalent_2010, li_pfaffian_2020}. 
$\mathrm{deg}[B^r]$ counts how many times the domain manifold ($\mathrm{T}^3$) wraps around the target manifold SU(2).
One can choose a {\it regular} value $q_{\rm ref}\in$ SU(2) and counts how many times $q_{\rm ref}$ is hit by the mapping $B^r$, as illustrated in \cref{fig:mapping}. 
Parity of the counting gives $\mathrm{deg}[B^r]$ mod 2, and hence determines $P_3$. 
We will prove that, if $(C_4T)^4=1$, there exists such a $q_{\rm ref}$ that the counting is always even, implying $P_3=0$.

We introduce the sewing matrix for $C_2=(C_4T)^2$
\begin{equation}
    D_{mn}(\kk) = \inn{ u_m(C_2\kk)| C_2 | u_n(\kk) }\ . 
\end{equation}
$D(\kk)$ is unitary and periodic over the Brillouin zone. 
It is related to the $C_4T$ sewing matrix via
\begin{align}
D_{mn}(\kk) = \inn{ u_m(C_2\kk)| C_4T \cdot C_4T | u_n(\kk) }
    = \bra{ u_m(C_2\kk) } C_4T \Big[  \ket{u_{l}(C_4T\cdot \kk}  B_{ln}(\kk)  \Big]
    = [B ( C_{4}T\cdot \kk) B^*(\kk) ]_{mn}\ ,
\end{align}
where \cref{eq:C4T-sewing} is inserted. 
Thus, the $C_4T$ sewing matrix at $C_4T \cdot \kk$ is related to that at $\kk$ by 
\begin{equation} \label{eq:B-D-relation}
    B(C_4T \cdot \kk) = D(\kk) \cdot B^T(\kk)\ . 
\end{equation}
Applying this equation twice, we obtain 
\begin{equation} \label{eq:B-D-relation2}
B(C_2 \cdot \kk) = D(C_{4}T\cdot \kk) \cdot B^T(C_{4}T \cdot \kk)
    =  D(C_{4}T\cdot \kk) \cdot B(\kk) \cdot D^T(\kk)\ . 
\end{equation}
For later convenience, we define the high symmetry momenta 
\begin{equation}
\begin{aligned}
&    \Gamma:\ (0,0,0),\qquad 
    \mathrm{X}:\ (\pi,0,0),\qquad 
    \mathrm{Y}:\ (0,\pi,0),\qquad 
    \mathrm{M}:\ (\pi,\pi,0) \nonumber\\
&   \mathrm{Z}:\ (0,0,\pi),\qquad 
    \mathrm{R}:\ (\pi,0,\pi),\qquad 
    \mathrm{T}:\ (0,\pi,\pi),\qquad 
    \mathrm{A}:\ (\pi,\pi,\pi)\ . 
\end{aligned}
\end{equation}
There are four $C_2$-invariant high symmetry lines: $\rm \Gamma Z$, $\rm XR$, $\rm YT$, $\rm MA$. 
For $\kk$ belonging to these lines, $D(\kk)$ is a diagonal matrix consisting of $C_2$ eigenvalues. 
Since $C_2^2=1$, the $C_2$ eigenvalues can only be $\pm 1$. 
There are four $C_4T$-invariant high symmetry momenta $K^4=\{ \mathrm{\Gamma, M, Z, A} \}$.
For $\kk_0 \in K^4$, the sewing matrix $B$ must satisfy $B(\kk_0) = D(\kk_0) B^T(\kk_0)$. 
A $D(\kk_0) =-1$ eigenvalue must be at least doubly degenerate because the minimal $B(\kk_0)$ matrix satisfying $B(\kk_0)=-B^T(\kk_0)$ is $ e^{i\alpha} \sigma_y$, where $\alpha$ is an arbitrary phase factor. 
States at $\rm X$ or $\rm Y$ are generally non-degenerate, and there must be  $D(\mathrm{X})=D^{*}(\mathrm{Y})$ according to \cref{eq:B-D-relation2}. 

For simplicity, we assume there is no accidental degeneracy within the occupied bands. 
If an accidental degeneracy happens, we add a symmetry-allowed perturbation term to lift it. This operation does not change the value of $P_3$ of the occupied bands.
(To be concrete, let us consider Weyl points between two nearby bands. According to the $C_4T$ symmetry, the irreducible Brillouin zone is $\frac14$ of the full Brillouin zone. Since the $C_4T$ symmetry preserves the topological charges of Weyl points, the irreducible Brillouin zone and its $C_4T$ partners must have the same charge. 
Thus, according to the Nielsen-Ninomiya theorem \cite{nielsen_absence_1981,nielsen_absence_1981-1}, which guarantees a vanishing total topological charge in lattice models, the irreducible Brillouin zone must also have a zero topological charge. 
The irreducible Brillouin zone could have pairs of Weyl points with opposite charges, but  one can design proper continuous deformation to annihilate them within the irreducible Brillouin zone.)
The only remained degeneracies are the $D(\kk_0)=-1$ states at $\kk_0\in K^4$. 
Then the occupied bands decompose into a set of disconnected groups, as shown in \cref{fig:sewing}(a) and (b).  
The sewing matrices then have block-diagonal forms:
\begin{equation}
D(\kk) = \bigoplus_{r} D^r(\kk),\qquad 
B(\kk) = \bigoplus_{r} B^r(\kk)\ .
\end{equation}
A 1D block must have $D^r(\Gamma,\mathrm{M})=1$ such that no Kramer degeneracy appears.
We then identify two types of 1D blocks in terms of $C_2$ eigenvalues 
\begin{equation}
    D^r(\Gamma) = 1,\qquad 
    D^r(\mathrm{M})=1,\qquad 
    D^r(\mathrm{X})=\pm1 . 
\end{equation}
Similarly, there are nine types of 2D blocks in terms of $C_2$ eigenvalues: 
\begin{equation} \label{eq:2Dblock-1}
    D^r(\Gamma) = -\sigma_0,\qquad 
    D^r(\mathrm{M}) = -\sigma_0,\qquad 
    D^r(\mathrm{X}) = \textcolor{red}{\sigma_0}, \textcolor{red}{-\sigma_0}, \sigma_z  \ ,
\end{equation}
\begin{equation} \label{eq:2Dblock-2}
    D^r(\Gamma) = -\sigma_0,\qquad 
    D^r(\mathrm{M}) = \sigma_0,\qquad 
    D^r(\mathrm{X}) = \sigma_0, -\sigma_0, \textcolor{red}{\sigma_z}  \ ,
\end{equation}
\begin{equation} \label{eq:2Dblock-3}
    D^r(\Gamma) = \sigma_0,\qquad 
    D^r(\mathrm{M}) = -\sigma_0,\qquad 
    D^r(\mathrm{X}) = \sigma_0, -\sigma_0, \textcolor{red}{\sigma_z}   \ . 
\end{equation}
There is no higher dimensional block. 
The blocks contribute to the winding number in \cref{eq:P3-winding} independently. 
Since the $C_2$ symmetry does not protect any stable band topology \cite{elcoro_magnetic_2021,peng_topological_2022}, and we are not considering 3D Chern insulators in this work, the $C_2$ sewing matrix $D(\kk)$ should take form of band representations (BRs) \cite{TCI8}  - symmetry-allowed atomic limits - of the space group $P2$, or represent fragile topological states of $P2$.
We will discuss the complexity from $C_2$ sewing matrix in \cref{sec:C2-eigenvalue} and for now only focus on the simplest $\kk$-independent choices
\begin{equation} \label{eq:constant-D}
        D^r(\kk) = 1,\qquad D^r(\kk) = -\sigma_0 \ 
\end{equation}
for 1D and 2D blocks, respectively. 
Since no block exhibits 3D Chern numbers, $B^r(\kk)$ can be smoothly defined over the Brillouin zone. 
We do {\it not} assume a diagonal structure of $B^r(\kk)$ on the energy eigenstates, which may lead to some singularities in $B^r(\kk)$.

We consider a U(1) gauge transformation to the $r$-th block: $\ket{u_{n}'(\kk)} = \ket{u_{n}(\kk)} e^{\ii\varphi(\kk)/d_r}$, where $\varphi(\kk)$ is smoothly defined over the Brillouin zone and $d_r=1$ or 2 is the dimension of the $r$-th block. 
$\varphi(\kk)$ should exhibit even (if nonzero) winding numbers across the Brillouin zone for $d_r=2$ to ensure the periodicity of $\ket{u_n'(\kk)}$. 
The $C_4T$ and $C_2$ sewing matrices on the transformed basis are 
\begin{equation} \label{eq:Br-prime}
B^{r\prime}(\kk) = e^{-\frac{\ii}{d_r}(\varphi(C_4T\cdot\kk) + \varphi(\kk))} B^r(\kk) ,\qquad 
D^{r\prime}(\kk) = e^{-\frac{\ii}{d_r}(\varphi(C_2\cdot\kk) - \varphi(\kk))} D^r(\kk) \ ,
\end{equation}
respectively. 
The purpose of this gauge transformation will become clear in the next paragraph. 
We require $\varphi(C_2\cdot\kk)=\varphi(\kk)$ such that the constant $D$ sewing matrices in \cref{eq:constant-D} remain unchanged under the transformation.
If $\varphi$ can be chosen to satisfy 
\begin{equation} \label{eq:varphi-det}
    \ii (\varphi(C_4T\kk) + \varphi(\kk)) =  \ln \det B^r(\kk)\ ,\qquad 
    \varphi(C_2\cdot\kk) = \varphi(\kk)\ ,
\end{equation}
then $\det B^{r\prime}(\kk)=1$ and $D^{r\prime}(\kk)=D^{r}(\kk)$ (\cref{eq:constant-D}) are fulfilled in the same gauge.  
We now examine whether there are topological obstructions to \cref{eq:varphi-det}:
\begin{enumerate}[label=(\roman*)]
\item Denote $\vartheta(\kk) = \varphi(C_4T\kk) + \varphi(\kk)$. As $\varphi$ is smoothly defined over the Brillouin zone, $\vartheta(\kk)$ must have vanishing winding numbers on arbitrary contractable loops. 
    $\ln \det B^r(\kk)$ must also have vanishing winding numbers on contractable loops as $B^r(\kk)$ is smoothly defined over the Brillouin zone (see discussions below \cref{eq:P3-winding}). 
\item $\vartheta(\kk)$ must have vanishing winding number on the non-contractable loop $\mathcal{C}_z: (0,0,-\pi) \to (0,0,\pi)$ because $\vartheta(0,0,k_z)=\vartheta(0,0,-k_z)$ by definition. 
    For the considered $C_2$ sewing matrices in \cref{eq:constant-D}, \cref{eq:B-D-relation} implies $\det B^r(0,0,k_z) = \det B^r(0,0,-k_z)$, meaning $\ln \det B^r(\kk)$ also has vanishing winding number along $\mathcal{C}_z$. 
\item The winding numbers of $\vartheta(\kk)$ along non-contractable loops  $\mathcal{C}_{x}: (-\pi,0,0)\to(\pi,0,0)$, $\mathcal{C}_{y}: (0,-\pi,0)\to (0,\pi,0)$ must be zero because $\vartheta(\kk)=\vartheta(C_2\cdot\kk)$ by definition. 
    For the considered sewing matrices in \cref{eq:constant-D}, \cref{eq:B-D-relation} also implies vanishing winding numbers of $\ln\det B^r(\kk) $ along $\mathcal{C}_{x,y}$. 
\end{enumerate}
Since there is no topological obstruction to \cref{eq:varphi-det}, one can find such a gauge where $\det B^{r\prime}(\kk) = 1$ and $D^{r\prime}(\kk)$ is given by \cref{eq:constant-D}. 
Moreover, as $\vartheta(\kk)$ has vanishing winding numbers along non-contractable loops, $B^{r\prime}(\kk)$ (\cref{eq:Br-prime}) remains periodic over the Brillouin zone \cite{wang_equivalent_2010}. 
In the following we will stick to this gauge and relabel $B^{r\prime}(\kk)$, $D^{r\prime}(\kk)$ as $B^{r}(\kk)$, $D^r(\kk)$, respectively.

A 1D block always has $B^r(\kk)=1$ and hence contributes trivially to the winding number. 
For a 2D block, $B^r$ realizes a mapping from $\mathrm{BZ}\sim \mathrm{T}^3$ to $\mathrm{SU(2)}\sim \mathrm{S}^3$. 
Then, the right hand side of \cref{eq:P3-winding} reduces to  \cite{wang_equivalent_2010, li_pfaffian_2020}
\begin{equation} \label{eq:Br-degree}
    \sum_{r} \mathrm{deg} [B^r] \mod 2
\end{equation}
with $\mathrm{deg}[B^r]$ being the degree of the mapping between oriented manifolds with the same dimension \cite{guillemin2010differential}. 
The degree can be computed via a simple counting 
\begin{equation} \label{eq:degree-counting}
     \mathrm{deg} [B^r] = \sum_{\kk \in (B^{r-1})[q_{\rm ref}]} \mathrm{sgn}_{\kk} [B^r]\ . 
\end{equation}
Here $q_{\rm ref}\in \mathrm{SU(2)}$ is a {\it regular value} of the mapping, meaning that at each preimage $\kk \in (B^{r-1})[q_{\rm ref}]$, the mapping has a full-rank Jacobian. 
($\kk$ is a regular (critical) point if the local Jacobian is (not) full-rank, and its image is a  regular (critical) value.)
See \cref{fig:mapping} for example. 
$\mathrm{sgn}_{\kk} [B^r]$ is the sign of the Jacobian determinant, which equals to $+1$ ($-1$) if the local mapping around $\kk$ keeps (reverses) the orientation of the manifold. 
We need to choose a convenient reference point $q_{\rm ref}$ to evaluate \cref{eq:Br-degree}. 
We notice that \cref{eq:B-D-relation} and $D^r(\kk)=-\sigma_0$ imply
\begin{equation}   \label{eq:Br-two-choices}
\forall \kk_0\in K^4,\qquad  B^r(\kk_0) = - B^{rT}(\kk_0)\qquad  \Rightarrow 
\qquad B^r(\kk_0) = \pm \ii \sigma_y\ . 
\end{equation}
Let us first try the reference point $q_{\rm ref} = \ii \sigma_y$. 
If there is $B^r(\kk)=\ii \sigma_y $ for some $\kk \not \in K^4$, then there must be $B^r(C_{4}T\cdot \kk) = - B^{rT}(\kk)=\ii \sigma_y$. 
We can reasonably assume that such $\kk \not\in K^4$ are regular points where the Jacobian is full-rank: Even if $\kk$ was critical, the singularity is not enforced by symmetry and hence can be removed.
In particular, a $C_2$-symmetric momentum $\kk_1$ must be critical.
This follows from \cref{eq:B-D-relation2} and $D^r(\kk_1)=-\sigma_0$, which together imply that $B^r(\kk_1+\pp)$ must be even in $p_x,p_y$, leading to a vanishing Jacobian determinant at $\kk_1$. 
While $B^r(\kk_1)=\ii\sigma_y$ could occur accidentally at a $C_2$-symmetric $\kk_1\not\in K^4$, we can avoid this by adding perturbation to the Hamiltonian or applying a gauge transformation to $B^r(\kk)$, moving the preimage of $\ii\sigma_y$ to one or more regular points in the neighborhood of $\kk_1$. 
Since this happens at $\kk_1$ and $C_4T \cdot \kk_1$ simultaneously, the contribution to $ \mathrm{deg} [B^r]$ (\cref{eq:degree-counting}) is always even.
Thus, we only need to examine contributions from $K^4$.
A $\kk_0\in K^4$ is automatically $C_2$-symmetric and hence critical.
However, unlike other $C_2$-symmetric points, if $\kk_0\in K^4$ is a preimage of $\ii\sigma_y$, one cannot continuously move the preimage to its neighborhood by a perturbation to the Hamiltonian or a gauge transformation because $B^r(\kk_0)$ can only take discrete values $\pm \ii \sigma_y$ due to \cref{eq:Br-two-choices}. 
Nevertheless, we can perturbatively change the reference point $q_{\rm ref}$ such that $\kk_0$ is no longer a preimage of the new $q_{\rm ref}$. 
In the next two paragraphs we will show that  $\kk_0$ will split into an even number of regular preimages upon the change of $q_{\rm ref}$, hence each $\kk_0$ also contributes trivially to $\mathrm{deg}[B^r]$ mod 2. 
Thus, $\mathrm{deg}[B^r]$ mod 2 is always zero. 

We parameterize $B^r(\kk)$ around a preimage $\kk_0\in K^4$ of $\ii\sigma_y$. 
Consider $\kk=\kk_0 + \pp$ with $\pp$ being a small quantity, \cref{eq:B-D-relation} and $D^r(\kk)=-\sigma_0$ imply 
\begin{align}
B^r(\kk_0 + \pp) & = \sum_{\mu=0,x,y,z} d_\mu(\pp) \sigma_\mu = 
     [\beta_{01} p_z  + \beta_{02} (p_x^2-p_y^2) + 2\beta_{03} p_xp_y ] \sigma_0 
    +[ \beta_{11} p_z + \beta_{12} (p_x^2-p_y^2) + 2\beta_{13} p_xp_y ] \ii \sigma_x \nonumber\\
  & \qquad \qquad +[1 + \alpha_{21} p_z^2  + \alpha_{22} (p_x^2+p_y^2)  ] \ii \sigma_y
    +[\beta_{31} p_z + \beta_{32} (p_x^2-p_y^2) + 2\beta_{33} p_xp_y ] \ii \sigma_z
    + \mathcal{O}(p^3)\ ,
\end{align}
where $\alpha,\beta$ are real coefficients.
Since
\begin{equation}
    B^{r\dagger} B^r = \sigma_0 [ 1 + 2\alpha_{22} (p_x^2+p_y^2)
    + (2\alpha_{21} + \beta_{01}^2+\beta_{11}^2+\beta_{31}^2) p_z^2
    ]
    + \mathcal{O}(p^3)\ . 
\end{equation}
The normalization condition requires $\alpha_{22}=0$, 
$\alpha_{21} = -(\beta_{01}^2+\beta_{11}^2+\beta_{31}^2)/2$. 
Using $d_{0,x,z}$ as local coordinates for SU(2), we obtain the Jacobian determinant 
\begin{equation} \label{eq:Jacobian}
\abs{ \frac{\partial d_{0,x,z}}{\partial p_{x,y,z}} } = 
4   \left| 
    \begin{array}{ccc}
         \beta_{01} & \beta_{02} & \beta_{03} \\
         \beta_{11} & \beta_{12} & \beta_{13} \\
         \beta_{31} & \beta_{32} & \beta_{33} 
    \end{array}
    \right| \cdot (p_x^2+p_y^2) 
    + \mathcal{O}(p^3)\ . 
\end{equation}
It vanishes at $\pp=0$. 

We perturbatively move the reference point away from the critical value $\ii\sigma_y$: 
\begin{equation} \label{eq:perturbation}
q_{\rm ref}(\varepsilon) = \ii e^{\ii\frac{\varepsilon}2 \sigma_z}\sigma_y e^{-\ii\frac{\varepsilon}2 \sigma_z} = \ii \sigma_y + \varepsilon\cdot \ii\sigma_x + \mathcal{O}(\varepsilon^2)\ ,   
\end{equation}
where $\varepsilon$ is a sufficiently small rotation along $\sigma_z$. 
(The following discussions also apply to other rotation axes than $\sigma_z$.)
For those regular $\kk \not\in K^4$ satisfying $B^r(C_4T\cdot\kk)=B^r(\kk)=\ii \sigma_y$, the  full-rank  local Jacobians ensure solutions $\pp',\pp\sim \varepsilon$ to $B^r(C_4T\cdot\kk+\pp')=B^r(\kk+\pp)=q_{\rm ref}(\varepsilon)$. 
Thus, the contribution to $\mathrm{deg}[B^r]$ from $\kk\not\in K^4$ remains even upon the perturbation, and we only need to examine the contribution from neighborhoods of $\kk_0\in K^4$. 
Suppose $B^r(\kk_0)=\ii\sigma_y$ ($\kk_0 \in K^4$), solutions to the equation $B^r(\kk_0+\pp)=q_{\rm ref}(\varepsilon)$, {\it i.e.}, 
\begin{equation} \label{eq:perturb1}
\beta_{01} p_z  + \beta_{02} (p_x^2-p_y^2) + 2\beta_{03} p_xp_y = \mathcal{O}(\varepsilon^2) \ ,
\end{equation}
\begin{equation}\label{eq:perturb2}
\beta_{11} p_z  + \beta_{12} (p_x^2-p_y^2) + 2\beta_{13} p_xp_y = \varepsilon + \mathcal{O}(\varepsilon^2) \ ,
\end{equation}
\begin{equation} \label{eq:perturb3}
\beta_{31} p_z  + \beta_{32} (p_x^2-p_y^2) + 2\beta_{33} p_xp_y = \mathcal{O}(\varepsilon^2) \ ,
\end{equation}
give the split preimages.
We expect the solutions to satisfy $|p_z|\sim \varepsilon$, $|p_{x,y}|\sim \varepsilon^{\frac12}$. 
\cref{eq:perturb1,eq:perturb3} imply
\begin{equation}
(p_x^2-p_y^2) \cdot \sin2\varphi  
+  2p_x p_y \cdot \cos2\varphi  = \mathcal{O}(\varepsilon^2)\ ,
\end{equation}
where $2\varphi = \arctan \frac{\beta_{02}\beta_{31} - \beta_{32}\beta_{01}}{\beta_{03}\beta_{31} - \beta_{33}\beta_{01}}\in (-\frac{\pi}2,\frac{\pi}2]$. 
We introduce a coordinate transformation 
$p_x=p_1 \cdot \cos\varphi  + p_2 \cdot \sin\varphi$, 
$p_y=-p_1 \cdot \sin\varphi  + p_2 \cdot \cos\varphi$ such that
$p_x^2-p_y^2=(p_1^2-p_2^2)\cdot \cos2\varphi + 2p_1p_2\cdot \sin2\varphi$, 
$2p_xp_y=-(p_1^2-p_2^2)\cdot \sin2\varphi + 2p_1 p_2 \cdot \cos2\varphi$. 
Then the above equation becomes 
\begin{equation}
    p_1 p_2 = 0 + \mathcal{O}(\varepsilon^2) \ . 
\end{equation}
Using the new coordinates, we find two branches of solutions to \cref{eq:perturb1,eq:perturb3}: 
\begin{equation}
\text{Curve-I:} \qquad  p_1=0+\mathcal{O}(\varepsilon^{\frac32}),\quad p_z = p_2^2 \cdot \frac{\cos2\varphi \cdot \beta_{02}  - \sin2\varphi\cdot\beta_{03} }{\beta_{01}} + \mathcal{O}(\varepsilon^2) \ ,
\end{equation}
\begin{equation}
\text{Curve-II:} \qquad  p_2=0+\mathcal{O}(\varepsilon^{\frac32}),\quad p_z = -p_1^2 \cdot \frac{\cos2\varphi \cdot \beta_{02}  - \sin2\varphi\cdot\beta_{03} }{\beta_{01}} {\beta_{01}} + \mathcal{O}(\varepsilon^2)\ . 
\end{equation}
The remaining \cref{eq:perturb2} gives a curve in the $p_1=0$ plane
\begin{equation}
    p_1=0+\mathcal{O}(\varepsilon^{\frac32}),\quad  p_z = \frac{\varepsilon}{\beta_{11}} + p_2^2 \frac{\cos2\varphi \cdot \beta_{12}  - \sin2\varphi\cdot\beta_{13} }{\beta_{11}} + \mathcal{O}(\varepsilon^2)\ ,
\end{equation}
which may have two or zero crossings with curve-I. 
\cref{eq:perturb2} also gives another curve in the $p_2=0$ plane
\begin{equation}
    p_2=0+\mathcal{O}(\varepsilon^{\frac32}),\quad  p_z = \frac{\varepsilon}{\beta_{11}} - p_1^2 \frac{\cos2\varphi \cdot \beta_{12}  - \sin2\varphi\cdot\beta_{13} }{\beta_{11}} + \mathcal{O}(\varepsilon^2)\ ,
\end{equation}
which may have zero or two crossings with curve-II. 
Thus, the number of solutions to 
\cref{eq:perturb1,eq:perturb2,eq:perturb3} is even, and they are regular points because in general the Jacobian determinant (\cref{eq:Jacobian}) is proportional to $p_x^2+p_y^2 \sim |\varepsilon| >0$.
Therefore, a critical point $\kk_0 \in K^4$ splits into an even number of regular points upon the perturbation \cref{eq:perturbation}, contributing trivially to $\mathrm{deg}[B^r]$ mod 2 (\cref{eq:Br-degree}). 

The proof for $P_3=0$ mod 1 is completed.

\subsection{Complexity from \texorpdfstring{$C_2$}{C2} sewing matrix}
\label{sec:C2-eigenvalue}

\begin{figure}[th]
\centering
\includegraphics[width=0.85\linewidth]{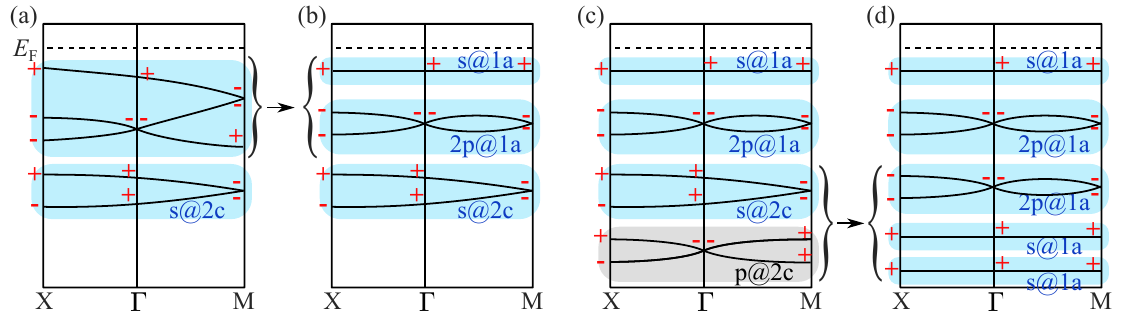}
\caption{Adiabatic deformation of the occupied bands.
    The red $\pm$ signs indicate the $C_2$ eigenvalues. 
    Bands shaded in blue form $C_2$-BRs, which are BRs of the space group $P2$ but may exhibit stable topology protected by $C_4T$ in principle. 
    Bands shaded in grey form $C_4T$-BRs, which are in trivial atomic limits. (a) and (b) demonstrate how the occupied bands are deformed into disconnected 1D and 2D blocks. 
    (c) and (d) demonstrate how a generic $C_2$-BR is deformed into a $C_2$-BR induced from the $1a$ position by adding auxiliary $C_4T$-BRs. 
}
\label{fig:sewing}
\end{figure}

Let us enumerate the elementary BRs - minimal symmetry-allowed atomic limits -  of space group $P2$ and derive their $C_2$ sewing matrices $D^r$. 
There are four $C_2$ invariant Wyckoff positions in real space:
$(0,0,z)$, $(\frac12,\frac12,z)$, $(\frac12,0,z)$, $(0,\frac12,z)$, 
where $z$ is a free parameter ranging from 0 to 1. 
Without loss of generality, in the following we choose $z=0$. 
We use $\ket{\RR +\tt, \xi}$ to represent a local orbital locating at $\RR+\tt$ with $C_2$ eigenvalue $\xi=\pm1$. 
Here $\RR$ represents the unit cell and $\tt$ is the corresponding Wyckoff position. 
$C_2$ acts on the local orbital as $C_{2} \ket{\RR +\tt, \xi} = \xi \ket{-\RR -\tt, \xi}$. 
The Bloch states are Fourier transformations
\begin{equation}
    \ket{\psi_{\xi}(\kk)} = \frac1{\sqrt{N}}\sum_{\RR} e^{\ii \RR\cdot\kk} \ket{\RR +\tt,\xi }\ ,
\end{equation}
where $N$ is the system size. 
They are periodic over the Brillouin zone. 
One can directly verify that 
\begin{align}
C_2 \ket{\psi_{\xi}(\kk)} = \xi \frac1{\sqrt{N}} \sum_{\RR} e^{\ii \RR\cdot\kk} \ket{-\RR -\tt,\xi }
    = \xi \frac1{\sqrt{N}} \sum_{\RR'} e^{\ii (-\RR'-2\tt)\cdot\kk} \ket{\RR' +\tt,\xi }
= \xi e^{-\ii 2\tt \cdot \kk} \ket{\psi_{\xi}(C_2\kk)}\ ,
\end{align}
where $\RR'=-\RR-2\tt$. 
We can read the $C_2$ sewing matrix as 
\begin{equation}
    D_{\tt,\xi}(\kk) = \xi e^{-\ii 2\tt \cdot\kk}\ . 
\end{equation}

For later convenience, we introduce Wyckoff position labels for the magnetic space group $P4'$: 
\begin{equation}
1a:\; (0,0,0),\qquad 
1b:\; (\frac12,\frac12,0),\qquad 
2c:\; (\frac12,0,0), \; (0,\frac12,0)\ ,
\end{equation}
where $1a$ and $1b$ is $C_4T$ symmetric, and the two positions in $2c$ transform to each under $C_4T$. 

As explained in the last subsection, since $C_2$ does not protect any stable topology, the nine $C_2$ sewing matrices in \cref{eq:2Dblock-1,eq:2Dblock-2,eq:2Dblock-3} should be identical to that of BRs or fragile topological bands of the space group $P2$.  
We hence refer to them as $C_2$-BRs (or $C_2$-fragile-bands). 
It might be worth emphasizing that such $C_2$-BRs are $C_4T$-symmetric, and the $C_4T$ sewing matrix $B^r(\kk)$ could be atomic or topological. 
For clarity, in the following we refer to the trivial atomic limits respecting the $C_4T$ symmetry as $C_4T$-BRs, whose $B^r(\kk)$ matrix is atomic. 
A $C_4T$-BR is by definition a $C_2$-BR, but the reverse is not necessarily true.  
We observe that four of the $C_2$ sewing matrices in  \cref{eq:2Dblock-1,eq:2Dblock-2,eq:2Dblock-3}  (marked red) are consistent with $C_2$-BRs.
The first (dubbed as $2p@1b$) can be obtained from two $\xi=-1$ states at $1b$. 
The corresponding sewing matrix is 
\begin{equation} \label{eq:BR_2p_1b}
    D^r(\kk) = -\sigma_0 \cdot e^{-\ii (k_x+k_y)} \ ,\qquad (2p@1b)\ .
\end{equation}
The second (dubbed as $2p@1a$) can be obtained from two $\xi=-1$ states at $1a$. 
The corresponding sewing matrix is 
\begin{equation} \label{eq:BR_2p_1a}
    D^r(\kk) = -\sigma_0 \ ,\qquad (2p@1a)\ .
\end{equation}
The third (dubbed as $p@2c$) can be obtained from two $\xi=-1$ states at $2c$, respectively. 
The corresponding sewing matrix is 
\begin{equation} \label{eq:BR_p_2c}
    D^r(\kk) = \begin{pmatrix}
        - e^{-\ii k_x} & 0 \\
        0 & - e^{-\ii k_y}
    \end{pmatrix}  \ ,\qquad (p@2c)\ .
\end{equation}
The fourth (dubbed as $s@2c$) can be obtained from two $\xi=1$ states at $2c$, respectively. 
The corresponding sewing matrix is 
\begin{equation} \label{eq:BR_s_2c}
    D^r(\kk) = \begin{pmatrix}
        e^{-\ii k_x} & 0 \\
        0 & e^{-\ii k_y}
    \end{pmatrix}  \ ,\qquad (s@2c)\ .
\end{equation}
One can similarly derive the two 1D $C_2$-BRs 
\begin{equation} \label{eq:BR_s_1a}
    D^r(\kk) = 1 \ ,\qquad (s@1a)\ ,
\end{equation}
\begin{equation} \label{eq:BR_s_1b}
    D^r(\kk) =  e^{-\ii (k_x+k_y)} \ ,\qquad (s@1b)\ .
\end{equation}
The other five sewing matrices (not marked red) in \cref{eq:2Dblock-1,eq:2Dblock-2,eq:2Dblock-3} must represent $C_2$-fragile-bands. 

In the last subsection, we have deformed the occupied bands into disconnected blocks where any possible accidental degeneracy is lifted by perturbation terms that do not change the topology. 
Then we proved that a block does not contribute to $P_3$ if it is a $C_2$-BR ($s@1a$ or $2p@1a$) induced from the $1a$ position, whose $C_2$ sewing matrix is a constant (\cref{eq:constant-D}). 
Here we show that other $C_2$-BRs cannot contribue to $P_3$ neither. 
First, as long as the occupied bands as a whole do not exhibit fragile topology, we can always deform them into disconnected blocks where each is a $C_2$-BR. 
Even the occupied bands exhibit a fragile topology as a whole, we can always add auxiliary atomic limits into the occupied bands - without changing the stable topology characterized by $P_3$ - to trivialize the fragile topology.
This is because the defining feature of fragile topology is that it can be trivialized by adding atomic limits, which, as completely localized states, do not contribute to $P_3$. 
Therefore, in order to calculate $P_3$, we can always safely assume that the occupied bands can be deformed into disconnected $C_2$-BRs.

Second, using the same trick of adding auxiliary atomic BRs, we can always deform the occupied bands into a set of 1D and 2D $C_2$-BRs induced from the $1a$ position. 
Now we explicitly construct the adiabatic deformations for the $C_2$-BRs $s@2c$, $p@2c$, $2p@1b$, $s@1b$.
\begin{enumerate}[label=(\roman*)]
\item For a 2D block forming a $C_2$-BR $s@2c$, whose $C_4T$ sewing matrix $B^r(\kk)$ may or may not characterize a stable topology, we couple it to an auxiliary $C_4T$-BR $p@2c$, whose $B^r(\kk)$ matrix is atomic, such that the four bands $[s@2c]\oplus [p@2c]$ are equivalent to a $C_2$-BR $[s@1a]\oplus [s@1a] \oplus [2p@1a]$. 
To be specific, we write the $C_2$ sewing matrix of $[s@2c]\oplus [p@2c]$ as
\begin{equation}
    D(\kk) = \mathrm{diag}\begin{pmatrix}
        e^{-\ii k_x} & e^{-\ii k_y} & - e^{-\ii k_x} &  - e^{-\ii k_y}
    \end{pmatrix}
\end{equation}
Notice that a gauge transformation to the Bloch state basis $\ket{\psi_{n}'(\kk)} = \ket{\psi_m(\kk)} U_{mn}(\kk)$ changes the $C_2$ sewing matrix to $ D'(\kk) = U^\dagger(C_2\kk) \cdot D(\kk) \cdot U(\kk)$. 
We find the gauge transformation  
\begin{equation} \label{eq:gauge-U1}
    U(\kk) = \begin{pmatrix}
        \frac{1+e^{\ii k_x}}2 & 0  & \frac{1-e^{\ii k_x}}2 & 0 \\
        0 & \frac{1+e^{\ii k_y}}2 & 0 & \frac{1-e^{\ii k_y}}2 \\
        \frac{1-e^{\ii k_x}}2 & 0  & \frac{1+e^{\ii k_x}}2 & 0 \\
        0 & \frac{1-e^{\ii k_y}}2 & 0 & \frac{1+e^{\ii k_y}}2
    \end{pmatrix}
\end{equation}
changes the $C_2$ sewing matrix to that of a $C_2$-BR $[s@1a]\oplus [s@1a] \oplus [2p@1a]$: 
\begin{equation}
    D'(\kk) = U^\dagger(C_2\cdot \kk) \cdot D(\kk) \cdot U(\kk) 
    = \mathrm{diag} \begin{pmatrix}
        1 & 1 & -1 & -1
    \end{pmatrix}\ .
\end{equation}
Thus, the adiabatic deformation shown in \cref{fig:sewing}(c) and (d) can be realized by the Hamiltonian
\begin{align} \label{eq:adiabatic}
 (1-\delta) \cdot \sum_{\kk} \sum_{n,m=1}^4 h_{nm}(\kk) \ket{\psi_{n}(\kk)}\bra{\psi_m(\kk)} 
 +\delta \cdot \sum_{\kk} \sum_{n,m=1}^4 h_{nm}'(\kk) \ket{\psi_{n}'(\kk)}\bra{\psi_m'(\kk)} \ ,
\end{align}
where $h(\kk)$ and $h'(\kk)$ give the bands shown in \cref{fig:sewing}(c) and (d), respectively, and $\delta$ is continuously turned from 0 to 1. 
\item For a 2D block forming a $C_2$-BR $p@2c$, we can add an auxiliary $C_4T$-BR  $s@2c$ and apply the same $U$ (\cref{eq:gauge-U1}) to transform them to a $C_2$-BR $[s@1a]\oplus [s@1a] \oplus [2p@1a]$. Then an adiabatic deformation similar to \cref{eq:adiabatic} can be introduced. 
\item For a 2D block forming a $C_2$-BR $2p@1b$, we can add an auxiliary $C_4T$-BR $[s@1b]\oplus [s@1b]$ and transform them to a $C_2$-BR $[2p@1a]\oplus [s@1a]\oplus [s@1a]$. 
To be specific, we write the $C_2$ sewing matrix of $[2p@1b]\oplus [s@1b]\oplus [s@1b]$ as
\begin{equation}
    D(\kk) = \mathrm{diag}\begin{pmatrix}
        -e^{-\ii (k_x+k_y)} & -e^{-\ii (k_x+k_y)} & e^{-\ii (k_x+k_y)} &  e^{-\ii (k_x+k_y)}
    \end{pmatrix}\ .
\end{equation}
We find the gauge transformation 
\begin{equation}\label{eq:gauge-U2}
    U(\kk) = \begin{pmatrix}
        \frac{1+e^{\ii (k_x+k_y)}}2 & 0  & \frac{1-e^{\ii (k_x+k_y)}}2 & 0 \\
        0 & \frac{1+e^{\ii (k_x+k_y)}}2 & 0 & \frac{1-e^{\ii (k_x+k_y)}}2 \\
        \frac{1-e^{\ii (k_x+k_y)}}2 & 0  & \frac{1+e^{\ii (k_x+k_y)}}2 & 0 \\
        0 & \frac{1-e^{\ii (k_x+k_y)}}2 & 0 & \frac{1+e^{\ii (k_x+k_y)}}2
    \end{pmatrix}
\end{equation}
changes the $C_2$ sewing matrix to that of a $C_2$-BR $[2p@1a]\oplus [s@1a]\oplus [s@1a]$: 
\begin{equation}
    D'(\kk) = U^\dagger(C_2\cdot \kk) \cdot D(\kk) \cdot U(\kk) 
    = \mathrm{diag} \begin{pmatrix}
        -1 & -1 & 1 & 1
    \end{pmatrix}\ .
\end{equation}
An adiabatic deformation similar to \cref{eq:adiabatic} can be introduced then. 
\item For a 1 D block forming a $C_2$-BR $s@1b$, we can add an auxiliary $C_4T$-BR $[s@1b]\oplus[2p@1b]$ and apply the same $U$ (\cref{eq:gauge-U2}) to transform into a $C_2$-BR $[2p@1a]\oplus [s@1a]\oplus [s@1a]$.  Then an adiabatic deformation similar to \cref{eq:adiabatic} can be introduced. 
\end{enumerate}

Therefore, in terms of the $C_2$ sewing matrix, the occupied bands can always be deformed into a set of $C_2$-BRs induced from the $1a$ position, whose $C_2$ sewing matrices are $\kk$-independent (\cref{eq:constant-D}).

\clearpage
\section{Topological crystal \label{sec:tcs}}

In this section, we argue that the single-valued magnetic space group $P4'$, where $(C_4T)^4=1$, protects no clean topological crystalline insulator (TCI) states following the topological crystal method.

\begin{figure}[h]
\centering
\includegraphics[width=1\linewidth]{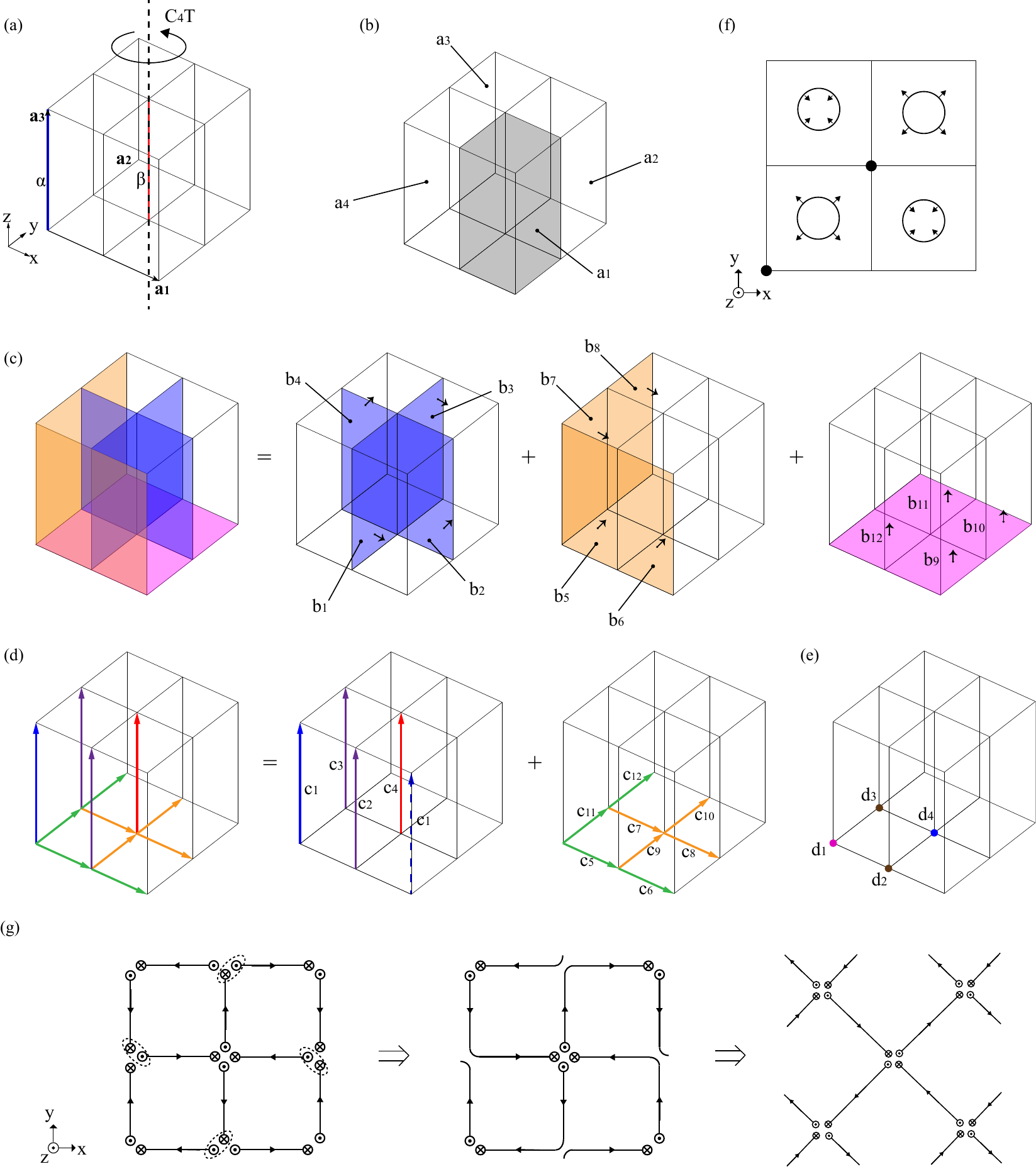}
\caption[]{ 
    (a) A unit cell of the cell-complex structure with magnetic space group $P4'$. $\boldsymbol{a_1}, \boldsymbol{a_2}, \boldsymbol{a_3}$ are the three basis vectors. $\alpha, \beta$ are the two $C_4T$ axes in the unit cell. In the paragraph, ``$C_4T$ operation'' refers to the one shown by the arrow unless otherwise specified.
    (b) The four 3-cells in the unit cell. The 3-cell $a_1$ is shown in grey, $a_{2, 3, 4}$ are transformed from $a_1$ by $C_4T$ around $\beta$ axis.
    (c) The twelve 2-cells in the unit cell.
    (d) The twelve 1-cells in the unit cell. The dashed blue line is the $\boldsymbol{a_1}$ translation of the solid blue line, so both are labeled $c_1$. The positive orientation of 2-cells and 1-cells are either $+x, +y$ or $+z$, as shown by the arrows in (c) and (d).
    (e) The four 0-cells in the unit cell.
    The $n$-cells in (c-e) with same color are equivalent up to crystalline symmetry operations.
    (f) (upper right) By creating a Chern bubble (the oriented circle) with $C = -1$ in the AU (3-cell $a_1$) and its symmetry partners and expanding the Chern bubbles to the boundaries of 3-cells, we arrive at the decoration with $C = 2$, demonstrating the ``bubble equivalence" of $C \in$ even to $C=0$. 
    (g) The $C=1$ decoration is equivalent to Fig.~1(c) in the main text up to gapping paired chiral edges and a spatial deformation.
   }
\label{fig:cell_complex_C4T}
\end{figure}

Let us first introduce the real space cell decomposition for $P4'$.
Following Ref.~\cite{TCS_1}, an asymmetric unit (AU) is defined as a largest connected open region in three dimensional space that has no overlap with its symmetry partners. The choice of AU is not unique, but we can always choose it as a convex polyhedron. The AU is then copied throughout space using crystalline symmetry operations, and the resulting nonoverlapping union of the copies is denoted $\mathcal{A}$. The complement of the open set $\mathcal{A}$ in three dimensional space is a two-skeleton denoted $X^2$. A 3-cell is one of the copies of AU in $\mathcal{A}$. A 2-cell is an oriented open segment of a plane separating two 3-cells, which does not overlap with its symmetry partners. Similarly, 1-cells are oriented open segments of lines between different 2-cells that does not overlap with its symmetry partners, and 0-cells are points joining different 1-cells. The construction of $n$-cells ($0 \leq n \leq 3$) gives the three dimensional space a cell complex structure. A unit cell with $C_4 T$ symmetry is shown in \cref{fig:cell_complex_C4T}(a). For simplicity, we define the positive orientation of each 2-cell and 1-cell to be either $+x, +y$ or $+z$. The red and blue lines $\alpha, \beta$ are the two $C_4T$ centers in the unit cell. The three basis lattice vectors are $\boldsymbol{a_1}, \boldsymbol{a_2}, \boldsymbol{a_3}$ respectively. \cref{fig:cell_complex_C4T}(b-e) show the cell decomposition of the unit cell. There are four 3-cells, twelve 2-cells, twelve 1-cells, and four 0-cells in each unit cell, which are labeled by $a_i, b_j, c_k, d_l, 1\leq i, l\leq 4, 1\leq j, k\leq 12$ respectively. There are also $n$-cells in the figure that belong to other unit cells, and are thus lattice translations of the labeled $n$-cells mentioned above. We assume translational symmetry in this section, and treat these $n$-cells in the same way as their translation partners. For example, the 1-cell shown by the dashed blue line in the second row of \cref{fig:cell_complex_C4T}(d) is also labeled $c_1$ since it can be obtained from the 1-cell $c_1$ in the unit cell (the solid blue line) by a lattice translation $\boldsymbol{a_1}$.

Now let us consider how the $n$-cells are transformed to each other by crystalline symmetries. 
Apparently, the $C_4T$ operation transforms $a_1 \to a_2 \to a_3 \to a_4 \to a_1$ (\cref{fig:cell_complex_C4T}(b)), so all 3-cells in the lattice are symmetry equivalent. 
For the 2-cells (\cref{fig:cell_complex_C4T}(c)), $C_4T$ operation transforms $b_1 \to b_2 \to -b_3 \to -b_4 \to b_1$, $b_5 \to -b_7 \to -b_6 \to b_8 \to b_5$, and $b_9 \to b_{10} \to b_{11} \to b_{12} \to b_9$. It is apparent that neither $C_4T$ nor lattice translation can transform the above three cycles to each other, so there are three distinct symmetry equivalent classes of 2-cells: $b_{1,2,3,4}, b_{5,6,7,8}$, and $ b_{9,10,11,12}$.
Now let us consider the 1-cells (\cref{fig:cell_complex_C4T}(d)). $c_1$ and $c_4$ lie on the two $C_4T$ axes $\alpha, \beta$ respectively. 
The relative position between them is $\frac {1} {2} (\boldsymbol{a_1} + \boldsymbol{a_2})$, which is not a lattice vector. It is also apparent that they cannot be transformed to each other by the $C_4T$ operation. They are also not symmetry equivalent to the other 1-cells since they lie on the only two $C_4T$ axes, so each of them forms a symmetry equivalent class by itself. The two other 1-cells parallel to $z$ direction are $c_2$ and $c_3$. $C_4T$ operation transforms $c_3 \to c_2$. Since they are obviously not symmetry equivalent to the 1-cells parallel to the $xy$ plane, $c_{2,3}$ form a symmetry equivalent class. Among the 1-cells parallel to the $xy$ plane, $C_4T$ operation transforms $c_5 \to c_{11} \to -c_6 \to -c_{12} \to c_5$, and $c_9 \to -c_8 \to -c_{10} \to c_7 \to c_9$, but cannot relate the above two cycles. Therefore, $c_{5,6,11,12}$ and $c_{7,8,9,10}$ form two distinct symmetry equivalent classes of 1-cells. 
In summary, the twelve 1-cells can be divided into five different symmetry equivalent classes: $c_1, c_4, c_{2,3}, c_{5,6,11,12}$ and $c_{7,8,9,10}$. The four 0-cells $d_{1,2,3,4}$ (\cref{fig:cell_complex_C4T}(e)) are ends of the four 1-cells that are parallel to $z$ direction $c_{1,2,3,4}$ respectively, so following above discussions, they can be divided into three different symmetry equivalent classes: $d_1, d_4, d_{2,3}$. 
In summary, there are $1, 3, 5, 3$ distinct symmetry equivalent classes of $3,2,1,0$-cells respectively. In \cref{fig:cell_complex_C4T}(c-e), symmetry equivalent $n$-cells are shown in the same color to emphasize their relationships.

Following Refs.~\cite{song_topological_2017,TCS_0,TCS_1}, a topological crystal is defined as a gapped decoration of lower-dimensional TIs on the $n$-cells ($n \leq 2$) that respects the crystalline symmetry. 
Now let us repeat the argument that a TCI state can always be adiabatically connected to a topological crystal. 
Since there is no 3D strong TIs protected by local symmetry in the considered symmetry class A, one can always trivialize the AU by deforming it into a set of occupied local orbitals.  
This is always accessible provided $\xi \ll a_{AU}$, where $\xi$ is the characteristic correlation length of the TCI ground state, and $a_{AU}$ is the linear size of the AU. 
Since TCI ground states are short range correlated, it is reasonable to expect that by adding a fine enough mesh of trivial degrees of freedom in the AU, $\xi$ can be made as small as required, while $a_{AU}$ remains unchanged. $\xi \ll a_{AU}$ is thus satisfied, and the desired adiabatic path can be found. The adiabatic path is then copied throughout space using crystalline symmetry to trivialize the open set $\mathcal{A}$. It is obvious that such a copy always respects the crystalline symmetry. After the above adiabatic deformation, the topologically nontrivial elements can only lie on the two-skeleton $X^2$, and can be decomposed to lower-dimensional TIs decorated on $n$-cells, $n\leq 2$. 
Ref.~\cite{TCS_1} shows that a full classification of all TCIs with spin-orbit coupling and time reversal symmetry can be obtained by topological crystals, proving the validity of the above argument. 
Since symmetry class A has no 0- and 1-dimensional TIs, we only consider decorations of Chern insulators on the 2-cells.

In a cell complex structure, the boundary of an $n$-cell can be expanded in terms of oriented $(n-1)$-cells. 
For examples: 
$\partial a_1 = b_7 + b_2 -b_1 - b_6$ for 3-cells, $\partial b_1 = -c_2 + c_4$, $\partial b_5 = c_1 - c_2$, and $\partial b_{9} = c_6 + c_{11} -c_8 -c_9$ for 2-cells, $\partial c_1 = d_1 - d_1 = 0$, $\partial c_2 = 0$, $\partial c_4 = 0$, $\partial c_5 = -d_1 + d_2$, and $\partial c_7 = -d_3 + d_4$ for 1-cells. 
Such boundary mappings will be used for the gluing conditions and bubble equivalence of the topological crystal.

Concerning decorations of Chern insulators on 2-cells, the gluing condition requires an equal number of chiral modes propagating in opposite directions on each 1-cell so that it is possible for them to be locally gapped. Here we show that such conditions, together with the crystalline symmetry, make restrictions on possible decorations. A decoration of Chern insulators on 2-cells in \cref{fig:cell_complex_C4T}(c) is described by twelve integers $C_{b_j} \in \mathbb{Z}, 1\leq j \leq 12$, which are the Chern numbers of decorated Chern insulators on the 2-cells $b_j$, with the same orientation as discussed above. Using the crystalline symmetry and the fact that time reversal operation reverses the Chern number, the description can be simplified to three integers $C_1, C_2, C_3 \in \mathbb{Z}$: $C_{b_1}= -C_{b_2} = -C_{b_3} = C_{b_4} = C_1$, $C_{b_5} = C_{b_7} = -C_{b_6} = -C_{b_8} = C_2$, and $C_{b_9} = -C_{b_{10}} = C_{b_{11}} = -C_{b_{12}} = C_3$. 
Since there are five different symmetry equivalent classes of 1-cells, there are also five symmetry inequivalent gluing conditions (for $c_1, c_2, c_4, c_5, c_7$ respectively):
\begin{equation}
    \begin{aligned}
        C_{b_5}-C_{b_7}-C_{b_6}+C_{b_8} = 0\\
        -C_{b_1}+ C_{b_3} + C_{b_6} - C_{b_5} =0\\
        C_{b_1} + C_{b_2} -C_{b_3} -C_{b_4} = 0\\
        C_{b_{12}} - C_{b_{11}} = 0\\
        -C_{b_{12}} + C_{b_{11}} = 0
    \end{aligned}
\end{equation}
where we have used translational symmetry implicitly. The first and third rows are automatically satisfied, the second row requires $C_1 = -C_2$, and the last two rows require $C_3 = 0$. 
 We hence conclude that decorations of Chern insulators with $C_{b_1}= -C_{b_2} = -C_{b_3} = C_{b_4} = -C_{b_5} = C_{b_8} = C_{b_6} = -C_{b_7} = C \in \mathbb{Z}$, $C_{b_9} = C_{b_{10}} = C_{b_{11}} = C_{b_{12}} = 0$ are the only possible decorations that satisfy both the crystalline symmetry and gluing conditions.

Since Chern insulators have a $\mathbb{Z}$ classification, one can make a decoration with $C$ any nonzero integer. However, here we show that some of the decorations can be adiabatically connected to the trivial $C = 0$ case by creating Chern bubbles from vacuum and locally annihilating Chern insulators with opposite Chern numbers, a process called ``bubble equivalence'' in Ref.~\cite{TCS_1}. 
Create a bubble of a $C=-1$ state within the AU as well as its symmetry partners, as shown in \cref{fig:cell_complex_C4T}(f). Expand these bubbles until they are touching the 2-cells, the resulted decoration is given by
\begin{equation}
-\partial a_1 + \partial a_2 - \partial a_3 + \partial a_4 =  2b_1 -2b_2 -2b_3 +2 b_4 - 2b_5 - 2b_7 + 2b_6 + 2b_8\ .
\end{equation}
Here the integer coefficients of the 2-cells represent the corresponding Chern numbers. 
The above equation gives a $C=2$ decoration.
Creating bubbles with other Chern numbers yield all $C\in$ even decorations. 
Therefore, after modulo the bubble equivalence, we obtain a $\mathbb{Z}_2$ classification represented by the $C=0,1$ decorations.

Now we show that the $C=1$ decoration is equivalent to the decoration in Fig.~1(c) in the main text. 
Unlike $c_1, c_4$, the four chiral modes on $c_2, c_3$ have only $C_2$ symmetry, and can be paired and gapped. Consider the dimerization shown in \cref{fig:cell_complex_C4T}(g), which effectively joins Chern insulators decorated on $b_5$ with $b_1$, $b_3$ with $b_6$, $b_4$ with $b_8$, and $b_2$ with $b_7$. After a deformation respecting $C_4T$ symmetry, we arrive at the same decoration as shown in Fig.~1(c) of the main text, whose low energy physics is realized by the lattice model.

The above discussion seems apply to both $(C_4T)^4=1$ and $-1$ cases. 
However, when $(C_4T)^4=1$, the $C=1$ decoration must be gapless according to the discussion in the main text. 
Therefore, no gapped topological state can be constructed in the framework of topological crystal, suggesting a trivial classification in the clean limit.

We now show that the $C=1$ decoration can be gapped if $(C_4T)^4=-1$. 
Consider four chiral modes $1,2,3,4$ on a hinge, which are transformed as $1 \to 2\to 3\to 4 \to 1$ by the $C_4T$ operation, and are propagating in $+z, -z, +z, -z$ direction respectively with velocity $v$. Let us span the four-by-four Hilbert space with two sets of Pauli matrices $\sigma_{0,x,y,z}, \tau_{0,x,y,z}$, where $\{\tau_z, \sigma_z\} = \{+,+\}, \{+,-\}, \{-,+\}, \{-,-\}$ denote chiral mode $1,2,3,4$ respectively. The four-by-four Hamiltonian of the four modes is
\begin{equation}
    H(k_z) = v k_z \tau_0 \sigma_z
\end{equation}
Consider the $(C_4T)^4 = -1$ symmetry $C_4 T = D(C_4T)K$, where $K$ is the complex conjugation, and
\begin{equation}
    D(C_4T) = \left(\begin{array}{cccc}
0 & 0 & 0 & 1 \\
i & 0 & 0 & 0 \\
0 & -1 & 0 & 0 \\
0 & 0 & -i & 0
\end{array}\right), \qquad
 D(C_4T){D^*(C_4T)} D(C_4T){D^*(C_4T)}= -1
\end{equation}
Since ${D(C_4T)}H^{*}(k_z) D^{\dagger}(C_4T) = H(-k_z)$, ${D(C_4T)}(\tau_x \sigma_x)^{*} D^{\dagger}(C_4T) = \tau_0\sigma_y$, and ${D(C_4T)}(\tau_0 \sigma_y)^{*} D^{\dagger}(C_4T) = \tau_x\sigma_x$, a mass term $m(\tau_x\sigma_x +\tau_0 \sigma_y)$ can be added to $H(k_z)$ without breaking the $(C_4T)^4 = -1$ symmetry. Since $\tau_0\sigma_z, \tau_x\sigma_x$ and $ \tau_0\sigma_y$ anti-commute with one another, the two pairs of energy bands are $E_{\pm}(k_z) = \pm \sqrt{(vk_z)^2 +  2m^2}$.
This allows the construction of a topological crystal for a clean axion insulator if $(C_4T)^4=-1$,  consistent with the previous discussions in Refs.~\cite{schindler_higher-order_2018,li_pfaffian_2020}.

\clearpage
\section{The lattice model}
\label{sec:lattice-model}

\subsection{Model and symmetry}

\begin{figure}[h]
\centering
\includegraphics[width=1\linewidth]{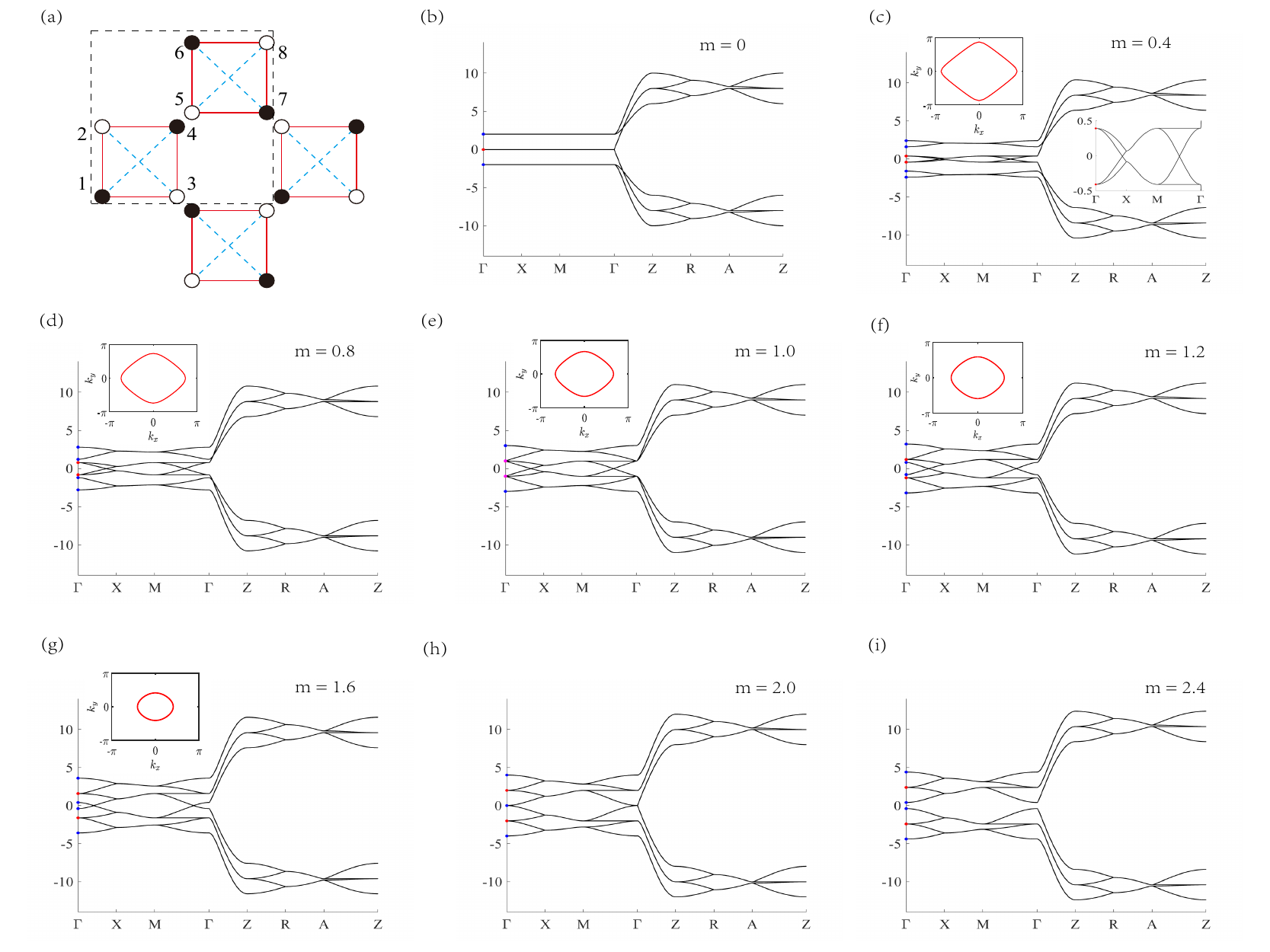}
\caption[]{ 
    (a). The lattice model with eight orbitals per unit cell. Orbitals 6, 7, 8 in unit cells $(010)$, $(100)$, and $(110)$ (rather than that in home unit cells) are shown for the convenience of illustration. In each cell, $\rho = \pm 1$ labels a square (e.g. $\rho = 1 \to$ sites $\{1, 2, 3, 4\}$), $\tau = \pm 1$ labels a column (e.g. $\rho = 1, \tau = 1 \to$ sites $\{1, 2\}$), and $\sigma = \pm 1$ labels a site (e.g. $\rho = 1, \tau = 1, \sigma = 1 \to$ site 1).
    (b-i). The band structure of the lattice model with $t=2, \gamma=1, \lambda=0.01$ and different $m$. Upper left insets of (c-g) show the position of the nodal-line on the $k_z = 0$ plane. Right inset of (c) shows the details of low-energy band structure at $k_z = 0$. Blue and red dots at $\Gamma$ label $C_2$ eigenstates with eigenvalue $\pm 1$ respectively. When $m \ll \gamma$, the low energy band structure is dominated by the $C_2 = -1$ sector.
    }
\label{fig:F_BAND}
\end{figure}

Here we give a complete description of the lattice model given in Eq.~(2) in the main text. As shown in \cref{fig:F_BAND}(a), a unit cell has 8 orbitals.
We use three sets of Pauli matrices $\rho_{0,x,y,z}$, $\tau_{0,x,y,z}$, $\sigma_{0,x,y,z}$ to span the eight-by-eight Hilbert space. 
$\rho_z=1,-1$ correspond to the two squares in a unit cell, $\tau_z=1,-1$ further labels the two columns in each square, and $\sigma_z=1,-1$ labels the two sites within each column. 
The Hamiltonian in momentum space takes the form
\begin{equation} \label{eq:Hk}
\begin{aligned}
 H(\kk) & = (m + 2t - 2t \cos{k_z}) \rho_x \tau_x \sigma_x + 2t\sin{k_z} \rho_z\tau_z\sigma_z \\
 & + \gamma (\cos{\frac{k_x} {2}} \rho_0 \tau_x \sigma_0 -\sin{\frac{k_x} {2}} \rho_0 \tau_y \sigma_0 +
 \cos{\frac{k_y} {2}} \rho_0 \tau_0 \sigma_x -
 \sin{\frac{k_y} {2}} \rho_0 \tau_0 \sigma_y) \\
 & + \lambda (\cos{\frac{k_x}{2}}\cos{\frac{k_y}{2}} \rho_0\tau_x\sigma_x + \sin{\frac{k_x}{2}}\sin{\frac{k_y}{2}} \rho_0\tau_y\sigma_y
 - \sin{\frac{k_x}{2}}\cos{\frac{k_y}{2}} \rho_0\tau_y\sigma_x
 - \cos{\frac{k_x}{2}}\sin{\frac{k_y}{2}} \rho_0\tau_x\sigma_y)
 \end{aligned}\ .
\end{equation}
The first row of \cref{eq:Hk} defines four pairs of chiral and anti-chiral states in the $z$-direction. 
For example, the 1st and 8th orbitals span the Hamiltonian $(m + 2t - 2t \cos k_z) \sigma_x + 2t \sin k_z \sigma_z $, where $\sigma_z=1,-1$ correspond to 1st and 8th orbitals, respectively. 
When $m$ is small, it reduces to a Dirac Hamiltonian around $k_z=0$:
$ (m + t k_z^2 ) \sigma_x + 2t k_z \sigma_z + \mathcal{O}(k_z^3)$, and has a large gap around $k_z=\pi$. 
Thus, the 1st and 8th orbitals simulate a pair of anti-chiral and chiral edge modes of a Chern block shown in Fig.~1(c) of the main text. 
$m$ is the coupling between two edges, and it should approach zero if the Chern block is sufficiently large. 
In this work we use $m$ as a tuning parameter. 
Similarly, 2nd and 7th orbitals, 3rd and 6th orbitals, 4th and 5th orbitals form the other three pairs of anti-chiral and chiral states, respectively. 
$\gamma$ in the second row of \cref{eq:Hk} is the hopping between nearest neighbors within each square. 
$\lambda$ in the third row of \cref{eq:Hk} is the hopping along the diagonals of each square. 

We have adopt a symmetric gauge for \cref{eq:Hk}.
In this gauge, the eight Bloch bases are defined as 
\begin{equation}
    \ket{\phi_{\alpha,\kk}} = \frac1{\sqrt{N}} \sum_{\RR} e^{\ii \kk\cdot(\RR+\tt_\alpha)} \ket{\RR,\alpha}
\end{equation}
where $\ket{\RR \alpha}$ ($\alpha=1\cdots 8$) is a local orbital locating at $\RR+\tt_\alpha$, $N$ is the number of unit cells, $\RR$ sums over lattice vector, and $\tt_\alpha$ is the sublattice vector for the $\alpha$-th orbital. 
We choose 
\begin{equation}
\begin{aligned}
& \tt_1=(\delta,\delta,0),\quad
\tt_2=(\delta,\frac12-\delta,0),\quad
\tt_3=(\frac12-\delta,\delta,0),\quad
\tt_4=(\frac12-\delta,\frac12-\delta,0), \\
& \tt_5=(\frac12+\delta,\frac12+\delta,0),\quad
\tt_6=(\frac12+\delta,-\delta,0),\quad
\tt_7=(-\delta,\frac12+\delta,0),\quad
\tt_8=(-\delta,-\delta,0)\ , 
\end{aligned}
\end{equation}
where $\delta>0$ is a small quantity. 
In \cref{fig:F_BAND} we use a finite $\delta$ for illustration. 
But in practical calculation we take the $\delta\to 0$ limit for simplicity. 
$H(\kk)$ is the Hamiltonian matrix on the Bloch basis, {\it i.e.}, $H_{\alpha\beta}(\kk) = \inn{\phi_{\alpha,\kk}|\hat{H}|\phi_{\beta,\kk}}$.
The $\cos\frac{k_{x,y}}2$ and $\sin\frac{k_{x,y}}2$ factors in \cref{eq:Hk} come from phase factors $e^{\ii \kk\cdot(\tt_\alpha - \tt_\beta)}$ appearing along hoppings between the $\alpha$-th and $\beta$-th orbitals. 
The Bloch basis satisfies a twisted boundary condition over the Brillouin zone
\begin{equation} \label{eq:embedding-matrix}
    \ket{\phi_{\alpha,\kk+\GG}} =  \ket{\phi_{\beta,\kk}} V_{\beta\alpha}(\GG),\qquad 
    V_{\beta\alpha}(\GG) = \delta_{\alpha\beta} e^{\ii \GG\cdot \tt_\alpha} \ ,
\end{equation}
where implicit summation over repeated indices is assumed. 
It follows that $H(\kk)$ satisfies 
\begin{equation}
    H(\kk+\GG) = V^{\dagger}(\GG) H(\kk) V(\GG)\ . 
\end{equation}
$V(\GG)$ is usually referred to as the embedding matrix. 

Now we summarize the symmetries of $H(\kk)$.
First, we show that $H(\kk)$ breaks the time-reversal symmetry (TRS). 
As explained in the above two paragraphs, the orbitals $\ket{\mathbf{0},1}$ $\ket{\mathbf{0},8}$ mimic a pair of anti-chiral and chiral modes at the same position. 
Thus, TRS, if existed, must interchange $\ket{\mathbf{0},1}$ and $\ket{\mathbf{0},8}$. 
For the same reason, TRS must send $\ket{\RR, \alpha}$ to $\ket{\RR, 9-\alpha}$. 
Consider the hopping $\lambda$ between $\ket{\mathbf{0},1}$, $\ket{\mathbf{0},2}$, TRS would transform it to a hopping between $\ket{\mathbf{0},8}$, $\ket{\mathbf{0},7}$, which does not exist in the model (\cref{fig:F_BAND}(a)). 
The model instead has a hopping between $\ket{\mathbf{0},7}$ and $\ket{\mathbf{010},8}$, wherein the latter is in another unit cell. 
We find that $H(\kk)$ enjoys a symmetry group generated by 
\begin{equation} \label{eq:inversion}
H(\kk) = \mD(P) \cdot H(-\kk) \cdot \mD^{\dagger}(P),\qquad \mD(P) = \rho_x\tau_x\sigma_x ,\qquad P = \{ -1 | \mathbf{0} \}\ ,
\end{equation}
\begin{equation}
H(-k_y, k_x, k_z) = \mD( \td C_4) \cdot H(\kk) \cdot \mD^{\dagger}(\td C_4),\qquad 
\mD(\td C_4) = \rho_x \left(
\begin{array}{cccc}
 0 & 0 & 1 & 0 \\
 1 & 0 & 0 & 0 \\
 0 & 0 & 0 & 1 \\
 0 & 1 & 0 & 0 \\
\end{array}
\right),\qquad 
\td C_4 = \{4_{001} | 0\frac12 0\}\ , 
\end{equation}
\begin{equation}
H(-k_x, k_y,k_z) = \mD( \td M_x) \cdot H(\kk) \cdot \mD^\dagger(\td{M}_x),\qquad 
\mD( \td M_x) = \rho_x\tau_x\sigma_0,\qquad 
\td M_x = \{m_{100}| 0 \frac12 0\}
\end{equation}
\begin{equation} \label{eq:magnetic-translation}
H(-\kk) = \mD(\td{T}) \cdot H^*(\kk) \cdot \mD^\dagger(\td{T}),\qquad 
\mD(\td{T}) = \rho_x \tau_0 \sigma_0,\qquad   
\td{T} = \{1'|\frac12\frac12 0\}\ . 
\end{equation}
Here $\{p^{(\prime)}|\boldsymbol{\tau}\}$ is the Seitz symbol for spatial operations, and a prime in the superscript of $p$ represent a time-reversal operation following the point group operation $p$. 
We adopt the convention of the \href{https://www.cryst.ehu.es/cgi-bin/cryst/programs/magget_gen.pl}{MGENPOS} program of the Bilbao Crystallographic Server \cite{gallego_magnetic_2012,perez-mato_symmetry-based_2015}. 
$P$ is an inversion centered at the origin, $\td C_4$ is a four-fold rotation centered at $(-\frac14 \frac14 0)$, $\td M_x$ is a glide mirror with respect to the plane $(0,y,z)$, and $\td{T}$ is a magnetic translation. 
These operations generate the single-valued magnetic space group $P_C 4/nbm$ (\#125.373 in the BNS setting). 
This group also has $C_4T = \{1| 0 \bar1 0 \} \cdot \td T \cdot \td C_4 = \{4_{001}'|\frac12 0 0\}$, $S_4 = P \cdot \td C_4^3 =\{ - 4_{001}^- | \frac12 0 0 \} $, $M_x T = \{ 1|0\bar1 0\} \cdot \td T \cdot \td M_x = \{m_{100}'| \frac12 0 0\}$, $M_zT = \{ 1|\bar1 0 0\} \cdot \td{T} \cdot P \cdot \td{C}_4^2 = \{m_{001}'|000\}$, $\td{M}_z = \td{T}\cdot M_z T = \{m_{001}|\frac12,\frac12,0\}$ symmetries. 
Both the $C_4T$ center and the $S_4$ center locate at $(\frac14,\frac14,0)$. 
The mirror planes of $M_xT$ and and $M_zT$ are $(\frac14,y,z)$ and $(x,y,0)$, respectively. 
$\td M_z$ is a glide mirror with respect to the $(x,y,0)$ plane. 
For later convenience, we also derive the representation of $S_4$ 
\begin{equation} \label{eq:S4}
    \mathcal{D}(S_4) = \mathcal{D}(P) \cdot \mathcal{D}^3(\td C_4)  
= \rho_0 \begin{pmatrix}
 0 & 0 & 1 & 0 \\
 1 & 0 & 0 & 0 \\
 0 & 0 & 0 & 1 \\
 0 & 1 & 0 & 0 \\
\end{pmatrix}\ .
\end{equation}

Here we also summarize the maximal Wykcoff positions and their magnetic point groups: 
\begin{equation}
2a:\quad (\frac14,\frac14,0) ,\quad (\frac34,\frac34,0),\quad 4'/m'm'm,
\qquad\qquad 
2b:\quad (\frac14,\frac14,\frac12) ,\quad (\frac34,\frac34,\frac12),\quad 4'/m'm'm\ ,
\end{equation}
\begin{equation}
2c:\quad (\frac14,\frac34,0) ,\quad (\frac34,\frac14,0),\quad 4/m'm'm',
\qquad\qquad 
2d:\quad (\frac14,\frac34,\frac12) ,\quad (\frac34,\frac14,\frac12),\quad 4/m'm'm'\ ,
\end{equation}
\begin{equation}
4e: \quad (0,0,\frac12),\quad (\frac12,0,\frac12),\quad (0,\frac12,\frac12),\quad (\frac12,\frac12,\frac12),\qquad m'.mm'\ ,
\end{equation}
\begin{equation}
4f: \quad (0,0,0),\quad (\frac12,0,0),\quad (0,\frac12,0),\quad (\frac12,\frac12,0),\qquad m'.mm'\ . 
\end{equation}
In our convention, the origin is shifted by $(\frac12,0,0)$ from that on the Bilbao Crystallographic Server.

Let us examine whether \cref{eq:Hk} has a chiral symmetry: $S\cdot H(\kk) \cdot S = - H(\kk)$ with $S$ being a unitary matrix  satisfying $S^2=1$. 
For $S$ to anti-commute with the $\rho_0\tau_0\sigma_{x,y}$ terms in the second row of \cref{eq:Hk}, $S$ must be proportional to $\sigma_z$. 
Similarly, for $S$ to anti-commute with the $\rho_0\tau_{x,y}\sigma_{0}$ terms in the second row, $S$ must be proportional to $\tau_z$. 
Then, $S$ could be a linear combination of $\rho_{0,x,y,z}\tau_{z}\sigma_z$, all of which commute with the $\lambda$ terms of \cref{eq:Hk}. 
If $\lambda=0$, $H(\kk)$ has an accidental chiral $S=\rho_y\tau_z\sigma_z$. 
Being off-diagonal in the $\rho_z=\pm1$ basis, $S$ is {\it non-local} in real space and hence will be broken by disorder potentials. 
Therefore, either a finite $\lambda$ or disorder potential will break the accidental chiral symmetry. 

Let us examine whether \cref{eq:Hk} has a particle-hole symmetry: $C\cdot H^*(\kk) \cdot C^\dagger = - H(-\kk)$ with $C$ being a unitary matrix satisfying $C\cdot C^*= \pm1$. 
Because of the $\td T$ symmetry of $H(\kk)$ (\cref{eq:magnetic-translation}), the existence of $C$ symmetry would automatically imply the presence of a chiral symmetry $S=\rho_x\cdot C$ (up to a U(1) phase factor), and vice versa. 
According to the discussions in the last paragraph, $H(\kk)$ would have an accidental $C=\rho_z\tau_z\sigma_z$ when $\lambda=0$. 
Either a finite $\lambda$ or on-site disorder potentials can break this accidental particle-hole symmetry.

In the following, we keep $t = 2, \gamma = 1, \lambda = 0.01$ fixed and discuss how $m$ changes the band structure of the system. For convenience, we define the high symmetry momenta
\begin{equation}
\begin{aligned}
&    \Gamma:\ (0,0,0),\qquad 
    \mathrm{X}:\ (\pi,0,0),\qquad 
    \mathrm{M}:\ (\pi,\pi,0) \nonumber\\
&   \mathrm{Z}:\ (0,0,\pi),\qquad 
    \mathrm{R}:\ (\pi,0,\pi),\qquad 
    \mathrm{A}:\ (\pi,\pi,\pi)\ . 
\end{aligned}
\end{equation}
and plot the energy bands along the high-symmetry line $\rm \Gamma\!-\!X\!-\!M\!-\!\Gamma\!-\!Z\!-\!R\!-\!A\!-\!Z$. Let us start with the $m = 0$ limit. 
As shown in \cref{fig:F_BAND}(b), the low-energy bands are quasi-1D around the $k_z = 0$ plane, which is consistent with the picture of decoupled helical modes explained below \cref{eq:Hk}. 
A small finite $m>0$ drives the system to a nodal line semi-metal phase, as illustrated in \cref{fig:F_BAND}(c). 
The nodal line is protected by the $P\td T$ symmetry and pinned in the $k_z=0$ plane by the glide symmetry $\td M_z = \{m_{001}|\frac12\frac12 0\}$. 
When $m$ increases, the nodal line shrinks towards the $\Gamma$ point (\cref{fig:F_BAND}(d-g)). At a critical value $m = 2\gamma$, the nodal line shrinks to a point at $\Gamma$ (\cref{fig:F_BAND}(h)). Further increasing $m$ will open a gap (\cref{fig:F_BAND}(i)). This gap is trivial since it is adiabatically connected to the $m \to \infty$ limit, which describes an atomic insulator with dimers formed by nearest neighbors. Now let us discuss the $m<0$ case. Rather similarly, a small $-2\gamma<m<0$ also drives the system to a nodal-line semi-metal, whose nodal-line is pinned on the $k_z = 0$ plane and shrinks as $|m|$ increases, finally becoming the $\Gamma$ point when $m = -2\gamma$. Tuning $m$ to $-2\gamma - \epsilon$, where $\epsilon$ is a small positive number, will also open a gap when $t>\gamma$. However, this state is not adiabatically connected to the $m\to -\infty$ limit. To see this, notice that when $k_z = \pi + \delta k_z, \delta k_z \ll \pi$, the first row of \cref{eq:Hk} can be expanded as $(m + 4t - t {\delta k_z}^2)\rho_x \tau_x \sigma_x - 2t \delta k_z \rho_z\tau_z\sigma_z$, which defines four pairs of chiral and anti-chiral states near $k_z = \pi$ when $m+4t$ is small (see the explanation below \cref{eq:Hk}). Similar to the above discussions, the band structure is quasi-1D on $k_z = \pi$ when $m = -4t$, and is a nodal-line semi-metal with a nodal line pinned on the $k_z = \pi$ plane when $0<|m+4t|<2\gamma$. In this paper, we take $t = 2 > \gamma=1$, which means there are no overlap between the two gapless regions $|m|<2\gamma$ and $|m+4t|<2\gamma$. Since the low energy physics of the above two nodal-line semi-metals are similar to each other, an intrinsic STI phase should emerge near $k_z = \pi$, when $|m+4t|\ll \gamma$ and $C_4T$ preserving disorder is introduced, similar to intrinsic STI phase near $k_z = 0$ when $|m|\ll \gamma$, as illustrated in Fig.~2(b) of the main text. Since our purpose is to demonstrate the existence of an intrinsic STI phase, we will focus on $|m|\ll \gamma$ and $k_z \approx 0$ in following discussions and numerical calculations.

\subsection{Low energy states}
\label{sec:low-energy-states}

When $m$ is zero, low energy states are the quasi-1D helical modes around $k_z=0$. 
In this subsection, we will derive a low energy k$\cdot$p expansion for the helical modes.
For a finite but small $m$, these modes form a nodal line semi-metal.  
We will also discuss the transition from the nodal line semi-metal to a trivial band insulator when $m$ becomes larger. 

We view $H(k_z=0)$ as a 2D Hamiltonian.
When $m=0$, it decomposes into decoupled molecules on the red squares shown in \cref{fig:F_BAND}(a). 
The Hamiltonian around $k_z=0$ reduces to 
{\small
\begin{align} \label{eq:Hk-decoupled}
H(\kk)=& 2t k_z \cdot \rho_z \tau_z \sigma_z + 
\rho_0 \begin{pmatrix}
    0 & \gamma e^{\ii \frac{k_y}2}  & \gamma e^{\ii \frac{k_x}2} & \lambda e^{\ii \frac{k_x+k_y}2} \\
    \gamma e^{-\ii\frac{k_y}2} & 0 & \lambda e^{\ii \frac{k_x-k_y}2} & \gamma e^{\ii \frac{k_x}2} \\
    \gamma e^{-\ii \frac{k_x}2} & \lambda e^{-\ii \frac{k_x-k_y}2} & 0 & \gamma e^{\ii \frac{k_y}2} \\
    \lambda e^{-\ii \frac{k_x+k_y}2} & \gamma e^{-\ii \frac{k_x}2} & \gamma e^{-\ii \frac{k_y}2} & 0 
\end{pmatrix} + \mathcal{O}(k_z^2) \nonumber\\
=& 2t k_z \cdot \rho_z \tau_z \sigma_z + \rho_0 \cdot \mathrm{diag}(\begin{array}{cccc} 1 & e^{\ii \frac{k_y}2}  & e^{\ii\frac{k_x}2} & e^{\ii   \frac{k_x+k_y}2} \end{array}) ^\dagger 
    \cdot \begin{pmatrix}
    0 & \gamma   & \gamma & \lambda \\
    \gamma  & 0 & \lambda & \gamma \\
    \gamma  & \lambda  & 0 & \gamma  \\
    \lambda  & \gamma  & \gamma  & 0 
    \end{pmatrix} \cdot 
    \mathrm{diag}(\begin{array}{cccc} 1 & e^{\ii \frac{k_y}2}  & e^{\ii\frac{k_x}2} & e^{\ii   \frac{k_x+k_y}2} \end{array})
    + \mathcal{O}(k_z^2) 
    \ ,
\end{align}}
where the $\kk$-independent matrix in the middle represent the local Hamiltonian of each molecule. 
We find the eigenvectors and eigenvalues at $k_z=0$: 
{\small
\begin{equation} \label{eq:eigenstate-1}
u_{\xi=-1,\eta=\pm1,n=1}(\kk) =
\frac12 
\begin{pmatrix}
    \frac{1+\eta}2 \\ \frac{1-\eta}2 \cdot e^{i\frac{k_x+k_y}2}
\end{pmatrix} \otimes 
\begin{pmatrix}
    1 \\ \ii  e^{-\ii \frac{k_y}2} \\ -\ii e^{-\ii \frac{k_x}2} \\ - e^{-\ii \frac{k_x+k_y}2}
\end{pmatrix},\qquad
E = -\lambda \ , 
\end{equation}
\begin{equation}
u_{\xi=-1,\eta=\pm1,n=2} (\kk)= 
\frac12
\begin{pmatrix}
    \frac{1+\eta}2 \\ \frac{1-\eta}2 \cdot e^{i\frac{k_x+k_y}2}
\end{pmatrix} \otimes 
\begin{pmatrix}
    1 \\ -\ii  e^{-\ii \frac{k_y}2} \\ \ii e^{-\ii \frac{k_x}2} \\ - e^{-\ii \frac{k_x+k_y}2}
\end{pmatrix},\qquad 
E = -\lambda \ ,
\end{equation}
\begin{equation}
u_{\xi=1,\eta=\pm1,n=1} (\kk)=
\frac12
\begin{pmatrix}
    \frac{1+\eta}2 \\ \frac{1-\eta}2\cdot e^{i\frac{k_x+k_y}2}
\end{pmatrix} \otimes 
\begin{pmatrix}
    1 \\ -  e^{-\ii \frac{k_y}2} \\ - e^{-\ii \frac{k_x}2} \\  e^{-\ii \frac{k_x+k_y}2}
\end{pmatrix},\qquad 
E = -2\gamma + \lambda \ , 
\end{equation}
\begin{equation}\label{eq:eigenstate-4}
u_{\xi=1,\eta=\pm1,n=2} (\kk)=
\frac12
\begin{pmatrix}
    \frac{1+\eta}2 \\ \frac{1-\eta}2\cdot e^{i\frac{k_x+k_y}2}
\end{pmatrix} \otimes 
\begin{pmatrix}
    1 \\  e^{-\ii \frac{k_y}2} \\  e^{-\ii \frac{k_x}2} \\  e^{-\ii \frac{k_x+k_y}2}
\end{pmatrix},\qquad 
E = 2\gamma + \lambda \ .
\end{equation}}
Here $\eta=\pm1$ indicates the $\rho_z=\pm1$ block, $\xi=\pm1$ is the $C_2$ ($=S_4^2$) eigenvalue of the molecule orbital. 
One can verify that these solutions satisfy 
\begin{equation}
    u_{\xi\eta n}(\kk+\GG) = V^\dagger(\GG) u_{\xi \eta n}(\kk)
\end{equation}
where $V(\GG)$ is the embedding matrix given in \cref{eq:embedding-matrix}. 

We have chosen $\gamma\gg \lambda>0$ such that $\xi=1$ states have a large gap $\sim 4\gamma$ around the $k_z=0$ plane provided $m=0$. 
Low energy physics at $m=0$ are hence dominated by the $\xi=-1$ states. 
We hence project the Dirac term $m \rho_x \tau_x \sigma_x + 2t k_z \rho_z \tau_z \sigma_z + \mathcal{O}(k_z^2)$ (given below \cref{eq:Hk}) onto the $\xi=-1$ subspace. 
We obtain the effective Hamiltonian 
\begin{align} \label{eq:H-proj1}
H^{(\xi=-1)}(\kk)
=& -\lambda \rho_0 \sigma_0 + 2t k_z  \rho_z\sigma_x 
-m \cdot \Re[f(\kk) + g(\kk)] \rho_x \sigma_0 
+ m \cdot \Im[f(\kk) + g(\kk)] \rho_y \sigma_0  \nonumber\\
&\qquad - m \cdot \Re[f(\kk) - g(\kk)] \rho_x \sigma_x 
+ m \cdot \Im[f(\kk) - g(\kk)] \rho_y \sigma_x  + \mathcal{O}(k_z^2) \ ,
\end{align}
where $f(\kk) = (1 + e^{\ii(k_x+k_y)})/4$, $g(\kk) = ( e^{\ii k_x} + e^{\ii k_y})/4$.
We use $\rho_z=\pm1$ for $\eta=\pm1$ states and $\sigma_z=\pm1$ for $n=1,2$ states, respectively, in the subspace of $\xi=-1$. 
Clearly, when $m=0$, $H^{(\xi=-1)}$ represent two decoupled helical modes (per unit cell), consistent with the topological crystal construction in Fig.~1(c) of the main text. 
For $|m|\ll \gamma$, we can omit the coupling between $\xi=\pm1$ subspaces. 
Then, the four energy bands in the $\xi=-1$ subspace are 
\begin{equation}
-\lambda \pm 2 \sqrt{  t^2 k_z^2 + m^2 |f(\kk)|^2 }, \qquad 
-\lambda \pm 2 \sqrt{ t^2 k_z^2 + m^2 |g(\kk)|^2 }\ . 
\end{equation}
Since $|f(\kk)|=0$ iff $k_y=-k_x \pm \pi$, $|g(\kk)|=0 $ iff $k_y = k_x \pm \pi$, a nodal line crossing is formed along the square 
\begin{equation}
(\pi,0,0)\to(0,\pi,0)\to(-\pi,0,0)\to(0,-\pi,0)\to (\pi,0,0)\ . 
\end{equation}
Being protected  the $P\td{T}$ symmetry, the nodal line is stable against coupling to the $\xi=1$ subspace. 
However, we find that the coupling lifts the four-fold degeneracy at X, which is not protected by the single-valued magnetic space group $P_C 4/nbm$, and eventually deforms the square-shaped nodal line into a circular shape (\cref{fig:F_BAND}(c)-(g)). 

For later convenience, we also derive the representation of $P$ (\cref{eq:inversion}) and $S_4$ (\cref{eq:S4}) at $\kk=0$ in the $\xi=-1$ subspace:
\begin{equation}
\mathcal{D}^{(\xi=-1)}(P) = - \rho_x \sigma_0 ,\qquad 
\mathcal{D}^{(\xi=-1)}(S_4) = -\ii \rho_0 \sigma_z \ . 
\end{equation}
At $\kk=0$, the doubly degenerate energy level $-\lambda + m$ has $P$ and $S_4$ eigenvalues $1$, $1$ and $\ii$, $-\ii$, respectively; and the other doubly degenerate  level $-\lambda - m$ has $P$ and $S_4$ eigenvalues $-1$, $-1$ and $\ii$, $-\ii$, respectively. 

The nodal line shrinks as $m$ increases. 
As shown in \cref{fig:F_BAND}, all the high symmetry momenta except $\Gamma$ remain gapped as $m$ changes from $0^+$ to values greater than $2\gamma$. 
A level crossing happens at $\Gamma$ when $m=2\gamma$, upon which the nodal line shrinks to zero. 
We now analyze this level crossing; it may help us understand the topology of the model. 
Thanks to the $C_2$ symmetry, the $\xi=\pm1$ subspaces are decoupled at $\Gamma$. 
Each of the energy levels $E=-\lambda \pm 2m $ in the $\xi=-1$ subspace is a monotonous function of $m$, hence, provided $\gamma\gg \lambda >0$, they do not participate in the level crossing around $\pm\lambda$ when $m=2\gamma$. 
Thus, we only need to study the $\xi=+1$ subspace. 
The effective Hamiltonian on the basis $u_{\xi=+1,\eta,n}(\mathbf{0})$ reads 
\begin{equation} \label{eq:H-proj2}
    H^{(\xi=+1)}(\kk=0) = \lambda \rho_0\sigma_0 - 2\gamma \rho_0 \sigma_z  +  m \rho_x \sigma_0 \ .
\end{equation}
The energy levels $\lambda + 2\gamma - m$ and $ \lambda - 2\gamma +m$ cross each other at $m=2\gamma$. 
We also notice that, at $\kk=0$, the $P$ (\cref{eq:inversion}) and $S_4$ (\cref{eq:S4}) symmetry operators act in the $\xi=+1$ subspace as
\begin{equation}
    \mathcal{D}^{(\xi=+1)}(P) = \rho_x \sigma_0,\qquad 
    \mathcal{D}^{(\xi=+1)}(S_4) = -\rho_0 \sigma_z \ . 
\end{equation}
One can immediately observe that the level  $\lambda + 2\gamma - m$ has $P$ and $S_4$ eigenvalues $-1$ and $1$, respectively, and the level  $\lambda - 2\gamma + m$ has $P$ and $S_4$ eigenvalues $1$ and $-1$, respectively. 

We now calculate the symmetry-based indicators [\onlinecite{TCI6},~\onlinecite{elcoro_magnetic_2021},~\onlinecite{peng_topological_2022}] given by $P$ and $S_4$ eigenvalues: 
\begin{equation} \label{eq:indicator-eta4}
    \eta_{4I} =  \sum_{\kk \in K^8} n_\kk (P=-1) \mod 4\ ,
\end{equation}
\begin{equation}\label{eq:indicator-z2}
    z_2 = \sum_{\kk \in K^4} \frac{n_\kk (S_4=1) - n_\kk(S_4=-1) }2 \mod 2\ , 
\end{equation}
\begin{equation}\label{eq:indicator-d2}
\delta_{2S} \!=\! \sum_{\kk\in K^4} e^{\ii k_z} 
    \pare{ n_\kk(S_4=-\ii) - n_\kk (S_4=-1) }
    \mod 2 . 
\end{equation}
Here $K^8=\{(0/\pi,0/\pi,0/\pi)\}$ are the eight inversion-invariant momenta, and $K^4=\{(0,0,0/\pi), (\pi,\pi,0/\pi)\}$ are the four $S_4$-invariant momenta. 
$n_\kk(P=-1)$, $n_\kk(S_4=\pm1,\pm\ii)$ denote the number of occupied levels at $\kk$ that have the corresponding symmetry eigenvalues.
For more details, as well as four additional indicators ($z_{2I,i}$, $z_{4S,\pi}$) representing 3D Chern numbers, readers may refer to discussions around Eqs.~(235) and (270) of the supplementary materials of Ref.~\cite{elcoro_magnetic_2021}. 
Since $z_{2I,i}$, $z_{4S,\pi}$ are always trivial in our model, we do not discuss them further here. 
Additionally, one need not worry about crossings enforced by the compatibility relations of $C_2=S_4^2$, which would invalidate $z_2,\delta_{4S}$, because there is no crossing along the high symmetry lines $\rm \Gamma Z$, $\rm M A$, $\rm XR$. 
Thus, $\eta_{4I}$, $z_2$, and $\delta_{2S}$ are all the meaningful indicators given by $P$ and $S_4$. 
Odd $\eta_{4I}$ and $\delta_{2}$ implies existence of Weyl points. 
In a gapped insulator there must be $\eta_{4}\in $ even, $\delta_{2S}=0$ and 
\begin{equation} \label{eq:P3-indicator}
    P_3 = \frac{\eta_{4I}}4 = \frac{z_2}2 \mod 1
\end{equation}
provided all 3D Chern numbers equal to zero. 

We now calculate $\eta_{4I}, z_2, \delta_{2S}$ for our model. 
Since the $m\to \infty$ limit must have trivial indicators, we can replace $n_\kk$ in \cref{eq:indicator-eta4,eq:indicator-z2,eq:indicator-d2} by $\delta n_\kk$ - the change of $n_\kk$ with respect to the $m\to \infty$ limit. 
According to the discussions around \cref{eq:H-proj2}, the nodal line semi-metal with $0<m<2\gamma$ has 
\begin{equation}
    \delta n_\Gamma(P=1) = 1,\quad 
    \delta n_\Gamma(P=-1) = -1,\quad 
    \delta n_\Gamma(S_4=1) = -1,\quad 
    \delta n_\Gamma(S_4=-1) = 1,\quad 
    \delta n_\Gamma(S_4=\pm\ii) = 0\ . 
\end{equation}
and hence 
\begin{equation}
\eta_{4I} = 3,\qquad z_2 = 1,\qquad \delta_{2S} = 1\ .
\end{equation}
$\eta_{4I}=3$ indicates that the considered state must have Weyl points. 
Similarly, $\delta_{2S}=1$ also implies that the considered state must have Weyl points. 
Due to the $P\td T$ symmetry, these Weyl points must be part of nodal lines, which are further pinned in the $k_z=0$ plane by the glide ($\td M_z = \{m_{001}|\frac12\frac12 0\}$) symmetry. 
It appears that the indicators only reveal the presence of nodal lines. 

For negative $m$, we find 
{\small
\begin{align}
-2\gamma<m<0:\quad &
    \delta n_\Gamma(P=1) = 3,\quad 
    \delta n_\Gamma(P=-1) = -3,\quad 
    \delta n_\Gamma(S_4=1) = -1,\quad 
    \delta n_\Gamma(S_4=-1) = 1,\quad 
    \delta n_\Gamma(S_4=\pm\ii) = 0 \nonumber\\
& \Rightarrow  
    \eta_{4I} = 1,\quad
    z_2 = 1,\quad 
    \delta_{2S} = 1\ ,
\end{align}
\begin{align}
-4t+2\gamma<m<-2\gamma:\quad &
    \delta n_\Gamma(P=1) = 4,\quad 
    \delta n_\Gamma(P=-1) = -4,\quad 
    \delta n_\Gamma(S_4=\pm1) = 0,\quad 
    \delta n_\Gamma(S_4=\pm\ii) = 0 \nonumber\\
& \Rightarrow  
    \eta_{4I} = 0,\quad
    z_2 = 0,\quad 
    \delta_{2S} = 0\ .
\end{align}}
Thus, the state with $-2\gamma<m<0$ is a nodal line semi-metal as the $0<m<2\gamma$ case, and the band insulator with $-4t+2\gamma <m<-2\gamma$ must have $P_3=0$ according to \cref{eq:P3-indicator}. 

\subsection{Random fluxes or imaginary hoppings}

As pointed out in the main text, one can introduce local gaps to the helical modes that only respect the $C_4T$ symmetry on average. 
These gaps localize the low energy states in the bulk  and give rise the intrinsic statistical topological insulator (STI) characterized by $P_3=1/2$. 
One choice of the local gap is random flux through the $2a$ and $2b$ Wyckoff positions. 
Consider the $m=0$ limit, where the low energy Hamiltonian around $k_z=0$ decomposes into decoupled helical modes (\cref{eq:Hk-decoupled}). 
In the $xy$ plane, these helical modes are molecule states localized on the squares centered at the 2a positions. 
Each molecule has a local Hamiltonian:
\begin{equation}
H^{\rm (loc)} (k_z) = 
2\eta t k_z \cdot \tau_z \sigma_z + 
    \begin{pmatrix}
    0 & \gamma e^{\ii \frac{\Phi}4 }  & \gamma e^{-\ii \frac{\Phi}4 } & \lambda \\
    \gamma e^{-\ii \frac{\Phi}4 } & 0 & \lambda & \gamma e^{\ii \frac{\Phi}4 } \\
    \gamma e^{\ii \frac{\Phi}4 }  & \lambda  & 0 & \gamma e^{-\ii \frac{\Phi}4 } \\
    \lambda  & \gamma  e^{-\ii \frac{\Phi}4 } & \gamma e^{\ii \frac{\Phi}4 } & 0 
    \end{pmatrix}
+ \mathcal{O}(k_z^2) \ ,
\end{equation}
where $\Phi$ is a flux passing through the molecule, $\eta=1$, $-1$ correspond to the molecule formed by orbitals 1, 2, 3, 4 and 5, 6, 7, 8, respectively. 
We have dropped the $\kk$-dependent phase factors in \cref{eq:Hk-decoupled}, which arise from the plane-wave-like Bloch basis, because we are now studying a local problem. 
The eigenstates at $k_z=0$ are identical to those in Eqs.~(\ref{eq:eigenstate-1})-(\ref{eq:eigenstate-4}) except that the $\kk$-dependent phase factors should be omitted here. 
Then we find that the low energy Hamiltonian in the $\xi=-1$ and $+1$ subpaces are 
\begin{equation}
H^{(\mathrm{loc},\xi=-1)} = - \lambda \cdot \sigma_0 + 2\eta t k_z \cdot \sigma_x - 2\gamma \sin \frac{\Phi}4 \cdot \sigma_z  + \mathcal{O}(k_z^2)
\end{equation}
and 
\begin{equation}
H^{(\mathrm{loc},\xi=1)} =  \lambda \cdot \sigma_0 + 2\eta t k_z \cdot \sigma_x - 2\gamma \cos \frac{\Phi}4 \cdot \sigma_z   + \mathcal{O}(k_z^2)
\end{equation}
respectively. 
Clearly, a finite flux will gap the helical modes in the $\xi=-1$ sector. 

We find it is more convenient to add random imaginary part to the $\gamma$ term (red bonds in \cref{fig:F_BAND}(a)). 
They introduce not only random fluxes, but also random hoping strengths. 
This disorder term can be written as 
\begin{align}
H^{(\rm dis)} =& \sum_{\RR} \ii \big( w_{\RR,1} c_{\RR, 1}^\dagger c_{\RR, 2}
    + w_{\RR,2} c_{\RR,2}^\dagger c_{\RR,3} 
    + w_{\RR,3} c_{\RR,3}^\dagger c_{\RR,1}
    + w_{\RR,4} c_{\RR,4}^\dagger c_{\RR,3}  \nonumber\\
&\qquad + w_{\RR,5} c_{\RR,5}^\dagger c_{\RR+(010), 6}
    + w_{\RR,6} c_{\RR,6}^\dagger c_{\RR,8}
    + w_{\RR,7} c_{\RR,7}^\dagger c_{\RR-(100),5} 
    + w_{\RR,8} c_{\RR,8}^\dagger c_{\RR-(010),7} \bigg) + h.c. 
\end{align}
Here $w_{\RR,\alpha}$ are quenched gaussian variables satisfying 
\begin{equation}
    \inn{ w_{\RR,\alpha} w_{\RR',\beta} } = W^2 \cdot \delta_{\RR,\RR'} \delta_{\alpha\beta}\ . 
\end{equation}
$W$ is the parameter controlling the disorder strength. 
For $m=0$, we expect the intrinsic STI phase at weak and intermediate $W$ because small random fluxes are able to gap out the helical modes.
The $W\to \infty$ limit must give the trivial Anderson insulator.

\subsection{Uniform \texorpdfstring{$C_4T$}{C4T}-breaking flux patterns}

In the last subsection we argued that random local fluxes can gap out the helical modes and drive the system into an intrinsic STI phase. 
We find that our model could become a clean axion insulator (protected by $S_4$ or $P$) in the presence of a properly designed $C_4T$-breaking flux pattern.

\begin{figure}[h]
\centering
\includegraphics[width=0.8\linewidth]{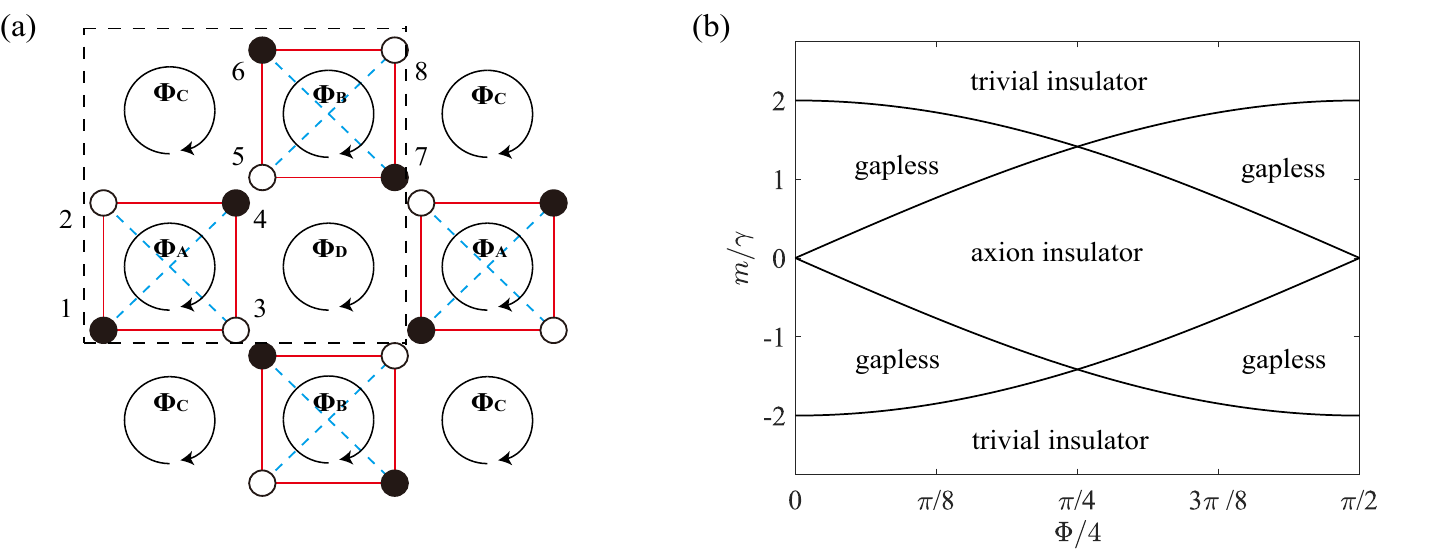}
\caption[]{ 
    (a) The position of the four fluxes $\Phi_A, \Phi_B, \Phi_C, \Phi_D$.
    (b) Phase diagram in the presence of $C_{4}T$-breaking flux patterns.  $0<\gamma<t$ is assumed for simplicity. Patterns (I) and (II) have the same phase diagram, but the gapless phases in the two patterns are physically different. 
   }
\label{fig:uniform_pattern}
\end{figure}

We assume the translation symmetry and denote the fluxes through the squares centered at $(\frac14,\frac14,0)$, $(\frac34,\frac34,0)$, $(\frac14,\frac34,0)$, $(\frac34,\frac14,0)$ and $\Phi_A$, $\Phi_B$, $\Phi_C$, $\Phi_D$, respectively (\cref{fig:uniform_pattern}(a)). 
We consider two patterns
\begin{equation}
\mathrm{(I)}:\qquad \Phi_A=\Phi_B = -\Phi_C = - \Phi_D = \Phi\ ,
\end{equation}
\begin{equation}
\mathrm{(II)}:\qquad \Phi_A=-\Phi_B = \Phi,\qquad \Phi_C = \Phi_D = 0 \ .
\end{equation}
Pattern-(I) respects both $S_4$ and $P$ symmetries, whereas pattern-(II) only respects $S_4$ but breaks $P$.
We find the corresponding Hamiltonians
\begin{align}
 H^{(\rm I)}(\kk) & = (m + 2t - 2t \cos{k_z}) \rho_x \tau_x \sigma_x + 2t\sin{k_z} \rho_z\tau_z\sigma_z \nonumber\\
 & + \gamma \cos{\frac{\Phi}{4}} (
   \cos{\frac{k_x} {2}}\rho_0\tau_x\sigma_0 
 - \sin{\frac{k_x} {2}}\rho_0\tau_y\sigma_0
 + \cos{\frac{k_y}{2}} \rho_0 \tau_0\sigma_x
 - \sin{\frac{k_y}{2}} \rho_0 \tau_0\sigma_y ) \nonumber \\
 & + \gamma \sin{\frac{\Phi}{4}}(
   \sin{\frac{k_x} {2}}\rho_0\tau_x\sigma_z 
 + \cos{\frac{k_x} {2}}\rho_0\tau_y\sigma_z
 - \sin{\frac{k_y}{2}} \rho_0 \tau_z\sigma_x
 - \cos{\frac{k_y}{2}} \rho_0 \tau_z\sigma_y 
  ) \nonumber\\
 & + \lambda (\cos{\frac{k_x}{2}}\cos{\frac{k_y}{2}} \rho_0\tau_x\sigma_x + \sin{\frac{k_x}{2}}\sin{\frac{k_y}{2}} \rho_0\tau_y\sigma_y
 - \sin{\frac{k_x}{2}}\cos{\frac{k_y}{2}} \rho_0\tau_y\sigma_x
 - \cos{\frac{k_x}{2}}\sin{\frac{k_y}{2}} \rho_0\tau_x\sigma_y)\ ,
\end{align}
\begin{align}
 H^{(\rm II)}(\kk) & = (m + 2t - 2t \cos{k_z}) \rho_x \tau_x \sigma_x + 2t\sin{k_z} \rho_z\tau_z\sigma_z \nonumber\\
 & + \gamma \cos{\frac{\Phi}{4}} (
   \cos{\frac{k_x} {2}}\rho_0\tau_x\sigma_0 
 - \sin{\frac{k_x} {2}}\rho_0\tau_y\sigma_0
 + \cos{\frac{k_y}{2}} \rho_0 \tau_0\sigma_x
 - \sin{\frac{k_y}{2}} \rho_0 \tau_0\sigma_y ) \nonumber \\
 & + \gamma \sin{\frac{\Phi}{4}}(
   \sin{\frac{k_x} {2}}\rho_z\tau_x\sigma_z 
 + \cos{\frac{k_x} {2}}\rho_z\tau_y\sigma_z
 - \sin{\frac{k_y}{2}} \rho_z \tau_z\sigma_x
 - \cos{\frac{k_y}{2}} \rho_z \tau_z\sigma_y 
  ) \nonumber\\
 & + \lambda (\cos{\frac{k_x}{2}}\cos{\frac{k_y}{2}} \rho_0\tau_x\sigma_x + \sin{\frac{k_x}{2}}\sin{\frac{k_y}{2}} \rho_0\tau_y\sigma_y
 - \sin{\frac{k_x}{2}}\cos{\frac{k_y}{2}} \rho_0\tau_y\sigma_x
 - \cos{\frac{k_x}{2}}\sin{\frac{k_y}{2}} \rho_0\tau_x\sigma_y)\ . 
\end{align}
$\gamma \cos \frac{\Phi}4$ replace the $\gamma$ parameter in \cref{eq:Hk}, and the $\gamma \sin \frac{\Phi}4$ terms are new. 
Symmetry-based indicators $\eta_{4I}, z_2, \delta_{2S}$ (\cref{eq:indicator-eta4,eq:indicator-z2,eq:indicator-d2} can be easily computed, and they suggest a phase diagram shown in (\cref{fig:uniform_pattern}(b)). 
$H^{(\rm I)}$ and $H^{(\rm II)}$ happen to have the same phase diagram, but their gapless phases are physically different.
Gapless phase in $H^{(\rm I)}$ is a nodal line semi-metal due to the $\td M_z = \{m_{001}|\frac12\frac12 0\}$ symmetry, whereas the gapless phase in $H^{(\rm II)}$ is a Weyl semi-metal. 

The phase diagram in \cref{fig:uniform_pattern}(b) may suggest that our model (\cref{eq:Hk}) is relatively ``close'' to a clean axion insulator. Perturbatively breaking $C_4T$ symmetry drives the gapless system into a clean axion insulator protected by $S_4$ or $P$. However, the intrinsic STI protected by $C_4T$ does {\it not} require the existence of (average or exact) $S_4$ or $P$. 

\clearpage
\section{Numerical methods and results}
\subsection{Quasi-1D localization length and the transfer matrix method}

The quasi-1D localization length $\xi_{\rm 1D}$ is a commonly used quantity to determine the localization behavior of 2D or 3D systems \cite{transfer_matrix,one_parameter_scaling}.
We focus on 3D systems in the following, while the 2D case is in principle similar.
$\xi_{\rm 1D}(L)$ is defined for a sample with a quasi-1D shape (for example, with geometry $M \times L \times L$ where $M \gg L$), and reflects the slowest decay rate of energy eigenstates in the longitudinal direction.
Since a quasi-1D sample can be viewed as a 1D system with many internal degrees of freedom, and electrons always localize in a 1D system with arbitrary non-zero disorder strength, $\xi_{\rm 1D}(L)$ always converges to a finite value when $M \to \infty$ with the exception of a perfectly clean sample.
The (de)localization of a normal shaped sample, whose sizes in different directions are comparable, can be inferred from the scaling behavior of the dimensionless normalized quasi-1D localization length $\Lambda(L) = \xi_{\rm 1D}(L)/L$.
For a metallic normal shaped sample of size $ L\times L\times L$, its 3D localization length $\xi_{3D}(L)$ is much larger than $L$ when $L$ is sufficiently large.
If the sample is prepared in a quasi-1D shape with longitudinal size $M \gg L$, $\xi_{\rm 1D}(L)$ grows faster than $L$, so $\Lambda(L) \to \infty$ as $L \to \infty$.
In contrast, for a localized phase, $\xi_{3D}(L)$ converges to a finite value $\xi_0$ as $L \to \infty$. In this case, $\xi_{\rm 1D}(L)$ also converges to $\xi_0$ in the large $L$ limit, leading to $\Lambda(L) \to 0$. 
For the critical state, $\xi_{\rm 1D}(L)$ increases linearly with $L$ for large $L$, meaning that $\Lambda(L)$ approaches a finite value $\Lambda_c$ when $L \to \infty$.

The transfer matrix method \cite{transfer_matrix} is widely used to numerically compute $\xi_{\rm 1D}(L)$.
A quasi-1D sample is devided into layers normal to the longitudinal direction, with the amplitudes of energy eigenstates on different layers related by the time-independent Schr\"odinger equation.
Specifically, the Hamiltonian of a tight binding lattice model can be written as
\begin{equation}
    H = \sum_{ij} t_{ij} {c_i}^{\dagger} c_j = \sum_{q = -r_1}^{r_2}\sum_{l=1}^{M}\sum_{\alpha, \beta = 1}^{s} t_{l \alpha,(l+q) \beta} c_{l \alpha}^{\dagger} c_{(l+q) \beta}
\end{equation}
where $i,j$ denote orbitals, $l$ denotes layers and $1 \leq \alpha, \beta \leq s$ denote the $s$ degrees of freedoms within each layer.
The condition $r_1 \leq q \leq r_2$ implies that orbitals in the $l$th layer can hop to at most the $(l - r_1)$th or $(l + r_2)$th layer in one step.
Since the transfer matrices have dimensions $(r_1+r_2)s$ (see below), the division into layers should be carefully designed to minimize $r_1, r_2$ and $s$ for optimal numerical efficiency.
$x$ direction is taken as the longitudianl direction for the lattice model introduced in \cref{sec:lattice-model}. 
Each layer consists of two orbitals in one ``column'' of a square, along with their partners generated by lattice translations in $y,z$ directions. 
For example, orbitals $\{3, 4\}$ in \cref{fig:F_BAND}(a), while orbitals $\{5, 6\}$ form the next layer.
In this setup, $r_1 = r_2 = 1$ and $s = 2L^2$.
The following derivation assumes $r_1 = r_2 = 1$ for simplicity. 

The Schr\"odinger equation $H|\psi\rangle = E |\psi\rangle$ can be written as
\begin{equation}
    \begin{aligned}
     & \sum_{q=-1}^{1} H_{l, l+q} \vec{\psi}_{l+q} = E \vec{\psi}_l \\
     \Longrightarrow &   \vec{\psi}_{l+1} = -H_{l, l+1}^{-1} ((H_{l,l}-E)\vec{\psi}_l + H_{l, l-1} \vec{\psi}_{l-1})
   \end{aligned}
\end{equation}
where $(H_{l, l'})_{\alpha, \beta} = t_{l\alpha, l'\beta}$, and $\vec \psi_l^T=(\psi_{l,1},\psi_{l,2},\cdots)$ are the amplitudes of an energy eigenstate with energy $E$ in the $l$th layer. 
If $H_{l, l+1}$ is singular, a different layer partition with an invertible $H_{l, l+1}$ must be used. The relation between amplitudes in adjacent layers is:
\begin{equation}
    \left[\begin{array}{c}
    \vec{\psi}_{l+1} \\
    \hline 
    \vec{\psi}_{l} \\
    \end{array}\right]
    = T_l 
    \left[\begin{array}{c}
    \vec{\psi}_{l} \\
    \hline 
    \vec{\psi}_{l-1} \\
    \end{array}\right]
\end{equation}
where 
\begin{equation}
    T_l = \left[\begin{array}{c|c}
    -H_{l, l+1}^{-1}(H_{l,l}- E) &  -H_{l, l+1}^{-1} H_{l,l-1} \\
    \hline 
    I_{s \times s} & 0_{s \times s} \\
    \end{array}\right]
\end{equation}
is the $l$th transfer matrix. 
In a disordered system, $T_l$'s contain random elements, and their consecutive product $O_M=\prod_{l=1}^M T_l$ transforms the amplitudes on the first two layers to those on the last two layers.
A theorem by Oseledec \cite{oseledec1968multiplicative} guarantees the existence of the limit $P = \lim_{M \to \infty} (O_M^{\dagger} O_M)^{1/2}$, whose eigenvalues $\left\{\exp \left(\nu_1\right), \exp \left(-\nu_1\right), \ldots \exp \left(\nu_s\right), \exp \left(-\nu_s\right)\right\}$ come in pairs. 
The positive exponents $\nu_1 > ... >\nu_i>\nu_{i+1}>... > \nu_s >0$ are called Lyapunov exponents (LEs). 
An eigenvector $\vec{\eta_i}$ of $P$ with eigenvalue $\exp(-\nu_i)$ satisfies $\left\|O_M \vec{\eta}_i\right\|^2 = \left\|\exp \left(-M \nu_i\right) \vec{\eta}_i\right\|^2$ for sufficiently large $M$.
Therefore, the smallest LE $\nu_s$ determines the slowest decay rate of energy eigenstates with energy $E$ along the longitudinal direction, and the quasi-1D localization length is defined as $\xi_{\rm 1D} = 1/(\nu_s l_0)$, where $l_0 = 4$ for the lattice model denotes the number of layers in a unit cell.

In practice, the matrix $O_M$ cannot be computed directly.
This is because when $M$ is large, $\exp(M\nu_i) \gg \exp(M\nu_j)$ for $\nu_i > \nu_j$, and the smallest LE will quickly pick up large round-off error since most computational resources are consumed by the large LEs.
To overcome this difficulty, we perform a $QR$ decomposition after each $q$ transfer matrices are multiplied:
\begin{equation}
    Q_j R_j = T_{qj}T_{qj-1}...T_{q(j-1)+1} Q_{j-1}
\end{equation}
where ${Q_j}^{\dagger}Q_j = I, Q_0 = I_{2s\times 2s}$, and $R_j$ is upper-triangular. 
The process is repeated for $j = 1,...,M/q$. 
During each step, the logarithms of diagonal elements of each $R_j$ are stored. 
We point out without proof that \cite{transfer_matrix}
\begin{equation}
    \nu_i=\lim _{M / q \rightarrow \infty} \frac{1}{M / q-n_0} \sum_{j=n_0+1}^{M / q} \frac{\ln \left(R_j\right)_{i, i}}{q}
\label{eq:qr_exponents}
\end{equation}
where we have excluded the first $n_0q$ layers to avoid possible boundary effects.

\cref{eq:qr_exponents} also provides us an unbiased estimation of the numerical precision of LEs: each LE can be viewed as an average of $M/q - n_0$ random samples, whose standard derivation can be estimated and interpreted as the numerical error of the LE.
Prudent readers may suspect that the ``random samples'' $\ln \left(R_j\right)_{i, i}/q$ with close $i$'s might not be independent for small $q$. In practice, we group up $r \sim 10$ adjacent ``random samples'' together and assume that different groups are independent. 

We have introduced the method to calculate $\nu_s$ and its numerical precision $\delta \nu_s$ for a quasi-1D sample with a given disorder configuration.
To further reduce the numerical uncertainty, we take the average of ${\nu_s}$ for $N_D$ different disorder configurations.
The averaged $\bar{\nu_s}$ and its uncertainty is given by
\begin{equation}
    \bar{\nu_s} = \frac{1}{N_D}\sum_{k=1}^{N_D} {\nu_{s}}^{(k)}, 
    \delta{\bar{\nu_s}} = \frac{1}{N_D} \sqrt{\sum_{k=1}^{N_D} ({\delta \nu_{s}}^{(k)})^2}
\end{equation}
where ${\nu_{s}}^{(k)}, 1\leq k \leq N_D$ denotes $\nu_s$ for the $k$th disorder configuration.
Numerical $\Lambda$'s in this paper are obtained by using $L\leq 40, M = 2.5\times10^{4}, N_D = 16, q = 4, r = 10$ and $n_0 = 100$.
The precision of $\Lambda = 1/(\bar{\nu_s} l_0 L)$ reaches $\delta \Lambda/\Lambda \leq 0.6\%$ for systems with periodic boundary conditions in both $y$ and $z$ directions, and $\delta \Lambda/\Lambda \leq 1.5\%$ for systems with open boundary condition in $z$ direction and periodic boundary condition in $y$ direction.

\subsection{Scaling analysis and Polynomial fitting of $\Lambda$}
As mentioned above, $\Lambda(L)$ converges to a finite value $\Lambda_c$ when $L \to \infty$ for a critical state.
Therefore, $\Lambda(L)$ should be scale invariant for large $L$'s at a metal-insulator phase transition point.
On the insulating side of the phase transition, the 3D localization length $\xi_{3D}$ diverges as $\xi_{3D} \sim \Tilde{r}^{-\nu}$, where $\Tilde{r} = |r-r_c|/r_c$, $r$ is a tuning parameter controlling the phase transition (not necessarily the disorder strength), and $r_c$ is the critical value of $r$. 
The phase transition is characterized by a universal critical exponent $\nu>0$, which only depends on the universality class.
For sufficiently large $L$, $\Lambda$ follows the one-parameter scaling law with scaling variable $L/\xi_{3D}$ \cite{one_parameter_scaling}, which requires that
\begin{equation}
    \Lambda(r,L) = f(L/\xi_{3D}(r)) = F(\Tilde{r} L^{1/\nu})
    \label{eq:Lambda1}
\end{equation}
where $f$ is the one-parameter scaling function.
However, numerically accessible $L$'s for 3D models are generally not large enough for us to neglect the finite size effects associated to the irrelevant scaling variables in the renormalization group (RG) theory, where the metal-insulator transition is described by a saddle-point fixed point with one relevant scaling variable and multiple irrelevant scaling variables.
The relevant scaling variable $\phi(\Tilde{r}, L)$ has scaling dimension $1/\nu>0$, while the irrelevant scaling variables all have negative scaling dimensions.
For simplicity, we neglect all irrelevant variables but the least irrelevant one, i.e., the one with the largest scaling dimension $y<0$, which is denoted $\psi(\Tilde{r}, L)$.
This approximation is justified by a relatively small irrelevant contribution to $\Lambda$ and a large $|y|$ (see below). 
Considering the irrelvant contribution and the non-linearality of variables $\phi$ and $\psi$ in $\Tilde{r}$, \cref{eq:Lambda1} is modified to \cite{scaling_irr}
\begin{equation}
    \Lambda(r, L) = F(\phi(\Tilde{r}, L), \psi(\Tilde{r}, L))
\label{eq:irr_considered}
\end{equation}
which can be Taylor expanded as
\begin{equation}
    \Lambda(r, L) = \sum_{j_1 = 0}^{n_1}\sum_{j_2 = 0}^{n_2} a_{j_1, j_2} \phi(\Tilde{r}, L)^{j_1} \psi(\Tilde{r}, L)^{j_2}
\label{eq:fit_expansion}
\end{equation}
where $\phi(\Tilde{r}, L) = u_1(\Tilde{r})L^{1/\nu}, \psi(\Tilde{r}, L) = u_2(\Tilde{r})L^{-y}$. When $\Tilde{r}$ is small, $u_1(\Tilde{r})$ and $u_2(\Tilde{r})$ can be further expanded as
\begin{equation}
    u_i(\Tilde{r}) = \sum_{j = 0}^{m_i} b_{ij}{\Tilde{r}}^j, i=1, 2
\end{equation}
Since $\Tilde{r} = 0$ corresponds to the phase transition point, $u_1(0) = 0$, which implies $b_{10} = 0$. 
We also require $a_{10} = a_{01} = 1$ to remove the ambiguity when defining $u_1, u_2$. 
Unknown parameters in the above expansion scheme include $a_{ij}, b_{ij}, r_c, \nu$ and $y$, with total number
\begin{equation}
    N_p = m_1 + m_2 + (n_1 + 1)(n_2 + 1) +2
\end{equation}
The phase transition point $r_c$ and the universal exponent $\nu$ are obtained by fitting the $N_P$ parameters to numerical $\Lambda$'s using the $\chi^2$ fitting method, which minimizes the residual
\begin{equation}
    \chi^2 = \sum_{j=1}^{N_d} (\frac{\Lambda_j - F_j}{\delta\Lambda_j})^2
\end{equation}
where $j=1,...,N_d$ denotes $N_d$ data points, $F_j$ denotes the value obtained from \cref{eq:fit_expansion}, and $\Lambda_j$ and $\delta\Lambda_j$ denote numerical $\Lambda$'s and their numerical error.
The goodness of fit $P$ is given by \cite{scaling_irr}
\begin{equation}
    P = 1 - \frac{1}{\Gamma(N_{DOF}/2)} \int_0^{\chi^2_{min}/2} \exp (-t) t^{\frac{N_{DOF}}{2}-1} \mathrm{~d} t
\end{equation}
where $N_{DOF} = N_d - N_P$, and $\chi^2_{min}$ is the minimal value of $\chi^2$. 
$P$ is the probability that $N_d$ data points randomly sampled from $F_j$ with standard derivations $\delta\Lambda_j$ give a larger $\chi^2$ than $\chi^2_{min}$.
We require $P \geq 0.1$ for a fit to be acceptable.
To avoid overfitting, we also require $N_d > 4 N_P$ and the irrelevant contribution to $F(\Tilde{r}, L)$ (sum of terms with $j_2>0$ in \cref{eq:fit_expansion}) to be small compared with $F_j$, which includes both the relevant and irrelevant contributions.

The error bars of the fitting parameters are determined using the Monte-Carlo method. 
This is done by generating synthetic data sets $\Tilde{\Lambda_j}'$, which are Gaussian random numbers with expectation values given by fitted $F_j$'s, and standard derivations given by $\delta \Lambda_j$'s.
We then fit \cref{eq:fit_expansion} to these synthetic data sets to obtain additional parameter sets.
The degree of certainty of the fitting parameters are estimated as their respective $95\%$ confidence intervals from 1000 different synthetic data sets.

We have focused on the polynomial fitting of Taylor expansions of $\Lambda(r, L)$ in above discussions.
Since $\Lambda$ is scale invariant at the phase transition point, fittings using Taylor expansions of $\Lambda, 1/\Lambda$ and $\ln{\Lambda}$ near the phase transition point are all in principle equivalent, provided that $\delta \Lambda/\Lambda$ is sufficiently small for the $\chi^2$ fitting method to be well justified.
In practice, one should choose the best fit with an acceptable goodness of fit, low expansion orders, small irrelavent contributions, and good numerical stability.
Most of the phase transition points in Fig.~2(d) in the main text are determined by fitting $\Lambda$, while some are determined by fitting $1/\Lambda$, as will be discussed in the next section.

\subsection{Numerical results}

Let us first discuss the Fermi energy used in the calculation of $\Lambda$.
As mentioned in the main text, we focus on the behavior of the model at the half-filled Fermi energy, with four electrons per unit cell.
As discussed in \cref{sec:low-energy-states}, this corresponds to $E_F = -\lambda$ for $m=0$, and a $m$-dependent $E_F$ for $m \neq 0$.
However, when $|m| \ll  \gamma$, the $m$ dependence of $E_F$ arises from couplings between subspaces with different $C_2$ eigenvalues, which involve high energy bands and is thus very weak. 
In this work, we keep $t = 2, \gamma = 1, \lambda = 0.01$ fixed.
As shown in Fig.~2(d) in the main text, the intrinsic STI phase occurs only when $|m| < 0.05$. 
It is explicit to numerically check that under these conditions, the difference between the half-filled Fermi energy (of the clean model) $E_F$ and $-\lambda$ satisfies $|E_F + \lambda|< 1.25 \times 10^{-5}$, which is much smaller than any of the model parameters.
Since such a small difference does not significantly change the phase boundary between the intrinsic STI and metal phases, we use $E_F = - \lambda = -0.01$ in all numerical calculations for simplicity.

We have shown the raw data and polynomial fitting of $\Lambda$ used to determine the phase boundary near $m = 0.01, W_{c2} \approx 1.125$ in Fig.~2(c) of the main text.
The Taylor expansion orders are chosen as $\{m_r, n_r, m_i, n_i\} = \{3, 2, 0, 1\}$. 
$N_P = 11$ parameters are used to fit $N_d = 96$ data points, yielding $N_{DOF} = N_d - N_P = 85$.
The minimum residual is $\chi_{min}^2 = 70.30$, resulting in $\chi_{min}^2 /N_{DOF} = 0.83$ and a goodness of fit $P = 0.71$.
The optimal fit gives $W_{c2} = 1.125 [1.015, 1.162 ] , \nu = 1.40 [1.24, 1.56]$, and $y = -2.67 [-4.07, -1.40]$, where brackets denote $95\%$ confidence intervals. 
Notably, the relevant contributions to $\Lambda$ (red dots) are close to the fitted values (lines with different colors), indicating that the irrelevant contributions are small.
This small irrelevant correction, together with a relatively large 
$|y|$, justifies neglecting scaling variables more irrelevant than $\psi$ in \cref{eq:irr_considered}.

\begin{figure}[h]
\centering
\includegraphics[width=1\linewidth]{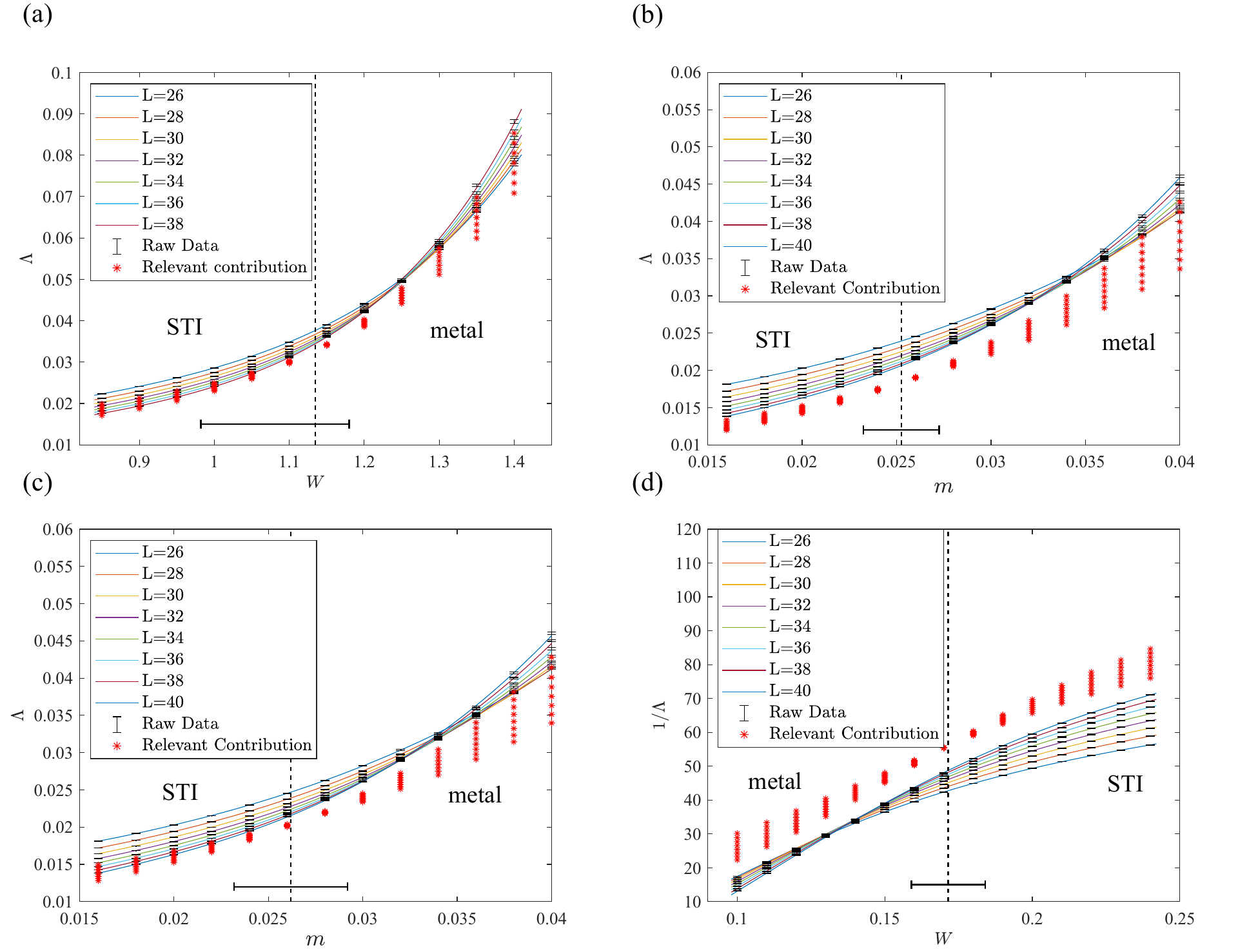}
\caption[]{ 
    Examples of other numerical data and polynomial fittings used to determine the phase diagram. Black dots with error bars show raw data, lines with different colors show the polynomial fitting, and red dots show the relevant contributions.
    (a) $m = -0.01$ fixed, $W_{c2} \approx 1.135$ controls the phase transition.
    (b-c) $W = 0.50$ fixed, $m$ controls the phase transition. (b) and (c) use same raw data but different expansion orders. Despite the discrepancy of the critical exponent $\nu$ as explained in the paragraph, their $m_c$'s are consistent with each other.
    (d) Polynomial fitting of $1/\Lambda$ with $m = 0.01$ fixed, and $W_{c1} \approx 0.171$ controls the phase transition.
   }
\label{fig:vfit}
\end{figure}

To obtain the phase diagram in Fig.~2(d) of the main text, both $W$ and $m$ are used to control the phase transition.
Since the models with $m>0$ and $m<0$ are not related to each other by symmetry, their phase boundaries must be determined separately.
Additionally, it turns out that near some points on the phase boundary, fitting $1/\Lambda$ is more suitable than fitting $\Lambda$. 
For completeness, here we present examples of fitting. 
In \cref{fig:vfit}(a), with $m = -0.01$ fixed, $W$ controls the phase transition. 
Polynomial fitting of $\Lambda$ using $\{m_r, n_r, m_i, n_i\} = \{3, 2, 0, 1\}$ gives $W_{c2} = 1.135 [0.982, 1.180]$ and $\nu = 1.48 [1.25, 1.77]$ with $P = 0.96$.
In \cref{fig:vfit}(b), $W = 0.50$ is fixed, and $m$ controls the phase transition. 
Polynomial fitting of $\Lambda$ with $\{m_r, n_r, m_i, n_i\} = \{1,3,2,1\}$ gives $m_c = 0.0255 [0.0232, 0.0273]$ and $\nu = 1.32 [1.20, 1.48]$ with $P = 0.99$. We note that another fit with slightly lower expansion order $\{m_r, n_r, m_i, n_i\} = \{1,3,1,1\}$ may also be considered acceptable (\cref{fig:vfit}(c)), yielding $m_c = 0.0271[0.0232, 0.0292], \nu = 1.15 [1.08, 1.25]$, and $P = 0.93$.
The critical exponent $\nu$ from this fit is inconsistent with the previously reported value $\nu = 1.443 [1.437, 1.449]$ for 3D Anderson transitions in the unitary universality class.
We attribute this discrepancy to large finite size effects, evidenced by the relatively large irrelevant contribution.
However, $m_c$ from \cref{fig:vfit}(b) and (c) are consistent with each other, suggesting that the estimation of $m_c$ and the phase boundary is less affected by finite size effects.
Finally, \cref{fig:vfit}(d) shows the phase transition near $m = 0.01, W_{c1} = 0.171$. The fit using $1/\Lambda$ rather than $\Lambda$ with expansion orders $\{m_r, n_r, m_i, n_i\} = \{4, 2, 1, 1\}$ yields $W_c = 0.171[0.159, 0.184], \nu = 1.32[1.20, 1.45]$ with $P = 1.00$.

\clearpage
\section{A lattice model with generic couplings and no additional symmetry}

\begin{figure}[h]
\centering
\includegraphics[width=0.8\linewidth]{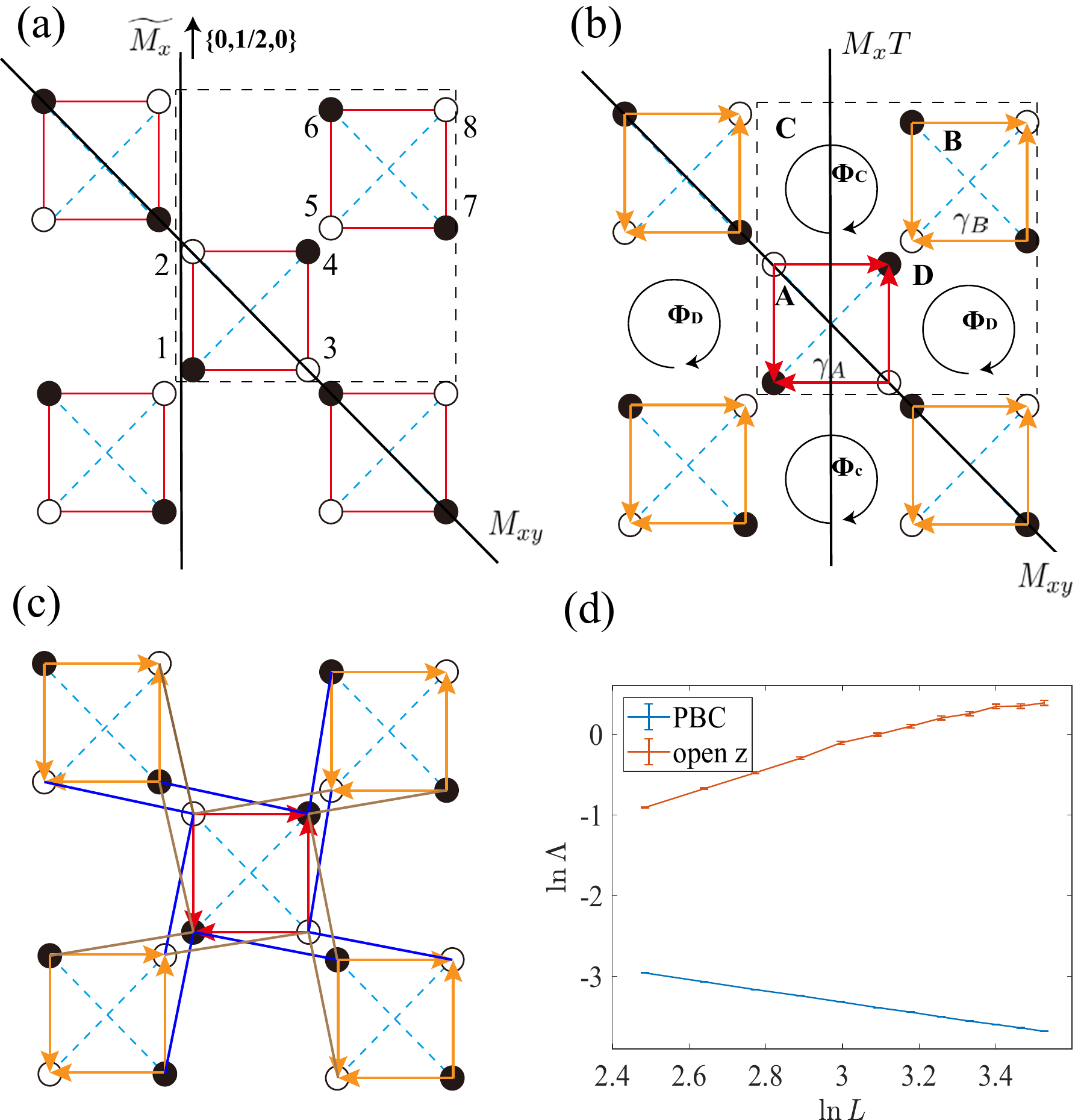}
\caption[]{ 
    (a) The original lattice model (Eq.2 in the main text). Two solid black lines show the glide $\td M_x$ and mirror $M_{xy}$ planes respectively. The dashed box shows a unit cell, with lattice sites labeled by $i\in \{1,...8\}$.
    (b) The lattice model with $|\gamma_A|\neq |\gamma_B|, \delta_A\neq \delta_B$. Red and orange arrows denote $\gamma_A$ and $\gamma_B$ respectively. Two solid black lines show the magnetic mirror $M_xT$ and mirror $M_{xy}$ planes. The magnetic fluxes through $C$ and $D$ regions are $\Phi_C = -\Phi_D = 2(\delta_B - \delta_A)\neq 0$.
    (c) The lattice model with no additional symmetry. Blue and brown solid lines denote hoppings with $\eta$ and $-\eta$ respectively. Only the hoppings with both ends within the figure range are shown, other hoppings can be determined by $C_4T$ and translational symmetry.
    (d) The normalized quasi-1D localization length $\Lambda$ with PBC in both $y,z$ directions, compared to that with OBC in $z$ direction. The parameters used here are $t = 2, \gamma_A = 1.2, \gamma_B = 0.8\cdot\exp({\ii\pi/4}), m=\lambda=\eta=0.02, E_F = -\lambda$ and $W = 1.0$. The disordered lattice model is clearly in an STI phase.
   }
\label{fig:onlyc4t}
\end{figure}

As discussed in \cref{sec:lattice-model}.A, the lattice model in the main text respects the single-valued magnetic space group $P_C 4/nbm$ (\#125.373 in the BNS setting) in the clean limit, which includes other crystalline symmetries than $C_4T$.
Among the additional symmetries, there are four symmetries reverse the sign of $\theta$ angle:  $S_4 = \{-4_{001}^-|\frac 1 2 00\}, \td{M}_z = \{m_{001}|\frac 1 2 \frac 1 2 0\}, \td{M}_x = \{m_{100}|0\frac 1 2 0\}$ and $M_{xy} = \{m_{110}|\frac 1 2\frac 1 20\}$.
$S_4$ is a fourfold improper rotation with the same axis as $C_4T$, $\td{M}_z$ and $\td{M}_z$ are glide planes, and $M_{xy}$ is a mirror symmetry.
The symmetry planes of $\td{M}_x$ and $M_{xy}$ are shown in \cref{fig:onlyc4t}(a), and the plane of $\td{M_z}$ coincide with the cross section of \cref{fig:onlyc4t}.
Under quenched disorder, these symmetries are respected on average, and can also protect axion STIs characterized by $\bar{\theta} = \pi$.
Nevertheless, these additional symmetries do not affect the validity of our theory of $C_4T$-protected intrinsic STI.
First, we have explicitly proved that exact $C_4T$ ($(C_4T)^4=1$) forbids a clean axion insulator with $\theta=\pi$ (\cref{sec:momentum_proof,sec:tcs}).
Crucially, this proof remains valid even in the presence of additional symmetries, thereby confirming that the numerically identified STI in the main text retains its intrinsic nature when these symmetries are taken into account.
Second, since the average $C_4T$ suffices to protect $\ovl{\theta}=\pi$, perturbatively breaking other average symmetries cannot trivialize the STI.

We can further provide a concrete example by constructing a lattice model with more generic couplings, which breaks all additional symmetries in the clean limit.
Starting from the lattice model in \cref{fig:onlyc4t}(a) (identical to Eq.(2) of the main text), we first note that the $C_4T$ symmetry does not enforce the ``red bonds'' ($\gamma$ hoppings) in the two squares (denoted A and B in \cref{fig:onlyc4t}(b)) to be equal. Also, $\gamma_A$ and $\gamma_B$ can be complex.
As shown in \cref{fig:onlyc4t}(b), in any of the two squares (taking A as an example), $C_4T$ symmetry is preserved if hoppings along the arrows are $\gamma_A$, while hoppings against the arrows are ${\gamma_A}^*$, where $\gamma_A = |\gamma_A|e^{i\delta_A}$.
Physically, when $\delta_{A}\neq \delta_B$, magnetic fluxes of $\pm2(\delta_B -\delta_A)$ threads through the C and D regions respectively, thereby breaking the $S_4$ symmetry, which requires $\Phi_D=\Phi_C$.
To be generic, we set $|\gamma_A|\neq|\gamma_B|, \delta_A\neq \delta_B$. By examining the 32 general positions \cite{gallego_magnetic_2012,perez-mato_symmetry-based_2015} of the $P_C 4/nbm$ group one by one, we find that the symmetry group is downgraded to $P4'm'm$ (\#99.165 in the BNS setting). In addition to $C_4T$ and $C_2$, the model still respects $M_{xy} = \{m_{110}|\frac 1 2\frac 1 20\}$ and $M_xT=\{m_{100}^{'}|\frac 1 2 0 0\}$ (\cref{fig:onlyc4t}(b)), with $M_{xy}$ reversing the $\theta$ angle.

To break these two symmetries, longer range hoppings are necessary. We therefore construct the model illustrated in \cref{fig:onlyc4t}(c), where blue and brown solid lines denote hoppings with strengths $\pm\eta$ respectively.
For simplicity, we require $\eta \in \mathbb{R}$.
It is clear that either $M_{xy}$ or $M_xT$ maps $\eta\to -\eta$, and is broken by $\eta\neq 0$.
The magnetic space group of the model is now $P4'$ (\#75.3 in the BNS setting), which includes only $C_4T, (C_4T)^{-1}$ and $C_2$ in addition to lattice translations.
As discussed in \cref{sec:lattice-model}.A, the model does not respect any local symmetry.

To show that our result is also independent of the choice of disorder, we now introduce another type of disorder. 
While in the main text we add random imaginary components to $\gamma \in \mathbb{R}$, here we add random real components to the hoppings along the $z$ direction. 
Formally, we modify the term $\sum_{\bold{r},i} \ii \zeta_i t c_{\bold{r},i}^\dagger c_{\bold{r+e_z}, i} + c.c.$ in the clean Hamiltonian (which simulates chiral wires with $2t\sin{k_z}\rho_z\tau_z\sigma_z$, see \cref{sec:lattice-model}.A and the main text) to $\sum_{\bold{r},i} (\ii \zeta_i t + t'_{\bold{r},i}) c_{\bold{r},i}^\dagger c_{\bold{r+e_z}, i} + c.c.$, where $\bold{r}$ denotes the unit cells, $\bold{e_z}$ is the lattice vector along $z$ direction, $i\in\{1,...8\}$ labels lattice sites in each unit cell, and $\zeta_i = \pm 1$ corresponds to black and white dots respectively. The random variables $t'_{\bold{r},i} \in \mathbb{R}$ satisfy an uncorrelated Gaussian distribution $\langle t'_{\bold{r},i} t'_{\bold{r'},j}\rangle = W^2 \delta_{\bold{r}, \bold{r'}}\delta_{ij}$, where $W$ is the disorder strength. The disordered system respects $C_4T,C_2$ and lattice translational symmetries on average, but does not respect any other exact or average symmetry.

To demonstrate the existence of an STI phase, we perform transfer matrix calculations along the $x$ direction.
As illustrated in \cref{fig:onlyc4t}(d), for a suitable set of parameters, the normalized quasi-1D localization length $\Lambda(L)$ with periodic boundary conditions in both $y$ and $z$ directions monotonically decreases as the transversal size $L$ increases, indicating that the bulk is localized.
In contrast, $\Lambda(L)$ with open boundary condition in $z$ direction increases with $L$ for small values of $L$, signaling the formation of topological surface states, and tends to converge to a constant value as $L$ further increases. 
This clearly shows that the model is in an STI phase with a nontrivial surface.
Since the clean limit with eact $(C_4T)^4 = 1$ symmetry must be either gapless or trivial, the STI phase is intrinsic.
We hence conclude that the validity of our results is independent of any average symmetries other than $C_4T$, as well as the specific type of disorder.

\clearpage
\section{More about electron-electron interactions}
An interesting topic is the influence of electron-electron interactions on the intrinsic axion STI.
In particular, for the intrinsic STI to be an intrinsic average average symmetry protected topological (ASPT) phase, both the intrinsic nature and topological property of the STI should be stable against symmetry-preserving short-range interaction. 
Here, ``stable'' means that the properties survive before the interaction closes the bulk mobility gap or induces spontaneous symmetry breaking. 
Correspondingly, we should show that both the $C_4T$ obstruction of axion insulator (see \cref{sec:momentum_proof}) and topological property of STI are stable against interaction.
In this section, we establish these stabilities based on the topological crystal construction (\cref{sec:tcs}) and existing literature.

\subsection{Absence of axion TI with exact $(C_4T)^4=1$ symmetry in presence of interaction}

According to \cref{sec:tcs}, nontrivial topological crystals allowed by  $P4'$ are equivalent to Fig.1(c) in the main text. When $(C_4T)^4=1$, one can recombine the four chiral edge states on a hinge according to the eigenvalue of $(C_4T)^2$. For the $(C_4T)^2=-1$ subspace, the two chiral states form a pair of helical modes protected by the Kramers degeneracy, which obstructs the system from opening a nontrivial bulk gap.
To see the stability of this obstruction under interaction, we first notice that, since the topological property (especially the chiral edge) of Chern insulator is stable against interaction, the low energy physics of the topological crystal is still dominated by these helical modes, so we can project interaction into these helical modes.
Hence, the interaction stability of such an obstruction is equivalent to the interaction stability of these helical modes.
If we further limit the short-range interaction to be effective only within each hinge, the problem is simplified to the interaction stability of a single pair of helical modes, which has been extensively verified in quantum spin Hall systems \cite{wu2006helical,xu2006stability,chou2018gapless}.
Even if we allow the interaction to involve finite topological crystal unit cells, as long as $C_4T$ is preserved, there must be an odd number of pairs of helical modes involved (one at the $C_4T$ center and $4n$ around it).
And it is also well known that odd pairs of helical modes remain gapless and delocalized in the presence of interaction \cite{xu2006stability}. 
Hence, we can conclude that the $C_4T$ obstruction for axion insulator is also stable against interaction.

In the remaining of this subsection, we sketch the stability of a helical mode under short-range repulsive interaction by bosonization (One can find detailed discussions in Refs.~\cite{chou2018gapless,wu2006helical}). We denote the field operator of chiral state in the Kramers pair going along the $+z$ ($-z$) direction by $\psi_{\uparrow}$ ($\psi_{\downarrow}$ ), and suppose the $C_4T$ action is $\psi_{\uparrow}\rightarrow\psi_{\downarrow},\,\psi_{\downarrow}\rightarrow-\psi_{\uparrow},\,i\rightarrow-i$. The free Hamiltonian and $C_4T$ invariant two-body interactions are
\begin{equation}
\label{eq:fermion-helical-luttinger}
    \begin{aligned}
        H&=H_0+H_{cfw}+H_{ncfw}+H_{um}\\
        H_0 &= v_f \int dz \left[\psi^\dagger_{\uparrow}(z)i\partial_z\psi_{\uparrow}(z)-\psi^\dagger_{\downarrow}(z)i\partial_z\psi_{\downarrow}(z)\right]\\
        H_{cfw} &= g_{4} \int dz \left[\psi^\dagger_{\uparrow}(z)\psi^\dagger_{\uparrow}(z+\alpha)\psi_{\uparrow}(z+\alpha)\psi_\uparrow(z)+\psi^\dagger_{\downarrow}(z)\psi^\dagger_{\downarrow}(z+\alpha)\psi_{\downarrow}(z+\alpha)\psi_\downarrow(z)\right]\\
        H_{ncfw} &= g_{2} \int dz \left[\psi^\dagger_{\uparrow}(z)\psi_{\uparrow}(z)\psi^\dagger_{\downarrow}(z)\psi_\downarrow(z)\right]\\
        H_{um}&=g_{3}\int dz [e^{-i\delta Qz}\psi^\dagger_{\uparrow}(z)\psi^\dagger_{\uparrow}(z+\alpha)\psi_{\downarrow}(z+\alpha)\psi_\downarrow(z)+\text{h.c.}]
    \end{aligned}
\end{equation}
where $H_{cfw}$ is the chiral forward scattering, $H_{ncfw}$ is the non-chiral forward scattering, and $H_{um}$ is the Umklapp backward scattering.
$\alpha$ is a microscopic cutoff of interaction range, and the phase factor $e^{-i \delta Qz}$ comes from the lattice translation symmetry of the corresponding Chern insulators. 
Concretely, $\delta Q=\frac{2\pi}{a}-4k_F$ measures the incommensuration between the helical electron Fermi vector ($k_F$) and lattice reciprocal vector ($2\pi/a$ with $a$ the
lattice constant of Chern insulators).
Here, we use nearly onsite interactions to simplify the discussion, reader who are not satisfied with this can consult Ref.~\cite{kainaris2014conductivity} for a perturbation calculation of more realistic interactions.

Following the standard abelian bosonization procedure \cite{senechal2004theoretical}, we define $\psi_{\uparrow/\downarrow}(z)=\frac{1}{\sqrt{2\pi\alpha}}e^{\mp i\sqrt{4\pi}\phi_{\uparrow/\downarrow}}$, $\varphi=\phi_{\uparrow}+\phi_{\downarrow}$, $\vartheta=\phi_{\uparrow}-\phi_{\downarrow}$, where $\phi_{\uparrow,\downarrow}$ are chiral boson fields and we have ignored the Klein factor. 
Then, \cref{eq:fermion-helical-luttinger} becomes
\begin{equation}
\label{eq:boson-helical-luttinger}
\begin{aligned}
    H&=\int dz \left\{\frac{v}{2}\left[K(\partial_z\vartheta)^2+\frac1K(\partial_z\varphi)^2\right]+\frac{g_3}{2\pi^2\alpha^2}\cos(\sqrt{16\pi}\varphi-\delta Qz)\right\}\\
    v&=\sqrt{\left(v_F+\frac{g_4}{\pi}\right)^2-\left(\frac{g_2}{\pi}\right)^2},\quad K=\sqrt{\frac{v_F-g_2/\pi+g_4/\pi}{v_F+g_2/\pi+g_4/\pi}}
\end{aligned}
\end{equation}
One can see that the forward scatterings are absorbed into the coefficients of bosonic kinetic terms, while the backward scattering becomes a sine-Gordon term. 
For non-interacting system the Luttinger parameter $K=1$. And the repulsive interaction ($g_{2,3,4}>0$) will give $K<1$. 

In commensurate filling ($\delta Q=0$), \cref{eq:boson-helical-luttinger} becomes the well-known sine-Gordon model except that $\cos(\sqrt{8\pi}\varphi)$ is now replaced by $\cos(\sqrt{16\pi}\varphi)$ \cite{senechal2004theoretical}.
Therefore, the interaction can open a charge gap only if $K<1/2$, where a spontaneous $C_4T$-symmetry-breaking is induced far from the non-interacting limit.
In general incommensurate filling ($\delta Q\neq 0$), the sine-Gordon term is more irrelevant, which can be understood by the quasi-momentum conservation imposed by the ion lattice (a detailed discussion on incommensuration of Luttinger liquid can be found in Ref.~\cite{giamarchi1992conductivity}).
In summary, before spontaneous symmetry breaking, symmetry-preserving short-range interaction will not localize the helical mode and remove the obstruction of the $C_4T$ topological crystal.

\subsection{Interaction stability of delocalized intrinsic STI surface}
With disorder degrades the $C_4T$ to an average symmetry, the bulk helical modes become localized.
Once we take open boundary condition in $z$-direction, according to Fig.~1(d) in the main text, there will be delocalized surface states dominating the low energy physics.
Also, as the decorated Chern insulators are stable against the interaction, this low-energy subspace remains intact with the presence of interaction.
Thus, one can project interaction onto the gapless surface network of STI. 
With the belief of bulk-boundary correspondence, we assume that the interaction stability of topological property of STI is equivalent to the stability of its delocalized surface. 

According to the topological crystal construction, the surface of the intrinsic STI can be described by a critical Chalker-Coddington network \cite{chalker}, which has the same universality class as the non-interacting quantum Hall transition (or topological transition of Chern insulators). We can now convert the question into the interaction stability of quantum Hall transition, which is an extensively studied topic \cite{lee1996effects,apalkov2003interplay,pruisken2008non}.
It turns out that short-range interaction is irrelevant for quantum Hall transition, i.e., interaction vanishes along the renormalization group (RG) flow.
Hence, the delocalized surface---manifesting nontrivial topology of intrinsic STI in this work---is stable against short-range interactions. 
Further, it is well known that, although unscreened Coulomb interaction changes the dynamical exponent of quantum Hall criticality from $2$ to $1$, it still cannot localize the transition point \cite{huckestein1999integer,wang2002electron,pruisken2008non,kumar2022interaction}.
In this perspective, the topological property of intrinsic STI may also be stable against long-range interactions. 

In the following, we sketch the perturbative proof for the irrelevance of short-range interaction on Hall criticality given by Ref.~\cite{lee1996effects}. 
We start with the action of interacting electrons under magnetic field ($B(\mathbf{r})=z\cdot(\nabla\times\mathbf{A}(\mathbf{r}))$) and disorder scalar potential $V(\mathbf{r})$.
\begin{equation}
\begin{aligned}
    S&=S_0+S_{int}\\
    S_0 &= \int d^2\mathbf{r}\sum_{n} \bar\psi_{\omega_n}(\mathbf{r})\left[-i\omega_n-\frac{(\nabla-i|e|\mathbf{A}(\mathbf{r}))^2}{2m}+V(\mathbf{r})\right]\psi_{\omega_n}(\mathbf{r})\\
    S_{int}&=\frac{1}{\beta}\sum_{\omega_{1,2,3,4}}\delta_{\omega_1+\omega_2,\omega_3+\omega_4}\int d^2\mathbf{r}_1 d^2\mathbf{r}_2 U(|\mathbf{r}_1-\mathbf{r}_2|)\bar\psi_{\omega_1}(\mathbf{r}_1)\bar\psi_{\omega_2}(\mathbf{r}_2)\psi_{\omega_4}(\mathbf{r}_2)\psi_{\omega_3}(\mathbf{r}_1)\\
    U(|\mathbf{r}|)&=\frac{g}{|\mathbf{r}|^{\lambda}}
\end{aligned}
\end{equation}
where $\psi_{\omega_n}$ is Grassmann field, $\frac{1}{\beta}$ is the scale of imaginary time dimension, and $\omega_n=\frac{(2n+1)\pi}{\beta}$ are the Matsubara frequency. 
For sufficiently weak interaction, one can compute the partition function perturbatively $\langle\langle e^{-S_{int}}\rangle\rangle= e^{-\langle\langle S_{int} \rangle\rangle+\frac{1}{2}\langle\langle S_{int}^2\rangle\rangle+\cdots}$, where $\langle\langle\cdots\rangle\rangle=\langle\Tr(e^{-S_0}\cdots)\rangle_{dis}$ stands for average over non-interacting action and disorder configurations.

To justify the irrelevance of interaction, corrections $\langle\langle S_{int}\rangle\rangle$ and $\langle\langle S_{int}^2\rangle\rangle$ should vanish in the thermodynamic limit and low-energy limit. 
However, due to the lack of analytic knowledge of quantum Hall criticality at half filling of disordered Landau level, one can only estimate $\langle\langle S_{int}\rangle\rangle$ and $\langle\langle S_{int}^2\rangle\rangle$ numerically on finite systems.
Also, since we focus on the universal behavior of quantum Hall criticality, we can replace $S_0$ by another system in the same universal class with a higher simulation efficiency.
In Ref.~\cite{lee1996effects}, the estimation is taken on the Chalker-Coddington model, which is further reformulated into a $U(2n)|_{n\rightarrow0}$ Hubbard model \cite{lee1996transitions}.

To estimate $\langle\langle S_{int}\rangle\rangle$, one can first estimate the symmetrized vertex $\Gamma^{(4)}$ defined by
\begin{equation}
\begin{aligned}
    \langle\langle S_{int}\rangle\rangle&=\frac{1}{4\beta}\sum_{\omega_{1,2,3,4}}\delta_{\omega_1+\omega_2,\omega_3+\omega_4}\int d^2\mathbf{r}_1 d^2\mathbf{r}_2 U(|\mathbf{r}_1-\mathbf{r}_2|)\delta_{\omega_1,\omega_3}\delta_{\omega_2,\omega_4}\Gamma^{(4)}(\mathbf{r}_1,\mathbf{r}_2,\omega_1,\omega_2)\\
    \Gamma^{(4)}(\mathbf{r}_1,\mathbf{r}_2,\omega_1,\omega_2)&=\left\langle\left\langle \bar B(\mathbf{r}_1,\mathbf{r}_2,\omega_1,\omega_2)B(\mathbf{r}_1,\mathbf{r}_2,\omega_1,\omega_2)\right\rangle\right\rangle\\
    B(\mathbf{r}_1,\mathbf{r}_2,\omega_1,\omega_2)&=\psi_{\omega_1}(\mathbf{r}_1)\psi_{\omega_2}(\mathbf{r}_2)+\psi_{\omega_2}(\mathbf{r}_1)\psi_{\omega_1}(\mathbf{r}_2)\\
\end{aligned}
\end{equation}

At the non-interacting fix point, we can take the scaling dimensions of $\bar\psi_{\omega_1}\psi_{\omega_2}$ and $\omega$ as 0 and 2, respectively. 
Hence, for finite system, $\Gamma^{(4)}$ has the scaling form
\begin{equation}
    \Gamma^{(4)}(\mathbf{r}_1,\mathbf{r}_2,\omega_1,\omega_2)=\mathcal{F}_1(|\mathbf{r}_1-\mathbf{r}_2|/L,\omega_1L^2,\omega_2L^2)
\end{equation}
In Ref.~\cite{lee1996effects}, it is numerically found that $\mathcal{F}_1\sim(|\mathbf{r}_1-\mathbf{r}_2|/L)^{0.65}$ as $L\rightarrow\infty$, and the leading order mainly comes from regions satisfying $|\mathbf{r}_1-\mathbf{r}_2|/L\ll 1$, $\omega_{1,2}L^2<O(1)$. Hence, for finite system, the first-order correction to $S_0$ is proportional to
\begin{equation}
\begin{aligned}
    \langle\langle S_{int} \rangle\rangle
    &= \frac{gL^{4-\lambda}}{4\beta}\sum_{\omega_{1,2}}\int_{\alpha/L}^1 d^2\left(\frac{\mathbf{r}_1+\mathbf{r}_2}{2L}\right)\int_{\alpha/L}^1 d^2\left(\frac{\mathbf{r}_1-\mathbf{r}_2}{L}\right) \left(\frac{L}{r}\right)^\lambda\mathcal{F}_1(|\mathbf{r}_1-\mathbf{r}_2|/L,\omega_1L^2,\omega_2L^2)\\
    &\sim(gL^{2-\lambda})[A+B(\alpha/L)^{2.65-\lambda}] 
\end{aligned}
\end{equation}
where $\alpha$ is the microscopic cutoff. 
As long as $\lambda>2.65$, the above equation approaches $(Bg\alpha^{2.65-\lambda}) L^{-0.65}$ in the thermodynamic limit, which eventually vanishes. 

One can perform a similar analysis for  $\langle\langle S_{int}^2 \rangle\rangle$ and finds that it also vanishes in the thermodynamic limit \cite{lee1996effects}.
We therefore conclude that the quantum Hall criticality and surface delocalization in our STI robustly survive against perturbative short-range interactions that decay faster than faster than $r^{-2.65}$.
Despite the absence of non-perturbative proofs---the strongest one we find is a two-loop RG analysis of Finkel’stein nonlinear sigma model \cite{pruisken2008non}---we believe that this stability persists to finite-strength interactions.

\clearpage
\section{Possible experimental signatures and realizations of the axion STI}
In this section, we discuss possible experimental signatures of axion STIs that distinguish them from axion TIs protected by exact symmetries.
We propose two different phenomena that are in principle observable -- a {\it statistical} Witten effect, which reflects the fluctuation of the local axion angle, and unique surface properties that can be probed by spectroscopy measurements.
We note that while the intrinsic nature of the STI is important theoretically, it is difficult to control the disorder strength in experiments.
Therefore, experimental measurements are usually limited to a given disorder strength, making it difficult to distinguish intrinsic and extrinsic STIs.
As a result, the following experimental signatures apply to both extrinsic and intrinsic axion STIs protected by average symmetries, but not axion TIs protected by exact symmetries.

The Witten effect is a natural result of the nontrivial axion angle $\theta$, which gives rise to the axion Lagrangian term (in SI unit)
\begin{equation}
    \mathcal{L}_\theta(\mathbf{r},t)=\frac{e^2}{4\pi^2 \hbar}\theta(\mathbf{r},t) \;\mathbf{E}(\mathbf{r},t) \cdot \mathbf{B}(\mathbf{r},t)
\end{equation}
where $\theta(\mathbf{r},t)$ can be dependent on both $\mathbf{r}$ and  $t$ in general. If we further postulate the existence of magnetic charges, the Maxwell equations are modified to \cite{wilczek1987two}
\begin{equation}
\label{eq:Maxwell-axion}
    \begin{aligned}
        \nabla\cdot\mathbf{E}&=\frac{\rho_e}{\epsilon_0}-\frac{c\alpha}{\pi}\nabla\theta\cdot \mathbf{B}\\
        \nabla\cdot \mathbf{B}&=\mu_0\rho_m\\
        \nabla\times\mathbf{E}&=-\frac{\partial\mathbf{B}}{\partial t}-\mu_0\mathbf{J}_m\\
        \nabla\times\mathbf{B}&=\frac{1}{c^2}\frac{\partial\mathbf{E}}{\partial t}+\frac{\alpha}{\pi c}\left(\frac{\partial\theta}{\partial t}\mathbf{B}+\nabla\theta\times\mathbf{E}\right)+\mu_0\mathbf{J}_e
    \end{aligned}
\end{equation}
where $\alpha=\frac{e^2}{4\pi\epsilon_0\hbar c}$ is the fine-structure constant, and $\rho_e$ and $\mathbf{J}_e$ ($\rho_m$ and $\mathbf{J}_m$) represent (normal) electric (magnetic) charge density and current, respectively. 
Also, we take the Ampere-meter convention for magnetic charge.
\cref{eq:Maxwell-axion} tells us that, apart from (non-axion) charge $\rho_e$, an ``axion charge'' term $-\frac{c\alpha}{\pi}\nabla\theta\cdot \mathbf{B}$ also contributes to the divergence $\nabla\cdot\mathbf{E}$.
If $\theta$ is a basic field within electrodynamics, this term is {\it not} realized by a real electric charge.
In condensed matter systems, however, $\theta$ is an emergent quantity, so this term describes the effect of an excess electron charge $\delta\rho_e = -\frac{c\epsilon_0\alpha}{\pi}\nabla\theta\cdot \mathbf{B}$ induced by the magnetic field, which concentrates in regions with considerable $|\nabla \theta|$.

Let us first consider an axion TI protected by exact $\theta$-odd symmetry, and place a magnetic monopole with unit strength $\Phi_0 = h/e$ in its bulk, leading to $\nabla\cdot\mathbf{B}=\Phi_0\mathbf{\delta}(\mathbf{r})$.
As pointed out in Ref.~\cite{PhysRevB.108.155104}, while most part of the the bulk has $\theta=\pi \mod 2\pi$, a ``trivial insulator island'' with $\theta=0 \mod 2\pi$ forms in a small region around the monopole due to the dense magnetic field, leading to a sharp domain wall with effective Chern number $C_{\text{eff}}=\frac \theta {2\pi} = (n+\frac 1 2)$ around the "island", where $n$ is a non-universal integer depending on microscopic details of the monopole.
Since $\nabla \theta$ is only nonzero on the domain wall, a half quantized charge $\delta Q_e  =(n+\frac 1 2)e$ is bound to the monopole, with the characteristic radius of the domain wall the effective monopole size.
This is the lattice version of the Witten effect, which was proposed in Ref.~\cite{TRS_AXION} and extensively studied in Refs.~\cite{PhysRevB.82.035105, PhysRevB.108.155104}.
Remarkably, the formation of the domain wall is verified analytically for the continuum Dirac fermion model and numerically for a lattice model in Ref.~\cite{PhysRevB.108.155104}.

In contrast to axion TIs protected by exact symmetries, axion STIs studied in this work respect $\theta$-odd symmetries only on average.
Therefore, their local axion angle $\theta(\mathbf{r})$ can fluctuate with $\mathbf{r}$, with only the average value $\bar\theta$ quantized. 
If the correlation length of the disorder (which also amounts to the correlation length of $\theta(\mathbf{r})$) is larger than the  effective monopole size (which is assumed to be larger than the lattice constant and the electron correlation length), a magnetic monopole at $\mathbf{r}_0$ in the STI should trap an \emph{unquantized} electric charge $\delta Q_e=(n+\frac{\theta(\mathbf{r}_0)}{2\pi})e$ around it.
In addition, there will be a nonzero excess charge distribution $\delta \rho_e$ even away from (but near) the monopole due to the nonzero $|\nabla\theta|$ caused by disorder fluctuations.
We hence term this phenomenon as \emph{statistical Witten effect}.
If one is able to randomly insert dilute identical magnetic monopoles to the STI and measure the charge response with high spatial resolution, one should find that the excess charge distributes with two characteristic lengths---the effective monopole size and the disorder correlation length.
$\delta Q_e$ bounded to each monopole (characterized by the monopole size) should average to a half-quantized value, while the distributing $\delta \rho_e$ (characterized by the disorder correlation length, and enveloped by a quadratic decay from the monopole) average to zero.
We stress that for axion TIs protected by exact symmetries, even with symmetry-preserving disorder, the local axion angle is quantized and uniform, and the monopole size should be the only characteristic length for monopole response.

While fundamental magnetic monopoles have not been observed in nature, there are a couple of proposals to experimentally realize {\it emergent} 
magnetic monopoles in condensed matter systems.
Indeed, emergent monopole-like behavior has been theoretically proposed \cite{castelnovo2008magnetic} and experimentally observed \cite{doi:10.1126/science.1177582, doi:10.1126/science.1178868} in frustrated magnetic systems called ``spin ice''.
Ref.~\cite{Sasaki_2014} proposed that the Witten effect may be observed on domain walls between spin ice compounds and axion TIs.
It is also theoretically possible to search for a material platform that realizes both STI and spin ice behaviors simultaneously, which is far beyond the scope of this work.
Effective ``planar magnetic monopole'' has also been proposed in thin films of axion TIs \cite{PhysRevLett.103.066402}.
When a finite electric bias is applied to the film, electron and hole Fermi surfaces form on the top and bottom surfaces respectively.
For a thin film, the attractive interaction between electron and hole on opposite surfaces can lead to exciton condensates, whose order parameter contains vortices---point-like topological defects indistinguishable from ``planar monopoles'' for electrons, which can also trap half-quantized charges in clean TIs.
Since our intrinsic STI model in the main text also has delocalized surface states, we expect the exciton condensation proposal to be valid for axion STIs, allowing the statistical Witten effect in thin films under electric bias.

Apart from the statistical Witten effect, axion STIs also have unique surface properties.
As illustrated Fig.~1d of the main text, the nontrivial surfaces of axion STIs consist of insulating regions with random half-quantized Hall conductance and chiral edge states between them.
In contrast, the surface of a TI protected by exact symmetry (without local symmetry breaking disorder) is characterized by gapless Dirac cone(s). 
Hence, for a spectroscopy measurement with spatial resolution smaller than the disorder correlation length, it is possible to observe both locally insulating and conducting regions on the STI surface, while only conducting regions can be observed on the conventional TI surface. 
Depending on the sample fabrication, the spatial resolution in need could be reached by nano-ARPES whose resolution has reached hundreds of nanometers \cite{iwasawa2019buried}. 

To close this section, we briefly discuss possible experimental realization of the intrinsic axion STI.
While we are not able to propose a specific material at this moment, we notice that there is one material that can realize similar phase transitions under disorder --- the Dirac semimetal $\text{Na}_3\text{Bi}$ \cite{PhysRevB.85.195320, doi:10.1126/science.1245085}.
In the clean limit, $\text{Na}_3\text{Bi}$ respects crystalline symmetry $P6_3/mmc$ and time reversal symmetry $T$ with significant spin-orbit coupling.
As discussed in Ref.~\cite{PhysRevB.85.195320}, the band structure of $\text{Na}_3\text{Bi}$ has two Dirac points on the $\Gamma-A$ line.
These Dirac points are protected by $C_3$ symmetry, since each Dirac node comprises two Weyl nodes with different $C_3$ eigenvalues, which prevents two Weyl nodes from coupling with each other and opening a gap.
The two Dirac nodes are related to each other by $T$, and both lie exactly at the Fermi level of an undoped $\text{Na}_3\text{Bi}$ sample.

With the presence of non-magnetic quenched disorder, $C_3$ is locally broken, yet $T$ is preserved. The two Weyl nodes can now locally gap each other, resulting in a localized state.
Since there is a band inversion between $\text{Na}-3s$ and $\text{Bi}-6p$ about $0.5eV$ away from the Fermi energy, as long as the disorder strength is much smaller than $0.5eV$, the gap opening will result in a nontrivial $\mathbb{Z}_2$ TI protected by $T$.
On the other hand, the system must be a trivial Anderson insulator at infinity disorder strength.
Hence, we expect successive (semi)metal-TI-metal-trivial insulator transitions as the disorder strength increases from zero to infinity.
If we further introduce quenched magnetic disorder without closing the bulk mobility gap, $T$ will become an average symmetry, and the TI phase at weak disorder becomes an axion STI characterized by fluctuating $\theta$ field with $\bar{\theta}=\pi$.
The resultant (semi)metal-STI-metal-trivial insulator transition is similar to our intrinsic axion STI model.
In addition, with fluctuating $\theta$ field, magnetically disordered $\text{Na}_3\text{Bi}$ should exhibit similar statistical Witten effect and surface spectroscopic signature as the intrinsic axion STI.

Despite these similarities to the intrinsic axion STI model, $\text{Na}_3\text{Bi}$ with magnetic disorder is actually an {\it extrinsic} STI --- there exists an adiabatic deformation connecting the STI to a clean system with exact $T$ and broken $C_3$ (for example, $\text{Na}_3\text{Bi}$ under strain, which is also a nontrivial TI).

\clearpage
\section{Generalization to $(C_2T)^2 = -1$ symmetry}

\begin{figure}[h]
\centering
\includegraphics[width=0.8\linewidth]{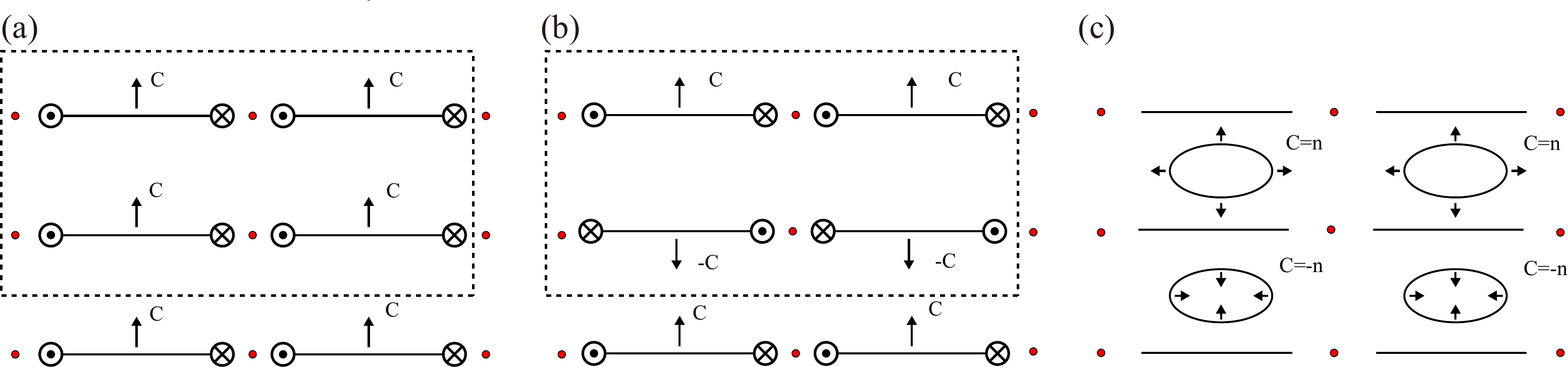}
\caption[]{ 
    Topological crystals with $(C_2T)^2 = -1$ symmetry. Red dots show the $C_2T$ axes, which are occupied by intersecting hinges of Chern insulators. The dashed boxes show unit cells.
    (a) The topological crystal for layered Chern insulators. Those with $C \in odd$ can be insulating only in presence of disorder.
    (b) The topological crystal with opposite Chern numbers for adjacent layers. The $C=1$ case become an intrinsic axion STI when average $C_2T$ preserving disorder is introduced on the hinges.
    (c) By creating $C_2T$-symmetric Chern bubbles, a $\Delta C = 2n$ change can be realized adiabatically for the topological crystal illustrated in (b).
   }
\label{fig:C2T}
\end{figure}

Now, we generalize our discussion to $(C_2T)^2 = -1$ symmetry following the topological crystal method.
Since $C_2T$ is an anti-unitary symmetry squaring to $-1$, it can also protect Kramers' pairs and forbid the opening of symmetric gaps, ruling out the existence of certain clean TCIs. 
Consider the topological crystal illustrated in \cref{fig:C2T}(a), which consists of layers of Chern insulators with the same Chern number $C$.
There are $|C|$ pairs of chiral and anti-chiral modes on each $C_2T$ axis. 
If $C$ is even, a $C_2T$-symmetric gap is allowed on each $C_2T$ axis due to the net chirality, resulting in a layered Chern insulator with even 3D Chern number $C \in 2 \mathbb{Z}$.
In contrast, when $C$ is odd, Kramers' degeneracy forbids the opening of $C_2T$-symmetric gaps on the $C_2T$ axes, so layered Chern insulators with odd 3D Chern numbers are forbidden by exact $C_2T$ symmetry.
Introducing disorder on the hinges which breaks $C_2T$ exactly but preserves it on average will lift the Kramers' degeneracy and allow the $C_2T$ axes with odd $C$ be localized. 
We have assumed the normals of layered Chern insulators to be $\pm \mathbf{e}_y$ in above discussions, while decorations of Chern insulators with normals $\pm \mathbf{e}_x$ can be made similarly, offering two 3D Chern numbers $C_x, C_y$ in total. 
Therefore, breaking $(C_2T)^2 = -1$ symmetry from exact to average enriches the classification of layered Chern insulators from $2 \mathbb{Z} \times 2 \mathbb{Z}$ to $\mathbb{Z} \times \mathbb{Z}$, with $C_x \in odd$ or $C_y \in odd$ cases being intrinsic STIs.

Similarly, consider the topological crystal illustrated in \cref{fig:C2T}(b), which consists of layers of Chern insulators with opposite Chern numbers $C$ and $-C$ for adjacent layers.
The bubble equivalence illustrated in \cref{fig:C2T}(c) adiabatically changes $C$ by $\Delta C = 2n$ without breaking the $C_2T$ symmetry, making $C\in even$ equivalent to $C=0$, and $C \in odd$ equivalent to $C=1$.
Since $C = 0$ is obviously trivial, we only have to consider the $C = 1$ case.
When $(C_2T)^2 = -1$ is exact, there is one pair of chiral and anti-chiral states on each $C_2T$ axis, and the Kramers' degeneracy forbids the opening of a symmetric gap as discussed above.
When average $C_2T$ preserving disorder is introduced on the hinges, the $C_2T$ axes can be localized, while the Chern insulators remain robust. 
The disordered construction resembles the one illustrated in Fig.~1(a) of the main text and forms an axion STI with $\bar{\theta} = \pi$. 
Therefore, average $C_2T$ symmetry with $(C_2T)^2 = -1$ can also protect intrinsic axion STIs with a $\mathbb{Z}_2$ invariant similar to the $(C_4T)^4 = 1$ symmetry.

In summary, average $(C_2T)^2 = -1$ symmetry protects a $\mathbb{Z}\times \mathbb{Z}\times \mathbb{Z}_2$ STI classification, where the first two topological indices $C_{x,y}$ denote 3D Chern numbers along $x$ and $y$ directions, and the third one $\delta$ indicates axion STI when $C_x = C_y = 0, \delta=1$.
Since states denoted by $\{C_x, C_y, \delta\} = \{2m, 2n, 0\}$ are extrinsic STIs, the classification of intrinsic STIs protected by average $(C_2T)^2 = -1$ symmetry is $\mathbb{Z}_2\times \mathbb{Z}_2\times \mathbb{Z}_2$.

\begin{thebibliography}{105}%
\makeatletter
\providecommand \@ifxundefined [1]{%
 \@ifx{#1\undefined}
}%
\providecommand \@ifnum [1]{%
 \ifnum #1\expandafter \@firstoftwo
 \else \expandafter \@secondoftwo
 \fi
}%
\providecommand \@ifx [1]{%
 \ifx #1\expandafter \@firstoftwo
 \else \expandafter \@secondoftwo
 \fi
}%
\providecommand \natexlab [1]{#1}%
\providecommand \enquote  [1]{``#1''}%
\providecommand \bibnamefont  [1]{#1}%
\providecommand \bibfnamefont [1]{#1}%
\providecommand \citenamefont [1]{#1}%
\providecommand \href@noop [0]{\@secondoftwo}%
\providecommand \href [0]{\begingroup \@sanitize@url \@href}%
\providecommand \@href[1]{\@@startlink{#1}\@@href}%
\providecommand \@@href[1]{\endgroup#1\@@endlink}%
\providecommand \@sanitize@url [0]{\catcode `\\12\catcode `\$12\catcode `\&12\catcode `\#12\catcode `\^12\catcode `\_12\catcode `\%12\relax}%
\providecommand \@@startlink[1]{}%
\providecommand \@@endlink[0]{}%
\providecommand \url  [0]{\begingroup\@sanitize@url \@url }%
\providecommand \@url [1]{\endgroup\@href {#1}{\urlprefix }}%
\providecommand \urlprefix  [0]{URL }%
\providecommand \Eprint [0]{\href }%
\providecommand \doibase [0]{http://dx.doi.org/}%
\providecommand \selectlanguage [0]{\@gobble}%
\providecommand \bibinfo  [0]{\@secondoftwo}%
\providecommand \bibfield  [0]{\@secondoftwo}%
\providecommand \translation [1]{[#1]}%
\providecommand \BibitemOpen [0]{}%
\providecommand \bibitemStop [0]{}%
\providecommand \bibitemNoStop [0]{.\EOS\space}%
\providecommand \EOS [0]{\spacefactor3000\relax}%
\providecommand \BibitemShut  [1]{\csname bibitem#1\endcsname}%
\let\auto@bib@innerbib\@empty
\bibitem [{\citenamefont {Kane}\ and\ \citenamefont {Mele}(2005)}]{TI1}%
  \BibitemOpen
  \bibfield  {author} {\bibinfo {author} {\bibfnamefont {C.~L.}\ \bibnamefont {Kane}}\ and\ \bibinfo {author} {\bibfnamefont {E.~J.}\ \bibnamefont {Mele}},\ }\bibfield  {title} {\enquote {\bibinfo {title} {${Z}_{2}$ topological order and the quantum spin hall effect},}\ }\href {\doibase 10.1103/PhysRevLett.95.146802} {\bibfield  {journal} {\bibinfo  {journal} {Phys. Rev. Lett.}\ }\textbf {\bibinfo {volume} {95}},\ \bibinfo {pages} {146802} (\bibinfo {year} {2005})}\BibitemShut {NoStop}%
\bibitem [{\citenamefont {Bernevig}\ \emph {et~al.}(2006)\citenamefont {Bernevig}, \citenamefont {Hughes},\ and\ \citenamefont {Zhang}}]{TI2}%
  \BibitemOpen
  \bibfield  {author} {\bibinfo {author} {\bibfnamefont {B.~Andrei}\ \bibnamefont {Bernevig}}, \bibinfo {author} {\bibfnamefont {Taylor~L.}\ \bibnamefont {Hughes}}, \ and\ \bibinfo {author} {\bibfnamefont {Shou-Cheng}\ \bibnamefont {Zhang}},\ }\bibfield  {title} {\enquote {\bibinfo {title} {Quantum spin hall effect and topological phase transition in hgte quantum wells},}\ }\href {\doibase 10.1126/science.1133734} {\bibfield  {journal} {\bibinfo  {journal} {Science}\ }\textbf {\bibinfo {volume} {314}},\ \bibinfo {pages} {1757--1761} (\bibinfo {year} {2006})}\BibitemShut {NoStop}%
\bibitem [{\citenamefont {Fu}\ \emph {et~al.}(2007)\citenamefont {Fu}, \citenamefont {Kane},\ and\ \citenamefont {Mele}}]{fu_topological_2007}%
  \BibitemOpen
  \bibfield  {author} {\bibinfo {author} {\bibfnamefont {Liang}\ \bibnamefont {Fu}}, \bibinfo {author} {\bibfnamefont {C.~L.}\ \bibnamefont {Kane}}, \ and\ \bibinfo {author} {\bibfnamefont {E.~J.}\ \bibnamefont {Mele}},\ }\bibfield  {title} {\enquote {\bibinfo {title} {Topological insulators in three dimensions},}\ }\href {\doibase 10.1103/PhysRevLett.98.106803} {\bibfield  {journal} {\bibinfo  {journal} {Phys. Rev. Lett.}\ }\textbf {\bibinfo {volume} {98}},\ \bibinfo {pages} {106803} (\bibinfo {year} {2007})}\BibitemShut {NoStop}%
\bibitem [{\citenamefont {Hasan}\ and\ \citenamefont {Kane}(2010)}]{TI3}%
  \BibitemOpen
  \bibfield  {author} {\bibinfo {author} {\bibfnamefont {M.~Z.}\ \bibnamefont {Hasan}}\ and\ \bibinfo {author} {\bibfnamefont {C.~L.}\ \bibnamefont {Kane}},\ }\bibfield  {title} {\enquote {\bibinfo {title} {Colloquium: Topological insulators},}\ }\href {\doibase 10.1103/RevModPhys.82.3045} {\bibfield  {journal} {\bibinfo  {journal} {Rev. Mod. Phys.}\ }\textbf {\bibinfo {volume} {82}},\ \bibinfo {pages} {3045--3067} (\bibinfo {year} {2010})}\BibitemShut {NoStop}%
\bibitem [{\citenamefont {Qi}\ and\ \citenamefont {Zhang}(2011)}]{TI4}%
  \BibitemOpen
  \bibfield  {author} {\bibinfo {author} {\bibfnamefont {Xiao-Liang}\ \bibnamefont {Qi}}\ and\ \bibinfo {author} {\bibfnamefont {Shou-Cheng}\ \bibnamefont {Zhang}},\ }\bibfield  {title} {\enquote {\bibinfo {title} {Topological insulators and superconductors},}\ }\href {\doibase 10.1103/RevModPhys.83.1057} {\bibfield  {journal} {\bibinfo  {journal} {Rev. Mod. Phys.}\ }\textbf {\bibinfo {volume} {83}},\ \bibinfo {pages} {1057--1110} (\bibinfo {year} {2011})}\BibitemShut {NoStop}%
\bibitem [{\citenamefont {Fu}(2011)}]{TCI1}%
  \BibitemOpen
  \bibfield  {author} {\bibinfo {author} {\bibfnamefont {Liang}\ \bibnamefont {Fu}},\ }\bibfield  {title} {\enquote {\bibinfo {title} {Topological crystalline insulators},}\ }\href {\doibase 10.1103/PhysRevLett.106.106802} {\bibfield  {journal} {\bibinfo  {journal} {Phys. Rev. Lett.}\ }\textbf {\bibinfo {volume} {106}},\ \bibinfo {pages} {106802} (\bibinfo {year} {2011})}\BibitemShut {NoStop}%
\bibitem [{\citenamefont {Hsieh}\ \emph {et~al.}(2012)\citenamefont {Hsieh}, \citenamefont {Lin}, \citenamefont {Liu}, \citenamefont {Duan}, \citenamefont {Bansil},\ and\ \citenamefont {Fu}}]{hsieh2012topological}%
  \BibitemOpen
  \bibfield  {author} {\bibinfo {author} {\bibfnamefont {Timothy~H}\ \bibnamefont {Hsieh}}, \bibinfo {author} {\bibfnamefont {Hsin}\ \bibnamefont {Lin}}, \bibinfo {author} {\bibfnamefont {Junwei}\ \bibnamefont {Liu}}, \bibinfo {author} {\bibfnamefont {Wenhui}\ \bibnamefont {Duan}}, \bibinfo {author} {\bibfnamefont {Arun}\ \bibnamefont {Bansil}}, \ and\ \bibinfo {author} {\bibfnamefont {Liang}\ \bibnamefont {Fu}},\ }\bibfield  {title} {\enquote {\bibinfo {title} {Topological crystalline insulators in the {S}n{T}e material class},}\ }\href@noop {} {\bibfield  {journal} {\bibinfo  {journal} {Nature communications}\ }\textbf {\bibinfo {volume} {3}},\ \bibinfo {pages} {982} (\bibinfo {year} {2012})}\BibitemShut {NoStop}%
\bibitem [{\citenamefont {Mong}\ \emph {et~al.}(2010)\citenamefont {Mong}, \citenamefont {Essin},\ and\ \citenamefont {Moore}}]{TCI2}%
  \BibitemOpen
  \bibfield  {author} {\bibinfo {author} {\bibfnamefont {Roger S.~K.}\ \bibnamefont {Mong}}, \bibinfo {author} {\bibfnamefont {Andrew~M.}\ \bibnamefont {Essin}}, \ and\ \bibinfo {author} {\bibfnamefont {Joel~E.}\ \bibnamefont {Moore}},\ }\bibfield  {title} {\enquote {\bibinfo {title} {Antiferromagnetic topological insulators},}\ }\href {\doibase 10.1103/PhysRevB.81.245209} {\bibfield  {journal} {\bibinfo  {journal} {Phys. Rev. B}\ }\textbf {\bibinfo {volume} {81}},\ \bibinfo {pages} {245209} (\bibinfo {year} {2010})}\BibitemShut {NoStop}%
\bibitem [{\citenamefont {Turner}\ \emph {et~al.}(2010)\citenamefont {Turner}, \citenamefont {Zhang},\ and\ \citenamefont {Vishwanath}}]{TCI3}%
  \BibitemOpen
  \bibfield  {author} {\bibinfo {author} {\bibfnamefont {Ari~M.}\ \bibnamefont {Turner}}, \bibinfo {author} {\bibfnamefont {Yi}~\bibnamefont {Zhang}}, \ and\ \bibinfo {author} {\bibfnamefont {Ashvin}\ \bibnamefont {Vishwanath}},\ }\bibfield  {title} {\enquote {\bibinfo {title} {Entanglement and inversion symmetry in topological insulators},}\ }\href {\doibase 10.1103/PhysRevB.82.241102} {\bibfield  {journal} {\bibinfo  {journal} {Phys. Rev. B}\ }\textbf {\bibinfo {volume} {82}},\ \bibinfo {pages} {241102} (\bibinfo {year} {2010})}\BibitemShut {NoStop}%
\bibitem [{\citenamefont {Hughes}\ \emph {et~al.}(2011)\citenamefont {Hughes}, \citenamefont {Prodan},\ and\ \citenamefont {Bernevig}}]{TCI4}%
  \BibitemOpen
  \bibfield  {author} {\bibinfo {author} {\bibfnamefont {Taylor~L.}\ \bibnamefont {Hughes}}, \bibinfo {author} {\bibfnamefont {Emil}\ \bibnamefont {Prodan}}, \ and\ \bibinfo {author} {\bibfnamefont {B.~Andrei}\ \bibnamefont {Bernevig}},\ }\bibfield  {title} {\enquote {\bibinfo {title} {Inversion-symmetric topological insulators},}\ }\href {\doibase 10.1103/PhysRevB.83.245132} {\bibfield  {journal} {\bibinfo  {journal} {Phys. Rev. B}\ }\textbf {\bibinfo {volume} {83}},\ \bibinfo {pages} {245132} (\bibinfo {year} {2011})}\BibitemShut {NoStop}%
\bibitem [{\citenamefont {Liu}\ \emph {et~al.}(2014{\natexlab{a}})\citenamefont {Liu}, \citenamefont {Zhang},\ and\ \citenamefont {VanLeeuwen}}]{TCI5}%
  \BibitemOpen
  \bibfield  {author} {\bibinfo {author} {\bibfnamefont {Chao-Xing}\ \bibnamefont {Liu}}, \bibinfo {author} {\bibfnamefont {Rui-Xing}\ \bibnamefont {Zhang}}, \ and\ \bibinfo {author} {\bibfnamefont {Brian~K.}\ \bibnamefont {VanLeeuwen}},\ }\bibfield  {title} {\enquote {\bibinfo {title} {Topological nonsymmorphic crystalline insulators},}\ }\href {\doibase 10.1103/PhysRevB.90.085304} {\bibfield  {journal} {\bibinfo  {journal} {Phys. Rev. B}\ }\textbf {\bibinfo {volume} {90}},\ \bibinfo {pages} {085304} (\bibinfo {year} {2014}{\natexlab{a}})}\BibitemShut {NoStop}%
\bibitem [{\citenamefont {Soluyanov}\ and\ \citenamefont {Vanderbilt}(2011)}]{soluyanov_wannier_2011}%
  \BibitemOpen
  \bibfield  {author} {\bibinfo {author} {\bibfnamefont {Alexey~A.}\ \bibnamefont {Soluyanov}}\ and\ \bibinfo {author} {\bibfnamefont {David}\ \bibnamefont {Vanderbilt}},\ }\bibfield  {title} {\enquote {\bibinfo {title} {Wannier representation of $\mathbb{Z}_{2}$ topological insulators},}\ }\href {\doibase 10.1103/PhysRevB.83.035108} {\bibfield  {journal} {\bibinfo  {journal} {Physical Review B}\ }\textbf {\bibinfo {volume} {83}},\ \bibinfo {pages} {035108} (\bibinfo {year} {2011})}\BibitemShut {NoStop}%
\bibitem [{\citenamefont {Bradlyn}\ \emph {et~al.}(2017)\citenamefont {Bradlyn}, \citenamefont {Elcoro}, \citenamefont {Cano}, \citenamefont {Vergniory}, \citenamefont {Wang}, \citenamefont {Felser}, \citenamefont {Aroyo},\ and\ \citenamefont {Bernevig}}]{TCI8}%
  \BibitemOpen
  \bibfield  {author} {\bibinfo {author} {\bibfnamefont {Barry}\ \bibnamefont {Bradlyn}}, \bibinfo {author} {\bibfnamefont {Luis}\ \bibnamefont {Elcoro}}, \bibinfo {author} {\bibfnamefont {Jennifer}\ \bibnamefont {Cano}}, \bibinfo {author} {\bibfnamefont {Maia~G}\ \bibnamefont {Vergniory}}, \bibinfo {author} {\bibfnamefont {Zhijun}\ \bibnamefont {Wang}}, \bibinfo {author} {\bibfnamefont {Claudia}\ \bibnamefont {Felser}}, \bibinfo {author} {\bibfnamefont {Mois~I}\ \bibnamefont {Aroyo}}, \ and\ \bibinfo {author} {\bibfnamefont {B~Andrei}\ \bibnamefont {Bernevig}},\ }\bibfield  {title} {\enquote {\bibinfo {title} {Topological quantum chemistry},}\ }\href@noop {} {\bibfield  {journal} {\bibinfo  {journal} {Nature}\ }\textbf {\bibinfo {volume} {547}},\ \bibinfo {pages} {298--305} (\bibinfo {year} {2017})}\BibitemShut {NoStop}%
\bibitem [{\citenamefont {Po}\ \emph {et~al.}(2017)\citenamefont {Po}, \citenamefont {Vishwanath},\ and\ \citenamefont {Watanabe}}]{TCI6}%
  \BibitemOpen
  \bibfield  {author} {\bibinfo {author} {\bibfnamefont {Hoi~Chun}\ \bibnamefont {Po}}, \bibinfo {author} {\bibfnamefont {Ashvin}\ \bibnamefont {Vishwanath}}, \ and\ \bibinfo {author} {\bibfnamefont {Haruki}\ \bibnamefont {Watanabe}},\ }\bibfield  {title} {\enquote {\bibinfo {title} {Symmetry-based indicators of band topology in the 230 space groups},}\ }\href@noop {} {\bibfield  {journal} {\bibinfo  {journal} {Nature communications}\ }\textbf {\bibinfo {volume} {8}},\ \bibinfo {pages} {50} (\bibinfo {year} {2017})}\BibitemShut {NoStop}%
\bibitem [{\citenamefont {Kruthoff}\ \emph {et~al.}(2017)\citenamefont {Kruthoff}, \citenamefont {De~Boer}, \citenamefont {Van~Wezel}, \citenamefont {Kane},\ and\ \citenamefont {Slager}}]{TCI7}%
  \BibitemOpen
  \bibfield  {author} {\bibinfo {author} {\bibfnamefont {Jorrit}\ \bibnamefont {Kruthoff}}, \bibinfo {author} {\bibfnamefont {Jan}\ \bibnamefont {De~Boer}}, \bibinfo {author} {\bibfnamefont {Jasper}\ \bibnamefont {Van~Wezel}}, \bibinfo {author} {\bibfnamefont {Charles~L}\ \bibnamefont {Kane}}, \ and\ \bibinfo {author} {\bibfnamefont {Robert-Jan}\ \bibnamefont {Slager}},\ }\bibfield  {title} {\enquote {\bibinfo {title} {Topological classification of crystalline insulators through band structure combinatorics},}\ }\href {\doibase 10.1103/PhysRevX.7.041069} {\bibfield  {journal} {\bibinfo  {journal} {Physical Review X}\ }\textbf {\bibinfo {volume} {7}},\ \bibinfo {pages} {041069} (\bibinfo {year} {2017})}\BibitemShut {NoStop}%
\bibitem [{\citenamefont {Kitaev}(2009)}]{AZ_CLASSIFICATION2}%
  \BibitemOpen
  \bibfield  {author} {\bibinfo {author} {\bibfnamefont {Alexei}\ \bibnamefont {Kitaev}},\ }\bibfield  {title} {\enquote {\bibinfo {title} {{Periodic table for topological insulators and superconductors}},}\ }\href {\doibase 10.1063/1.3149495} {\bibfield  {journal} {\bibinfo  {journal} {AIP Conference Proceedings}\ }\textbf {\bibinfo {volume} {1134}},\ \bibinfo {pages} {22--30} (\bibinfo {year} {2009})}\BibitemShut {NoStop}%
\bibitem [{\citenamefont {Schnyder}\ \emph {et~al.}(2008)\citenamefont {Schnyder}, \citenamefont {Ryu}, \citenamefont {Furusaki},\ and\ \citenamefont {Ludwig}}]{AZ_CLASSIFICATION1}%
  \BibitemOpen
  \bibfield  {author} {\bibinfo {author} {\bibfnamefont {Andreas~P.}\ \bibnamefont {Schnyder}}, \bibinfo {author} {\bibfnamefont {Shinsei}\ \bibnamefont {Ryu}}, \bibinfo {author} {\bibfnamefont {Akira}\ \bibnamefont {Furusaki}}, \ and\ \bibinfo {author} {\bibfnamefont {Andreas W.~W.}\ \bibnamefont {Ludwig}},\ }\bibfield  {title} {\enquote {\bibinfo {title} {Classification of topological insulators and superconductors in three spatial dimensions},}\ }\href {\doibase 10.1103/PhysRevB.78.195125} {\bibfield  {journal} {\bibinfo  {journal} {Phys. Rev. B}\ }\textbf {\bibinfo {volume} {78}},\ \bibinfo {pages} {195125} (\bibinfo {year} {2008})}\BibitemShut {NoStop}%
\bibitem [{\citenamefont {Ryu}\ \emph {et~al.}(2010)\citenamefont {Ryu}, \citenamefont {Schnyder}, \citenamefont {Furusaki},\ and\ \citenamefont {Ludwig}}]{AZ_CLASSIFICATION3}%
  \BibitemOpen
  \bibfield  {author} {\bibinfo {author} {\bibfnamefont {Shinsei}\ \bibnamefont {Ryu}}, \bibinfo {author} {\bibfnamefont {Andreas~P}\ \bibnamefont {Schnyder}}, \bibinfo {author} {\bibfnamefont {Akira}\ \bibnamefont {Furusaki}}, \ and\ \bibinfo {author} {\bibfnamefont {Andreas W~W}\ \bibnamefont {Ludwig}},\ }\bibfield  {title} {\enquote {\bibinfo {title} {Topological insulators and superconductors: tenfold way and dimensional hierarchy},}\ }\href {\doibase 10.1088/1367-2630/12/6/065010} {\bibfield  {journal} {\bibinfo  {journal} {New Journal of Physics}\ }\textbf {\bibinfo {volume} {12}},\ \bibinfo {pages} {065010} (\bibinfo {year} {2010})}\BibitemShut {NoStop}%
\bibitem [{\citenamefont {Chiu}\ \emph {et~al.}(2013)\citenamefont {Chiu}, \citenamefont {Yao},\ and\ \citenamefont {Ryu}}]{TCI_CLASSIFICATION_1}%
  \BibitemOpen
  \bibfield  {author} {\bibinfo {author} {\bibfnamefont {Ching-Kai}\ \bibnamefont {Chiu}}, \bibinfo {author} {\bibfnamefont {Hong}\ \bibnamefont {Yao}}, \ and\ \bibinfo {author} {\bibfnamefont {Shinsei}\ \bibnamefont {Ryu}},\ }\bibfield  {title} {\enquote {\bibinfo {title} {Classification of topological insulators and superconductors in the presence of reflection symmetry},}\ }\href {\doibase 10.1103/PhysRevB.88.075142} {\bibfield  {journal} {\bibinfo  {journal} {Phys. Rev. B}\ }\textbf {\bibinfo {volume} {88}},\ \bibinfo {pages} {075142} (\bibinfo {year} {2013})}\BibitemShut {NoStop}%
\bibitem [{\citenamefont {Shiozaki}\ \emph {et~al.}(2015)\citenamefont {Shiozaki}, \citenamefont {Sato},\ and\ \citenamefont {Gomi}}]{TCI_CLASSIFICATION_3}%
  \BibitemOpen
  \bibfield  {author} {\bibinfo {author} {\bibfnamefont {Ken}\ \bibnamefont {Shiozaki}}, \bibinfo {author} {\bibfnamefont {Masatoshi}\ \bibnamefont {Sato}}, \ and\ \bibinfo {author} {\bibfnamefont {Kiyonori}\ \bibnamefont {Gomi}},\ }\bibfield  {title} {\enquote {\bibinfo {title} {${Z}_{2}$ topology in nonsymmorphic crystalline insulators: M\"obius twist in surface states},}\ }\href {\doibase 10.1103/PhysRevB.91.155120} {\bibfield  {journal} {\bibinfo  {journal} {Phys. Rev. B}\ }\textbf {\bibinfo {volume} {91}},\ \bibinfo {pages} {155120} (\bibinfo {year} {2015})}\BibitemShut {NoStop}%
\bibitem [{\citenamefont {Shiozaki}\ \emph {et~al.}(2016)\citenamefont {Shiozaki}, \citenamefont {Sato},\ and\ \citenamefont {Gomi}}]{TCI_CLASSIFICATION_4}%
  \BibitemOpen
  \bibfield  {author} {\bibinfo {author} {\bibfnamefont {Ken}\ \bibnamefont {Shiozaki}}, \bibinfo {author} {\bibfnamefont {Masatoshi}\ \bibnamefont {Sato}}, \ and\ \bibinfo {author} {\bibfnamefont {Kiyonori}\ \bibnamefont {Gomi}},\ }\bibfield  {title} {\enquote {\bibinfo {title} {Topology of nonsymmorphic crystalline insulators and superconductors},}\ }\href {\doibase 10.1103/PhysRevB.93.195413} {\bibfield  {journal} {\bibinfo  {journal} {Phys. Rev. B}\ }\textbf {\bibinfo {volume} {93}},\ \bibinfo {pages} {195413} (\bibinfo {year} {2016})}\BibitemShut {NoStop}%
\bibitem [{\citenamefont {Song}\ \emph {et~al.}(2017)\citenamefont {Song}, \citenamefont {Huang}, \citenamefont {Fu},\ and\ \citenamefont {Hermele}}]{song_topological_2017}%
  \BibitemOpen
  \bibfield  {author} {\bibinfo {author} {\bibfnamefont {Hao}\ \bibnamefont {Song}}, \bibinfo {author} {\bibfnamefont {Sheng-Jie}\ \bibnamefont {Huang}}, \bibinfo {author} {\bibfnamefont {Liang}\ \bibnamefont {Fu}}, \ and\ \bibinfo {author} {\bibfnamefont {Michael}\ \bibnamefont {Hermele}},\ }\bibfield  {title} {\enquote {\bibinfo {title} {Topological {Phases} {Protected} by {Point} {Group} {Symmetry}},}\ }\href {\doibase 10.1103/PhysRevX.7.011020} {\bibfield  {journal} {\bibinfo  {journal} {Physical Review X}\ }\textbf {\bibinfo {volume} {7}},\ \bibinfo {pages} {011020} (\bibinfo {year} {2017})}\BibitemShut {NoStop}%
\bibitem [{\citenamefont {Huang}\ \emph {et~al.}(2017)\citenamefont {Huang}, \citenamefont {Song}, \citenamefont {Huang},\ and\ \citenamefont {Hermele}}]{TCS_0}%
  \BibitemOpen
  \bibfield  {author} {\bibinfo {author} {\bibfnamefont {Sheng-Jie}\ \bibnamefont {Huang}}, \bibinfo {author} {\bibfnamefont {Hao}\ \bibnamefont {Song}}, \bibinfo {author} {\bibfnamefont {Yi-Ping}\ \bibnamefont {Huang}}, \ and\ \bibinfo {author} {\bibfnamefont {Michael}\ \bibnamefont {Hermele}},\ }\bibfield  {title} {\enquote {\bibinfo {title} {Building crystalline topological phases from lower-dimensional states},}\ }\href {\doibase 10.1103/PhysRevB.96.205106} {\bibfield  {journal} {\bibinfo  {journal} {Phys. Rev. B}\ }\textbf {\bibinfo {volume} {96}},\ \bibinfo {pages} {205106} (\bibinfo {year} {2017})}\BibitemShut {NoStop}%
\bibitem [{\citenamefont {Khalaf}\ \emph {et~al.}(2018)\citenamefont {Khalaf}, \citenamefont {Po}, \citenamefont {Vishwanath},\ and\ \citenamefont {Watanabe}}]{TCI9}%
  \BibitemOpen
  \bibfield  {author} {\bibinfo {author} {\bibfnamefont {Eslam}\ \bibnamefont {Khalaf}}, \bibinfo {author} {\bibfnamefont {Hoi~Chun}\ \bibnamefont {Po}}, \bibinfo {author} {\bibfnamefont {Ashvin}\ \bibnamefont {Vishwanath}}, \ and\ \bibinfo {author} {\bibfnamefont {Haruki}\ \bibnamefont {Watanabe}},\ }\bibfield  {title} {\enquote {\bibinfo {title} {Symmetry indicators and anomalous surface states of topological crystalline insulators},}\ }\href {\doibase 10.1103/PhysRevX.8.031070} {\bibfield  {journal} {\bibinfo  {journal} {Phys. Rev. X}\ }\textbf {\bibinfo {volume} {8}},\ \bibinfo {pages} {031070} (\bibinfo {year} {2018})}\BibitemShut {NoStop}%
\bibitem [{\citenamefont {Song}\ \emph {et~al.}(2019)\citenamefont {Song}, \citenamefont {Huang}, \citenamefont {Qi}, \citenamefont {Fang},\ and\ \citenamefont {Hermele}}]{TCS_1}%
  \BibitemOpen
  \bibfield  {author} {\bibinfo {author} {\bibfnamefont {Zhida}\ \bibnamefont {Song}}, \bibinfo {author} {\bibfnamefont {Sheng-Jie}\ \bibnamefont {Huang}}, \bibinfo {author} {\bibfnamefont {Yang}\ \bibnamefont {Qi}}, \bibinfo {author} {\bibfnamefont {Chen}\ \bibnamefont {Fang}}, \ and\ \bibinfo {author} {\bibfnamefont {Michael}\ \bibnamefont {Hermele}},\ }\bibfield  {title} {\enquote {\bibinfo {title} {Topological states from topological crystals},}\ }\href {\doibase 10.1126/sciadv.aax2007} {\bibfield  {journal} {\bibinfo  {journal} {Science Advances}\ }\textbf {\bibinfo {volume} {5}},\ \bibinfo {pages} {eaax2007} (\bibinfo {year} {2019})}\BibitemShut {NoStop}%
\bibitem [{\citenamefont {Cornfeld}\ and\ \citenamefont {Chapman}(2019)}]{TCI11}%
  \BibitemOpen
  \bibfield  {author} {\bibinfo {author} {\bibfnamefont {Eyal}\ \bibnamefont {Cornfeld}}\ and\ \bibinfo {author} {\bibfnamefont {Adam}\ \bibnamefont {Chapman}},\ }\bibfield  {title} {\enquote {\bibinfo {title} {Classification of crystalline topological insulators and superconductors with point group symmetries},}\ }\href {\doibase 10.1103/PhysRevB.99.075105} {\bibfield  {journal} {\bibinfo  {journal} {Phys. Rev. B}\ }\textbf {\bibinfo {volume} {99}},\ \bibinfo {pages} {075105} (\bibinfo {year} {2019})}\BibitemShut {NoStop}%
\bibitem [{\citenamefont {Shiozaki}(2022)}]{TCI12}%
  \BibitemOpen
  \bibfield  {author} {\bibinfo {author} {\bibfnamefont {Ken}\ \bibnamefont {Shiozaki}},\ }\bibfield  {title} {\enquote {\bibinfo {title} {The classification of surface states of topological insulators and superconductors with magnetic point group symmetry},}\ }\href@noop {} {\bibfield  {journal} {\bibinfo  {journal} {Progress of Theoretical and Experimental Physics}\ }\textbf {\bibinfo {volume} {2022}},\ \bibinfo {pages} {04A104} (\bibinfo {year} {2022})}\BibitemShut {NoStop}%
\bibitem [{\citenamefont {Huang}\ \emph {et~al.}(2022)\citenamefont {Huang}, \citenamefont {Hsieh},\ and\ \citenamefont {Yu}}]{PhysRevB.105.045112}%
  \BibitemOpen
  \bibfield  {author} {\bibinfo {author} {\bibfnamefont {Sheng-Jie}\ \bibnamefont {Huang}}, \bibinfo {author} {\bibfnamefont {Chang-Tse}\ \bibnamefont {Hsieh}}, \ and\ \bibinfo {author} {\bibfnamefont {Jiabin}\ \bibnamefont {Yu}},\ }\bibfield  {title} {\enquote {\bibinfo {title} {Effective field theories of topological crystalline insulators and topological crystals},}\ }\href {\doibase 10.1103/PhysRevB.105.045112} {\bibfield  {journal} {\bibinfo  {journal} {Phys. Rev. B}\ }\textbf {\bibinfo {volume} {105}},\ \bibinfo {pages} {045112} (\bibinfo {year} {2022})}\BibitemShut {NoStop}%
\bibitem [{\citenamefont {Li}\ \emph {et~al.}(2009)\citenamefont {Li}, \citenamefont {Chu}, \citenamefont {Jain},\ and\ \citenamefont {Shen}}]{TAI}%
  \BibitemOpen
  \bibfield  {author} {\bibinfo {author} {\bibfnamefont {Jian}\ \bibnamefont {Li}}, \bibinfo {author} {\bibfnamefont {Rui-Lin}\ \bibnamefont {Chu}}, \bibinfo {author} {\bibfnamefont {J.~K.}\ \bibnamefont {Jain}}, \ and\ \bibinfo {author} {\bibfnamefont {Shun-Qing}\ \bibnamefont {Shen}},\ }\bibfield  {title} {\enquote {\bibinfo {title} {Topological anderson insulator},}\ }\href {\doibase 10.1103/PhysRevLett.102.136806} {\bibfield  {journal} {\bibinfo  {journal} {Phys. Rev. Lett.}\ }\textbf {\bibinfo {volume} {102}},\ \bibinfo {pages} {136806} (\bibinfo {year} {2009})}\BibitemShut {NoStop}%
\bibitem [{\citenamefont {Fu}\ and\ \citenamefont {Kane}(2012)}]{fu_topology_2012}%
  \BibitemOpen
  \bibfield  {author} {\bibinfo {author} {\bibfnamefont {Liang}\ \bibnamefont {Fu}}\ and\ \bibinfo {author} {\bibfnamefont {C.~L.}\ \bibnamefont {Kane}},\ }\bibfield  {title} {\enquote {\bibinfo {title} {Topology, {Delocalization} via {Average} {Symmetry} and the {Symplectic} {Anderson} {Transition}},}\ }\href {\doibase 10.1103/PhysRevLett.109.246605} {\bibfield  {journal} {\bibinfo  {journal} {Phys. Rev. Lett.}\ }\textbf {\bibinfo {volume} {109}},\ \bibinfo {pages} {246605} (\bibinfo {year} {2012})}\BibitemShut {NoStop}%
\bibitem [{\citenamefont {Ringel}\ \emph {et~al.}(2012)\citenamefont {Ringel}, \citenamefont {Kraus},\ and\ \citenamefont {Stern}}]{ringel_strong_2012}%
  \BibitemOpen
  \bibfield  {author} {\bibinfo {author} {\bibfnamefont {Zohar}\ \bibnamefont {Ringel}}, \bibinfo {author} {\bibfnamefont {Yaacov~E.}\ \bibnamefont {Kraus}}, \ and\ \bibinfo {author} {\bibfnamefont {Ady}\ \bibnamefont {Stern}},\ }\bibfield  {title} {\enquote {\bibinfo {title} {Strong side of weak topological insulators},}\ }\href {\doibase 10.1103/PhysRevB.86.045102} {\bibfield  {journal} {\bibinfo  {journal} {Physical Review B}\ }\textbf {\bibinfo {volume} {86}},\ \bibinfo {pages} {045102} (\bibinfo {year} {2012})}\BibitemShut {NoStop}%
\bibitem [{\citenamefont {Mong}\ \emph {et~al.}(2012)\citenamefont {Mong}, \citenamefont {Bardarson},\ and\ \citenamefont {Moore}}]{mong_quantum_2012}%
  \BibitemOpen
  \bibfield  {author} {\bibinfo {author} {\bibfnamefont {Roger S.~K.}\ \bibnamefont {Mong}}, \bibinfo {author} {\bibfnamefont {Jens~H.}\ \bibnamefont {Bardarson}}, \ and\ \bibinfo {author} {\bibfnamefont {Joel~E.}\ \bibnamefont {Moore}},\ }\bibfield  {title} {\enquote {\bibinfo {title} {Quantum transport and two-parameter scaling at the surface of a weak topological insulator},}\ }\href {\doibase 10.1103/PhysRevLett.108.076804} {\bibfield  {journal} {\bibinfo  {journal} {Phys. Rev. Lett.}\ }\textbf {\bibinfo {volume} {108}},\ \bibinfo {pages} {076804} (\bibinfo {year} {2012})}\BibitemShut {NoStop}%
\bibitem [{\citenamefont {Fulga}\ \emph {et~al.}(2014)\citenamefont {Fulga}, \citenamefont {van Heck}, \citenamefont {Edge},\ and\ \citenamefont {Akhmerov}}]{STI}%
  \BibitemOpen
  \bibfield  {author} {\bibinfo {author} {\bibfnamefont {I.~C.}\ \bibnamefont {Fulga}}, \bibinfo {author} {\bibfnamefont {B.}~\bibnamefont {van Heck}}, \bibinfo {author} {\bibfnamefont {J.~M.}\ \bibnamefont {Edge}}, \ and\ \bibinfo {author} {\bibfnamefont {A.~R.}\ \bibnamefont {Akhmerov}},\ }\bibfield  {title} {\enquote {\bibinfo {title} {Statistical topological insulators},}\ }\href {\doibase 10.1103/PhysRevB.89.155424} {\bibfield  {journal} {\bibinfo  {journal} {Phys. Rev. B}\ }\textbf {\bibinfo {volume} {89}},\ \bibinfo {pages} {155424} (\bibinfo {year} {2014})},\ \bibinfo {note} {{A}verage symmetry is referred to as statistical symmetry in this work.}\BibitemShut {Stop}%
\bibitem [{\citenamefont {Chaou}\ \emph {et~al.}(2024)\citenamefont {Chaou}, \citenamefont {Moreno-Gonzalez}, \citenamefont {Altland},\ and\ \citenamefont {Brouwer}}]{chaou2024disordered}%
  \BibitemOpen
  \bibfield  {author} {\bibinfo {author} {\bibfnamefont {Adam~Yanis}\ \bibnamefont {Chaou}}, \bibinfo {author} {\bibfnamefont {Mateo}\ \bibnamefont {Moreno-Gonzalez}}, \bibinfo {author} {\bibfnamefont {Alexander}\ \bibnamefont {Altland}}, \ and\ \bibinfo {author} {\bibfnamefont {Piet~W}\ \bibnamefont {Brouwer}},\ }\bibfield  {title} {\enquote {\bibinfo {title} {Disordered topological crystalline phases},}\ }\href {https://arxiv.org/abs/2412.01883} {\bibfield  {journal} {\bibinfo  {journal} {arXiv preprint arXiv:2412.01883}\ } (\bibinfo {year} {2024})}\BibitemShut {NoStop}%
\bibitem [{\citenamefont {Ma}\ and\ \citenamefont {Wang}(2023)}]{ASPT_PRX}%
  \BibitemOpen
  \bibfield  {author} {\bibinfo {author} {\bibfnamefont {Ruochen}\ \bibnamefont {Ma}}\ and\ \bibinfo {author} {\bibfnamefont {Chong}\ \bibnamefont {Wang}},\ }\bibfield  {title} {\enquote {\bibinfo {title} {Average symmetry-protected topological phases},}\ }\href {\doibase 10.1103/PhysRevX.13.031016} {\bibfield  {journal} {\bibinfo  {journal} {Phys. Rev. X}\ }\textbf {\bibinfo {volume} {13}},\ \bibinfo {pages} {031016} (\bibinfo {year} {2023})}\BibitemShut {NoStop}%
\bibitem [{\citenamefont {Ma}\ \emph {et~al.}(2023)\citenamefont {Ma}, \citenamefont {Zhang}, \citenamefont {Bi}, \citenamefont {Cheng},\ and\ \citenamefont {Wang}}]{ASPT_INTRINSIC}%
  \BibitemOpen
  \bibfield  {author} {\bibinfo {author} {\bibfnamefont {Ruochen}\ \bibnamefont {Ma}}, \bibinfo {author} {\bibfnamefont {Jian-Hao}\ \bibnamefont {Zhang}}, \bibinfo {author} {\bibfnamefont {Zhen}\ \bibnamefont {Bi}}, \bibinfo {author} {\bibfnamefont {Meng}\ \bibnamefont {Cheng}}, \ and\ \bibinfo {author} {\bibfnamefont {Chong}\ \bibnamefont {Wang}},\ }\bibfield  {title} {\enquote {\bibinfo {title} {Topological phases with average symmetries: the decohered, the disordered, and the intrinsic},}\ }\href {https://arxiv.org/abs/2305.16399} {\bibfield  {journal} {\bibinfo  {journal} {arXiv preprint arXiv:2305.16399}\ } (\bibinfo {year} {2023})}\BibitemShut {NoStop}%
\bibitem [{\citenamefont {Qi}\ \emph {et~al.}(2008)\citenamefont {Qi}, \citenamefont {Hughes},\ and\ \citenamefont {Zhang}}]{TRS_AXION}%
  \BibitemOpen
  \bibfield  {author} {\bibinfo {author} {\bibfnamefont {Xiao-Liang}\ \bibnamefont {Qi}}, \bibinfo {author} {\bibfnamefont {Taylor~L.}\ \bibnamefont {Hughes}}, \ and\ \bibinfo {author} {\bibfnamefont {Shou-Cheng}\ \bibnamefont {Zhang}},\ }\bibfield  {title} {\enquote {\bibinfo {title} {Topological field theory of time-reversal invariant insulators},}\ }\href {\doibase 10.1103/PhysRevB.78.195424} {\bibfield  {journal} {\bibinfo  {journal} {Phys. Rev. B}\ }\textbf {\bibinfo {volume} {78}},\ \bibinfo {pages} {195424} (\bibinfo {year} {2008})}\BibitemShut {NoStop}%
\bibitem [{\citenamefont {Wang}\ \emph {et~al.}(2010)\citenamefont {Wang}, \citenamefont {Qi},\ and\ \citenamefont {Zhang}}]{wang_equivalent_2010}%
  \BibitemOpen
  \bibfield  {author} {\bibinfo {author} {\bibfnamefont {Zhong}\ \bibnamefont {Wang}}, \bibinfo {author} {\bibfnamefont {Xiao-Liang}\ \bibnamefont {Qi}}, \ and\ \bibinfo {author} {\bibfnamefont {Shou-Cheng}\ \bibnamefont {Zhang}},\ }\bibfield  {title} {\enquote {\bibinfo {title} {Equivalent topological invariants of topological insulators},}\ }\href {\doibase 10.1088/1367-2630/12/6/065007} {\bibfield  {journal} {\bibinfo  {journal} {New Journal of Physics}\ }\textbf {\bibinfo {volume} {12}},\ \bibinfo {pages} {065007} (\bibinfo {year} {2010})}\BibitemShut {NoStop}%
\bibitem [{\citenamefont {Turner}\ \emph {et~al.}(2012)\citenamefont {Turner}, \citenamefont {Zhang}, \citenamefont {Mong},\ and\ \citenamefont {Vishwanath}}]{AXION4}%
  \BibitemOpen
  \bibfield  {author} {\bibinfo {author} {\bibfnamefont {Ari~M.}\ \bibnamefont {Turner}}, \bibinfo {author} {\bibfnamefont {Yi}~\bibnamefont {Zhang}}, \bibinfo {author} {\bibfnamefont {Roger S.~K.}\ \bibnamefont {Mong}}, \ and\ \bibinfo {author} {\bibfnamefont {Ashvin}\ \bibnamefont {Vishwanath}},\ }\bibfield  {title} {\enquote {\bibinfo {title} {Quantized response and topology of magnetic insulators with inversion symmetry},}\ }\href {\doibase 10.1103/PhysRevB.85.165120} {\bibfield  {journal} {\bibinfo  {journal} {Phys. Rev. B}\ }\textbf {\bibinfo {volume} {85}},\ \bibinfo {pages} {165120} (\bibinfo {year} {2012})}\BibitemShut {NoStop}%
\bibitem [{\citenamefont {Khalaf}(2018)}]{khalaf_higher-order_2018}%
  \BibitemOpen
  \bibfield  {author} {\bibinfo {author} {\bibfnamefont {Eslam}\ \bibnamefont {Khalaf}},\ }\bibfield  {title} {\enquote {\bibinfo {title} {Higher-order topological insulators and superconductors protected by inversion symmetry},}\ }\href {\doibase 10.1103/PhysRevB.97.205136} {\bibfield  {journal} {\bibinfo  {journal} {Phys. Rev. B}\ }\textbf {\bibinfo {volume} {97}},\ \bibinfo {pages} {205136} (\bibinfo {year} {2018})}\BibitemShut {NoStop}%
\bibitem [{\citenamefont {Kooi}\ \emph {et~al.}(2018)\citenamefont {Kooi}, \citenamefont {van Miert},\ and\ \citenamefont {Ortix}}]{kooi_inversion-symmetry_2018}%
  \BibitemOpen
  \bibfield  {author} {\bibinfo {author} {\bibfnamefont {Sander~H.}\ \bibnamefont {Kooi}}, \bibinfo {author} {\bibfnamefont {Guido}\ \bibnamefont {van Miert}}, \ and\ \bibinfo {author} {\bibfnamefont {Carmine}\ \bibnamefont {Ortix}},\ }\bibfield  {title} {\enquote {\bibinfo {title} {Inversion-symmetry protected chiral hinge states in stacks of doped quantum hall layers},}\ }\href {\doibase 10.1103/PhysRevB.98.245102} {\bibfield  {journal} {\bibinfo  {journal} {Phys. Rev. B}\ }\textbf {\bibinfo {volume} {98}},\ \bibinfo {pages} {245102} (\bibinfo {year} {2018})}\BibitemShut {NoStop}%
\bibitem [{\citenamefont {Essin}\ \emph {et~al.}(2009)\citenamefont {Essin}, \citenamefont {Moore},\ and\ \citenamefont {Vanderbilt}}]{AXION2}%
  \BibitemOpen
  \bibfield  {author} {\bibinfo {author} {\bibfnamefont {Andrew~M.}\ \bibnamefont {Essin}}, \bibinfo {author} {\bibfnamefont {Joel~E.}\ \bibnamefont {Moore}}, \ and\ \bibinfo {author} {\bibfnamefont {David}\ \bibnamefont {Vanderbilt}},\ }\bibfield  {title} {\enquote {\bibinfo {title} {Magnetoelectric polarizability and axion electrodynamics in crystalline insulators},}\ }\href {\doibase 10.1103/PhysRevLett.102.146805} {\bibfield  {journal} {\bibinfo  {journal} {Phys. Rev. Lett.}\ }\textbf {\bibinfo {volume} {102}},\ \bibinfo {pages} {146805} (\bibinfo {year} {2009})}\BibitemShut {NoStop}%
\bibitem [{\citenamefont {Fang}\ \emph {et~al.}(2012)\citenamefont {Fang}, \citenamefont {Gilbert},\ and\ \citenamefont {Bernevig}}]{fang_bulk_2012}%
  \BibitemOpen
  \bibfield  {author} {\bibinfo {author} {\bibfnamefont {Chen}\ \bibnamefont {Fang}}, \bibinfo {author} {\bibfnamefont {Matthew~J.}\ \bibnamefont {Gilbert}}, \ and\ \bibinfo {author} {\bibfnamefont {B.~Andrei}\ \bibnamefont {Bernevig}},\ }\bibfield  {title} {\enquote {\bibinfo {title} {Bulk topological invariants in noninteracting point group symmetric insulators},}\ }\href {\doibase 10.1103/PhysRevB.86.115112} {\bibfield  {journal} {\bibinfo  {journal} {Phys. Rev. B}\ }\textbf {\bibinfo {volume} {86}},\ \bibinfo {pages} {115112} (\bibinfo {year} {2012})}\BibitemShut {NoStop}%
\bibitem [{\citenamefont {Varjas}\ \emph {et~al.}(2015)\citenamefont {Varjas}, \citenamefont {de~Juan},\ and\ \citenamefont {Lu}}]{varjas_bulk_2015}%
  \BibitemOpen
  \bibfield  {author} {\bibinfo {author} {\bibfnamefont {D\'aniel}\ \bibnamefont {Varjas}}, \bibinfo {author} {\bibfnamefont {Fernando}\ \bibnamefont {de~Juan}}, \ and\ \bibinfo {author} {\bibfnamefont {Yuan-Ming}\ \bibnamefont {Lu}},\ }\bibfield  {title} {\enquote {\bibinfo {title} {Bulk invariants and topological response in insulators and superconductors with nonsymmorphic symmetries},}\ }\href {\doibase 10.1103/PhysRevB.92.195116} {\bibfield  {journal} {\bibinfo  {journal} {Phys. Rev. B}\ }\textbf {\bibinfo {volume} {92}},\ \bibinfo {pages} {195116} (\bibinfo {year} {2015})}\BibitemShut {NoStop}%
\bibitem [{\citenamefont {van Miert}\ and\ \citenamefont {Ortix}(2018)}]{miert_higher-order_2018}%
  \BibitemOpen
  \bibfield  {author} {\bibinfo {author} {\bibfnamefont {Guido}\ \bibnamefont {van Miert}}\ and\ \bibinfo {author} {\bibfnamefont {Carmine}\ \bibnamefont {Ortix}},\ }\bibfield  {title} {\enquote {\bibinfo {title} {Higher-order topological insulators protected by inversion and rotoinversion symmetries},}\ }\href {\doibase 10.1103/PhysRevB.98.081110} {\bibfield  {journal} {\bibinfo  {journal} {Phys. Rev. B}\ }\textbf {\bibinfo {volume} {98}},\ \bibinfo {pages} {081110} (\bibinfo {year} {2018})}\BibitemShut {NoStop}%
\bibitem [{\citenamefont {Zhang}\ and\ \citenamefont {Liu}(2015)}]{zhang_topological_2015}%
  \BibitemOpen
  \bibfield  {author} {\bibinfo {author} {\bibfnamefont {Rui-Xing}\ \bibnamefont {Zhang}}\ and\ \bibinfo {author} {\bibfnamefont {Chao-Xing}\ \bibnamefont {Liu}},\ }\bibfield  {title} {\enquote {\bibinfo {title} {Topological magnetic crystalline insulators and corepresentation theory},}\ }\href {\doibase 10.1103/PhysRevB.91.115317} {\bibfield  {journal} {\bibinfo  {journal} {Physical Review B}\ }\textbf {\bibinfo {volume} {91}},\ \bibinfo {pages} {115317} (\bibinfo {year} {2015})}\BibitemShut {NoStop}%
\bibitem [{\citenamefont {Schindler}\ \emph {et~al.}(2018)\citenamefont {Schindler}, \citenamefont {Cook}, \citenamefont {Vergniory}, \citenamefont {Wang}, \citenamefont {Parkin}, \citenamefont {Bernevig},\ and\ \citenamefont {Neupert}}]{schindler_higher-order_2018}%
  \BibitemOpen
  \bibfield  {author} {\bibinfo {author} {\bibfnamefont {Frank}\ \bibnamefont {Schindler}}, \bibinfo {author} {\bibfnamefont {Ashley~M.}\ \bibnamefont {Cook}}, \bibinfo {author} {\bibfnamefont {Maia~G.}\ \bibnamefont {Vergniory}}, \bibinfo {author} {\bibfnamefont {Zhijun}\ \bibnamefont {Wang}}, \bibinfo {author} {\bibfnamefont {Stuart S.~P.}\ \bibnamefont {Parkin}}, \bibinfo {author} {\bibfnamefont {B.~Andrei}\ \bibnamefont {Bernevig}}, \ and\ \bibinfo {author} {\bibfnamefont {Titus}\ \bibnamefont {Neupert}},\ }\bibfield  {title} {\enquote {\bibinfo {title} {Higher-order topological insulators},}\ }\href {\doibase 10.1126/sciadv.aat0346} {\bibfield  {journal} {\bibinfo  {journal} {Science Advances}\ }\textbf {\bibinfo {volume} {4}},\ \bibinfo {pages} {eaat0346} (\bibinfo {year} {2018})}\BibitemShut {NoStop}%
\bibitem [{\citenamefont {Varnava}\ and\ \citenamefont {Vanderbilt}(2018)}]{varnava_surfaces_2018}%
  \BibitemOpen
  \bibfield  {author} {\bibinfo {author} {\bibfnamefont {Nicodemos}\ \bibnamefont {Varnava}}\ and\ \bibinfo {author} {\bibfnamefont {David}\ \bibnamefont {Vanderbilt}},\ }\bibfield  {title} {\enquote {\bibinfo {title} {Surfaces of axion insulators},}\ }\href {\doibase 10.1103/PhysRevB.98.245117} {\bibfield  {journal} {\bibinfo  {journal} {Phys. Rev. B}\ }\textbf {\bibinfo {volume} {98}},\ \bibinfo {pages} {245117} (\bibinfo {year} {2018})}\BibitemShut {NoStop}%
\bibitem [{\citenamefont {Wieder}\ and\ \citenamefont {Bernevig}(2018)}]{wieder2018axion}%
  \BibitemOpen
  \bibfield  {author} {\bibinfo {author} {\bibfnamefont {Benjamin~J}\ \bibnamefont {Wieder}}\ and\ \bibinfo {author} {\bibfnamefont {B~Andrei}\ \bibnamefont {Bernevig}},\ }\bibfield  {title} {\enquote {\bibinfo {title} {The axion insulator as a pump of fragile topology},}\ }\href {https://arxiv.org/abs/1810.02373} {\bibfield  {journal} {\bibinfo  {journal} {arXiv preprint arXiv:1810.02373}\ } (\bibinfo {year} {2018})}\BibitemShut {NoStop}%
\bibitem [{\citenamefont {Ezawa}(2018)}]{ezawa_magnetic_2018}%
  \BibitemOpen
  \bibfield  {author} {\bibinfo {author} {\bibfnamefont {Motohiko}\ \bibnamefont {Ezawa}},\ }\bibfield  {title} {\enquote {\bibinfo {title} {Magnetic second-order topological insulators and semimetals},}\ }\href {\doibase 10.1103/PhysRevB.97.155305} {\bibfield  {journal} {\bibinfo  {journal} {Phys. Rev. B}\ }\textbf {\bibinfo {volume} {97}},\ \bibinfo {pages} {155305} (\bibinfo {year} {2018})}\BibitemShut {NoStop}%
\bibitem [{\citenamefont {Ahn}\ and\ \citenamefont {Yang}(2019)}]{ahn_symmetry_2019}%
  \BibitemOpen
  \bibfield  {author} {\bibinfo {author} {\bibfnamefont {Junyeong}\ \bibnamefont {Ahn}}\ and\ \bibinfo {author} {\bibfnamefont {Bohm-Jung}\ \bibnamefont {Yang}},\ }\bibfield  {title} {\enquote {\bibinfo {title} {Symmetry representation approach to topological invariants in ${C}_{2z} {T}$-symmetric systems},}\ }\href {\doibase 10.1103/PhysRevB.99.235125} {\bibfield  {journal} {\bibinfo  {journal} {Phys. Rev. B}\ }\textbf {\bibinfo {volume} {99}},\ \bibinfo {pages} {235125} (\bibinfo {year} {2019})}\BibitemShut {NoStop}%
\bibitem [{\citenamefont {Li}\ and\ \citenamefont {Sun}(2020)}]{li_pfaffian_2020}%
  \BibitemOpen
  \bibfield  {author} {\bibinfo {author} {\bibfnamefont {Heqiu}\ \bibnamefont {Li}}\ and\ \bibinfo {author} {\bibfnamefont {Kai}\ \bibnamefont {Sun}},\ }\bibfield  {title} {\enquote {\bibinfo {title} {Pfaffian formalism for higher-order topological insulators},}\ }\href {\doibase 10.1103/PhysRevLett.124.036401} {\bibfield  {journal} {\bibinfo  {journal} {Phys. Rev. Lett.}\ }\textbf {\bibinfo {volume} {124}},\ \bibinfo {pages} {036401} (\bibinfo {year} {2020})}\BibitemShut {NoStop}%
\bibitem [{\citenamefont {Gonz\'alez-Hern\'andez}\ \emph {et~al.}(2022)\citenamefont {Gonz\'alez-Hern\'andez}, \citenamefont {Pinilla},\ and\ \citenamefont {Uribe}}]{PhysRevB.106.195144}%
  \BibitemOpen
  \bibfield  {author} {\bibinfo {author} {\bibfnamefont {Rafael}\ \bibnamefont {Gonz\'alez-Hern\'andez}}, \bibinfo {author} {\bibfnamefont {Carlos}\ \bibnamefont {Pinilla}}, \ and\ \bibinfo {author} {\bibfnamefont {Bernardo}\ \bibnamefont {Uribe}},\ }\bibfield  {title} {\enquote {\bibinfo {title} {Axion insulators protected by ${C}_{2}\mathbb{T}$ symmetry, their $k$-theory invariants, and material realizations},}\ }\href {\doibase 10.1103/PhysRevB.106.195144} {\bibfield  {journal} {\bibinfo  {journal} {Phys. Rev. B}\ }\textbf {\bibinfo {volume} {106}},\ \bibinfo {pages} {195144} (\bibinfo {year} {2022})}\BibitemShut {NoStop}%
\bibitem [{\citenamefont {Nenno}\ \emph {et~al.}(2020)\citenamefont {Nenno}, \citenamefont {Garcia}, \citenamefont {Gooth}, \citenamefont {Felser},\ and\ \citenamefont {Narang}}]{nenno2020axion}%
  \BibitemOpen
  \bibfield  {author} {\bibinfo {author} {\bibfnamefont {Dennis~M}\ \bibnamefont {Nenno}}, \bibinfo {author} {\bibfnamefont {Christina~AC}\ \bibnamefont {Garcia}}, \bibinfo {author} {\bibfnamefont {Johannes}\ \bibnamefont {Gooth}}, \bibinfo {author} {\bibfnamefont {Claudia}\ \bibnamefont {Felser}}, \ and\ \bibinfo {author} {\bibfnamefont {Prineha}\ \bibnamefont {Narang}},\ }\bibfield  {title} {\enquote {\bibinfo {title} {Axion physics in condensed-matter systems},}\ }\href@noop {} {\bibfield  {journal} {\bibinfo  {journal} {Nature Reviews Physics}\ }\textbf {\bibinfo {volume} {2}},\ \bibinfo {pages} {682--696} (\bibinfo {year} {2020})}\BibitemShut {NoStop}%
\bibitem [{\citenamefont {Sekine}\ and\ \citenamefont {Nomura}(2021)}]{10.1063/5.0038804}%
  \BibitemOpen
  \bibfield  {author} {\bibinfo {author} {\bibfnamefont {Akihiko}\ \bibnamefont {Sekine}}\ and\ \bibinfo {author} {\bibfnamefont {Kentaro}\ \bibnamefont {Nomura}},\ }\bibfield  {title} {\enquote {\bibinfo {title} {Axion electrodynamics in topological materials},}\ }\href {\doibase 10.1063/5.0038804} {\bibfield  {journal} {\bibinfo  {journal} {Journal of Applied Physics}\ }\textbf {\bibinfo {volume} {129}},\ \bibinfo {pages} {141101} (\bibinfo {year} {2021})}\BibitemShut {NoStop}%
\bibitem [{\citenamefont {Varnava}\ \emph {et~al.}(2020)\citenamefont {Varnava}, \citenamefont {Souza},\ and\ \citenamefont {Vanderbilt}}]{PhysRevB.101.155130}%
  \BibitemOpen
  \bibfield  {author} {\bibinfo {author} {\bibfnamefont {Nicodemos}\ \bibnamefont {Varnava}}, \bibinfo {author} {\bibfnamefont {Ivo}\ \bibnamefont {Souza}}, \ and\ \bibinfo {author} {\bibfnamefont {David}\ \bibnamefont {Vanderbilt}},\ }\bibfield  {title} {\enquote {\bibinfo {title} {Axion coupling in the hybrid wannier representation},}\ }\href {\doibase 10.1103/PhysRevB.101.155130} {\bibfield  {journal} {\bibinfo  {journal} {Phys. Rev. B}\ }\textbf {\bibinfo {volume} {101}},\ \bibinfo {pages} {155130} (\bibinfo {year} {2020})}\BibitemShut {NoStop}%
\bibitem [{\citenamefont {Leung}\ and\ \citenamefont {Prodan}(2020)}]{Leung_2020}%
  \BibitemOpen
  \bibfield  {author} {\bibinfo {author} {\bibfnamefont {Bryan}\ \bibnamefont {Leung}}\ and\ \bibinfo {author} {\bibfnamefont {Emil}\ \bibnamefont {Prodan}},\ }\bibfield  {title} {\enquote {\bibinfo {title} {Bulk-boundary correspondence for topological insulators with quantized magneto-electric effect},}\ }\href {\doibase 10.1088/1751-8121/ab8415} {\bibfield  {journal} {\bibinfo  {journal} {Journal of Physics A: Mathematical and Theoretical}\ }\textbf {\bibinfo {volume} {53}},\ \bibinfo {pages} {205203} (\bibinfo {year} {2020})}\BibitemShut {NoStop}%
\bibitem [{\citenamefont {You}\ \emph {et~al.}(2016)\citenamefont {You}, \citenamefont {Cho},\ and\ \citenamefont {Hughes}}]{PhysRevB.94.085102}%
  \BibitemOpen
  \bibfield  {author} {\bibinfo {author} {\bibfnamefont {Yizhi}\ \bibnamefont {You}}, \bibinfo {author} {\bibfnamefont {Gil~Young}\ \bibnamefont {Cho}}, \ and\ \bibinfo {author} {\bibfnamefont {Taylor~L.}\ \bibnamefont {Hughes}},\ }\bibfield  {title} {\enquote {\bibinfo {title} {Response properties of axion insulators and weyl semimetals driven by screw dislocations and dynamical axion strings},}\ }\href {\doibase 10.1103/PhysRevB.94.085102} {\bibfield  {journal} {\bibinfo  {journal} {Phys. Rev. B}\ }\textbf {\bibinfo {volume} {94}},\ \bibinfo {pages} {085102} (\bibinfo {year} {2016})}\BibitemShut {NoStop}%
\bibitem [{\citenamefont {Alexandradinata}\ \emph {et~al.}(2020)\citenamefont {Alexandradinata}, \citenamefont {H\"oller}, \citenamefont {Wang}, \citenamefont {Cheng},\ and\ \citenamefont {Lu}}]{PhysRevB.102.115117}%
  \BibitemOpen
  \bibfield  {author} {\bibinfo {author} {\bibfnamefont {A.}~\bibnamefont {Alexandradinata}}, \bibinfo {author} {\bibfnamefont {J.}~\bibnamefont {H\"oller}}, \bibinfo {author} {\bibfnamefont {Chong}\ \bibnamefont {Wang}}, \bibinfo {author} {\bibfnamefont {Hengbin}\ \bibnamefont {Cheng}}, \ and\ \bibinfo {author} {\bibfnamefont {Ling}\ \bibnamefont {Lu}},\ }\bibfield  {title} {\enquote {\bibinfo {title} {Crystallographic splitting theorem for band representations and fragile topological photonic crystals},}\ }\href {\doibase 10.1103/PhysRevB.102.115117} {\bibfield  {journal} {\bibinfo  {journal} {Phys. Rev. B}\ }\textbf {\bibinfo {volume} {102}},\ \bibinfo {pages} {115117} (\bibinfo {year} {2020})}\BibitemShut {NoStop}%
\bibitem [{\citenamefont {Chalker}\ and\ \citenamefont {Coddington}(1988)}]{chalker}%
  \BibitemOpen
  \bibfield  {author} {\bibinfo {author} {\bibfnamefont {JT}~\bibnamefont {Chalker}}\ and\ \bibinfo {author} {\bibfnamefont {PD}~\bibnamefont {Coddington}},\ }\bibfield  {title} {\enquote {\bibinfo {title} {Percolation, quantum tunnelling and the integer hall effect},}\ }\href@noop {} {\bibfield  {journal} {\bibinfo  {journal} {Journal of Physics C: Solid State Physics}\ }\textbf {\bibinfo {volume} {21}},\ \bibinfo {pages} {2665} (\bibinfo {year} {1988})}\BibitemShut {NoStop}%
\bibitem [{\citenamefont {Pichard}\ and\ \citenamefont {Sarma}(1981)}]{transfer_matrix}%
  \BibitemOpen
  \bibfield  {author} {\bibinfo {author} {\bibfnamefont {Jean-Louis}\ \bibnamefont {Pichard}}\ and\ \bibinfo {author} {\bibfnamefont {G}~\bibnamefont {Sarma}},\ }\bibfield  {title} {\enquote {\bibinfo {title} {Finite size scaling approach to anderson localisation},}\ }\href@noop {} {\bibfield  {journal} {\bibinfo  {journal} {Journal of Physics C: Solid State Physics}\ }\textbf {\bibinfo {volume} {14}},\ \bibinfo {pages} {L127} (\bibinfo {year} {1981})}\BibitemShut {NoStop}%
\bibitem [{\citenamefont {MacKinnon}\ and\ \citenamefont {Kramer}(1981)}]{one_parameter_scaling}%
  \BibitemOpen
  \bibfield  {author} {\bibinfo {author} {\bibfnamefont {A.}~\bibnamefont {MacKinnon}}\ and\ \bibinfo {author} {\bibfnamefont {B.}~\bibnamefont {Kramer}},\ }\bibfield  {title} {\enquote {\bibinfo {title} {One-parameter scaling of localization length and conductance in disordered systems},}\ }\href {\doibase 10.1103/PhysRevLett.47.1546} {\bibfield  {journal} {\bibinfo  {journal} {Phys. Rev. Lett.}\ }\textbf {\bibinfo {volume} {47}},\ \bibinfo {pages} {1546--1549} (\bibinfo {year} {1981})}\BibitemShut {NoStop}%
\bibitem [{\citenamefont {Song}\ \emph {et~al.}(2021)\citenamefont {Song}, \citenamefont {Lian}, \citenamefont {Queiroz}, \citenamefont {Ilan}, \citenamefont {Bernevig},\ and\ \citenamefont {Stern}}]{song_delocalization_2021}%
  \BibitemOpen
  \bibfield  {author} {\bibinfo {author} {\bibfnamefont {Zhi-Da}\ \bibnamefont {Song}}, \bibinfo {author} {\bibfnamefont {Biao}\ \bibnamefont {Lian}}, \bibinfo {author} {\bibfnamefont {Raquel}\ \bibnamefont {Queiroz}}, \bibinfo {author} {\bibfnamefont {Roni}\ \bibnamefont {Ilan}}, \bibinfo {author} {\bibfnamefont {B.~Andrei}\ \bibnamefont {Bernevig}}, \ and\ \bibinfo {author} {\bibfnamefont {Ady}\ \bibnamefont {Stern}},\ }\bibfield  {title} {\enquote {\bibinfo {title} {Delocalization transition of a disordered axion insulator},}\ }\href {\doibase 10.1103/PhysRevLett.127.016602} {\bibfield  {journal} {\bibinfo  {journal} {Phys. Rev. Lett.}\ }\textbf {\bibinfo {volume} {127}},\ \bibinfo {pages} {016602} (\bibinfo {year} {2021})}\BibitemShut {NoStop}%
\bibitem [{\citenamefont {Li}\ \emph {et~al.}(2021)\citenamefont {Li}, \citenamefont {Jiang}, \citenamefont {Chen},\ and\ \citenamefont {Xie}}]{li_critical_2021}%
  \BibitemOpen
  \bibfield  {author} {\bibinfo {author} {\bibfnamefont {Hailong}\ \bibnamefont {Li}}, \bibinfo {author} {\bibfnamefont {Hua}\ \bibnamefont {Jiang}}, \bibinfo {author} {\bibfnamefont {Chui-Zhen}\ \bibnamefont {Chen}}, \ and\ \bibinfo {author} {\bibfnamefont {X.~C.}\ \bibnamefont {Xie}},\ }\bibfield  {title} {\enquote {\bibinfo {title} {Critical behavior and universal signature of an axion insulator state},}\ }\href {\doibase 10.1103/PhysRevLett.126.156601} {\bibfield  {journal} {\bibinfo  {journal} {Phys. Rev. Lett.}\ }\textbf {\bibinfo {volume} {126}},\ \bibinfo {pages} {156601} (\bibinfo {year} {2021})}\BibitemShut {NoStop}%
\bibitem [{\citenamefont {St{\v{r}}eda}(1982)}]{streda}%
  \BibitemOpen
  \bibfield  {author} {\bibinfo {author} {\bibfnamefont {P}~\bibnamefont {St{\v{r}}eda}},\ }\bibfield  {title} {\enquote {\bibinfo {title} {Theory of quantised hall conductivity in two dimensions},}\ }\href@noop {} {\bibfield  {journal} {\bibinfo  {journal} {Perspectives in Condensed Matter Physics}\ }\textbf {\bibinfo {volume} {2}},\ \bibinfo {pages} {161--165} (\bibinfo {year} {1982})}\BibitemShut {NoStop}%
\bibitem [{\citenamefont {Peng}\ \emph {et~al.}(2022)\citenamefont {Peng}, \citenamefont {Jiang}, \citenamefont {Fang}, \citenamefont {Weng},\ and\ \citenamefont {Fang}}]{peng_topological_2022}%
  \BibitemOpen
  \bibfield  {author} {\bibinfo {author} {\bibfnamefont {Bingrui}\ \bibnamefont {Peng}}, \bibinfo {author} {\bibfnamefont {Yi}~\bibnamefont {Jiang}}, \bibinfo {author} {\bibfnamefont {Zhong}\ \bibnamefont {Fang}}, \bibinfo {author} {\bibfnamefont {Hongming}\ \bibnamefont {Weng}}, \ and\ \bibinfo {author} {\bibfnamefont {Chen}\ \bibnamefont {Fang}},\ }\bibfield  {title} {\enquote {\bibinfo {title} {Topological classification and diagnosis in magnetically ordered electronic materials},}\ }\href {\doibase 10.1103/PhysRevB.105.235138} {\bibfield  {journal} {\bibinfo  {journal} {Phys. Rev. B}\ }\textbf {\bibinfo {volume} {105}},\ \bibinfo {pages} {235138} (\bibinfo {year} {2022})}\BibitemShut {NoStop}%
\bibitem [{sup()}]{sup}%
  \BibitemOpen
  \href@noop {} {}\bibinfo {note} {See \url{...} for {S}upplemental {M}aterials, where {R}efs.~\cite{guillemin2010differential,nielsen_absence_1981,nielsen_absence_1981-1, gallego_magnetic_2012, perez-mato_symmetry-based_2015, elcoro_magnetic_2021, oseledec1968multiplicative, kainaris2014conductivity, senechal2004theoretical, giamarchi1992conductivity, lee1996effects, apalkov2003interplay, lee1996transitions, wilczek1987two, castelnovo2008magnetic, doi:10.1126/science.1177582, doi:10.1126/science.1178868, Sasaki_2014, PhysRevLett.103.066402, iwasawa2019buried, PhysRevB.85.195320, doi:10.1126/science.1245085} are cited}\BibitemShut {NoStop}%
\bibitem [{\citenamefont {Slevin}\ and\ \citenamefont {Ohtsuki}(1999)}]{scaling_irr}%
  \BibitemOpen
  \bibfield  {author} {\bibinfo {author} {\bibfnamefont {Keith}\ \bibnamefont {Slevin}}\ and\ \bibinfo {author} {\bibfnamefont {Tomi}\ \bibnamefont {Ohtsuki}},\ }\bibfield  {title} {\enquote {\bibinfo {title} {Corrections to scaling at the anderson transition},}\ }\href {\doibase 10.1103/PhysRevLett.82.382} {\bibfield  {journal} {\bibinfo  {journal} {Phys. Rev. Lett.}\ }\textbf {\bibinfo {volume} {82}},\ \bibinfo {pages} {382--385} (\bibinfo {year} {1999})}\BibitemShut {NoStop}%
\bibitem [{\citenamefont {Slevin}\ and\ \citenamefont {Ohtsuki}(2016)}]{slevin2016estimate}%
  \BibitemOpen
  \bibfield  {author} {\bibinfo {author} {\bibfnamefont {Keith}\ \bibnamefont {Slevin}}\ and\ \bibinfo {author} {\bibfnamefont {Tomi}\ \bibnamefont {Ohtsuki}},\ }\bibfield  {title} {\enquote {\bibinfo {title} {Estimate of the critical exponent of the anderson transition in the three and four-dimensional unitary universality classes},}\ }\href@noop {} {\bibfield  {journal} {\bibinfo  {journal} {Journal of the Physical Society of Japan}\ }\textbf {\bibinfo {volume} {85}},\ \bibinfo {pages} {104712} (\bibinfo {year} {2016})}\BibitemShut {NoStop}%
\bibitem [{\citenamefont {Burkov}\ \emph {et~al.}(2011)\citenamefont {Burkov}, \citenamefont {Hook},\ and\ \citenamefont {Balents}}]{burkov_topological_2011}%
  \BibitemOpen
  \bibfield  {author} {\bibinfo {author} {\bibfnamefont {A.~A.}\ \bibnamefont {Burkov}}, \bibinfo {author} {\bibfnamefont {M.~D.}\ \bibnamefont {Hook}}, \ and\ \bibinfo {author} {\bibfnamefont {Leon}\ \bibnamefont {Balents}},\ }\bibfield  {title} {\enquote {\bibinfo {title} {Topological nodal semimetals},}\ }\href {\doibase 10.1103/PhysRevB.84.235126} {\bibfield  {journal} {\bibinfo  {journal} {Physical Review B}\ }\textbf {\bibinfo {volume} {84}},\ \bibinfo {pages} {235126} (\bibinfo {year} {2011})}\BibitemShut {NoStop}%
\bibitem [{\citenamefont {Luo}\ \emph {et~al.}(2020)\citenamefont {Luo}, \citenamefont {Xu}, \citenamefont {Ohtsuki},\ and\ \citenamefont {Shindou}}]{shindou_critical_2020}%
  \BibitemOpen
  \bibfield  {author} {\bibinfo {author} {\bibfnamefont {Xunlong}\ \bibnamefont {Luo}}, \bibinfo {author} {\bibfnamefont {Baolong}\ \bibnamefont {Xu}}, \bibinfo {author} {\bibfnamefont {Tomi}\ \bibnamefont {Ohtsuki}}, \ and\ \bibinfo {author} {\bibfnamefont {Ryuichi}\ \bibnamefont {Shindou}},\ }\bibfield  {title} {\enquote {\bibinfo {title} {Critical behavior of anderson transitions in three-dimensional orthogonal classes with particle-hole symmetries},}\ }\href {\doibase 10.1103/PhysRevB.101.020202} {\bibfield  {journal} {\bibinfo  {journal} {Phys. Rev. B}\ }\textbf {\bibinfo {volume} {101}},\ \bibinfo {pages} {020202} (\bibinfo {year} {2020})}\BibitemShut {NoStop}%
\bibitem [{\citenamefont {Jia}\ \emph {et~al.}(2023)\citenamefont {Jia}, \citenamefont {Liu}, \citenamefont {Ma}, \citenamefont {Geng}, \citenamefont {Sheng},\ and\ \citenamefont {Xing}}]{jia2023phase}%
  \BibitemOpen
  \bibfield  {author} {\bibinfo {author} {\bibfnamefont {KX}~\bibnamefont {Jia}}, \bibinfo {author} {\bibfnamefont {XY}~\bibnamefont {Liu}}, \bibinfo {author} {\bibfnamefont {R}~\bibnamefont {Ma}}, \bibinfo {author} {\bibfnamefont {H}~\bibnamefont {Geng}}, \bibinfo {author} {\bibfnamefont {L}~\bibnamefont {Sheng}}, \ and\ \bibinfo {author} {\bibfnamefont {DY}~\bibnamefont {Xing}},\ }\bibfield  {title} {\enquote {\bibinfo {title} {Phase diagram of three dimensional disordered nodal-line semimetals: weak localization to anderson localization},}\ }\href {\doibase 10.1088/1367-2630/ad08f4} {\bibfield  {journal} {\bibinfo  {journal} {New Journal of Physics}\ }\textbf {\bibinfo {volume} {25}},\ \bibinfo {pages} {113033} (\bibinfo {year} {2023})}\BibitemShut {NoStop}%
\bibitem [{\citenamefont {Gon\ifmmode~\mbox{\c{c}}\else \c{c}\fi{}alves}\ \emph {et~al.}(2020)\citenamefont {Gon\ifmmode~\mbox{\c{c}}\else \c{c}\fi{}alves}, \citenamefont {Ribeiro}, \citenamefont {Castro},\ and\ \citenamefont {Ara\'ujo}}]{goncalves_disorder_nodal_2020}%
  \BibitemOpen
  \bibfield  {author} {\bibinfo {author} {\bibfnamefont {Miguel}\ \bibnamefont {Gon\ifmmode~\mbox{\c{c}}\else \c{c}\fi{}alves}}, \bibinfo {author} {\bibfnamefont {Pedro}\ \bibnamefont {Ribeiro}}, \bibinfo {author} {\bibfnamefont {Eduardo~V.}\ \bibnamefont {Castro}}, \ and\ \bibinfo {author} {\bibfnamefont {Miguel A.~N.}\ \bibnamefont {Ara\'ujo}},\ }\bibfield  {title} {\enquote {\bibinfo {title} {Disorder-driven multifractality transition in weyl nodal loops},}\ }\href {\doibase 10.1103/PhysRevLett.124.136405} {\bibfield  {journal} {\bibinfo  {journal} {Phys. Rev. Lett.}\ }\textbf {\bibinfo {volume} {124}},\ \bibinfo {pages} {136405} (\bibinfo {year} {2020})}\BibitemShut {NoStop}%
\bibitem [{\citenamefont {Wu}\ \emph {et~al.}(2006)\citenamefont {Wu}, \citenamefont {Bernevig},\ and\ \citenamefont {Zhang}}]{wu2006helical}%
  \BibitemOpen
  \bibfield  {author} {\bibinfo {author} {\bibfnamefont {Congjun}\ \bibnamefont {Wu}}, \bibinfo {author} {\bibfnamefont {B.~Andrei}\ \bibnamefont {Bernevig}}, \ and\ \bibinfo {author} {\bibfnamefont {Shou-Cheng}\ \bibnamefont {Zhang}},\ }\bibfield  {title} {\enquote {\bibinfo {title} {Helical liquid and the edge of quantum spin hall systems},}\ }\href {\doibase 10.1103/PhysRevLett.96.106401} {\bibfield  {journal} {\bibinfo  {journal} {Phys. Rev. Lett.}\ }\textbf {\bibinfo {volume} {96}},\ \bibinfo {pages} {106401} (\bibinfo {year} {2006})}\BibitemShut {NoStop}%
\bibitem [{\citenamefont {Xu}\ and\ \citenamefont {Moore}(2006)}]{xu2006stability}%
  \BibitemOpen
  \bibfield  {author} {\bibinfo {author} {\bibfnamefont {Cenke}\ \bibnamefont {Xu}}\ and\ \bibinfo {author} {\bibfnamefont {J.~E.}\ \bibnamefont {Moore}},\ }\bibfield  {title} {\enquote {\bibinfo {title} {Stability of the quantum spin hall effect: Effects of interactions, disorder, and ${Z}_{2}$ topology},}\ }\href {\doibase 10.1103/PhysRevB.73.045322} {\bibfield  {journal} {\bibinfo  {journal} {Phys. Rev. B}\ }\textbf {\bibinfo {volume} {73}},\ \bibinfo {pages} {045322} (\bibinfo {year} {2006})}\BibitemShut {NoStop}%
\bibitem [{\citenamefont {Chou}\ \emph {et~al.}(2018)\citenamefont {Chou}, \citenamefont {Nandkishore},\ and\ \citenamefont {Radzihovsky}}]{chou2018gapless}%
  \BibitemOpen
  \bibfield  {author} {\bibinfo {author} {\bibfnamefont {Yang-Zhi}\ \bibnamefont {Chou}}, \bibinfo {author} {\bibfnamefont {Rahul~M}\ \bibnamefont {Nandkishore}}, \ and\ \bibinfo {author} {\bibfnamefont {Leo}\ \bibnamefont {Radzihovsky}},\ }\bibfield  {title} {\enquote {\bibinfo {title} {Gapless insulating edges of dirty interacting topological insulators},}\ }\href@noop {} {\bibfield  {journal} {\bibinfo  {journal} {Physical Review B}\ }\textbf {\bibinfo {volume} {98}},\ \bibinfo {pages} {054205} (\bibinfo {year} {2018})}\BibitemShut {NoStop}%
\bibitem [{\citenamefont {Huckestein}\ and\ \citenamefont {Backhaus}(1999)}]{huckestein1999integer}%
  \BibitemOpen
  \bibfield  {author} {\bibinfo {author} {\bibfnamefont {Bodo}\ \bibnamefont {Huckestein}}\ and\ \bibinfo {author} {\bibfnamefont {Michael}\ \bibnamefont {Backhaus}},\ }\bibfield  {title} {\enquote {\bibinfo {title} {Integer quantum hall effect of interacting electrons: Dynamical scaling and critical conductivity},}\ }\href@noop {} {\bibfield  {journal} {\bibinfo  {journal} {Physical review letters}\ }\textbf {\bibinfo {volume} {82}},\ \bibinfo {pages} {5100} (\bibinfo {year} {1999})}\BibitemShut {NoStop}%
\bibitem [{\citenamefont {Wang}\ and\ \citenamefont {Xiong}(2002)}]{wang2002electron}%
  \BibitemOpen
  \bibfield  {author} {\bibinfo {author} {\bibfnamefont {Ziqiang}\ \bibnamefont {Wang}}\ and\ \bibinfo {author} {\bibfnamefont {Shanhui}\ \bibnamefont {Xiong}},\ }\bibfield  {title} {\enquote {\bibinfo {title} {Electron-electron interactions, quantum coulomb gap, and dynamical scaling near integer quantum hall transitions},}\ }\href@noop {} {\bibfield  {journal} {\bibinfo  {journal} {Physical Review B}\ }\textbf {\bibinfo {volume} {65}},\ \bibinfo {pages} {195316} (\bibinfo {year} {2002})}\BibitemShut {NoStop}%
\bibitem [{\citenamefont {Pruisken}\ and\ \citenamefont {Burmistrov}(2008)}]{pruisken2008non}%
  \BibitemOpen
  \bibfield  {author} {\bibinfo {author} {\bibfnamefont {Adrianus~MM}\ \bibnamefont {Pruisken}}\ and\ \bibinfo {author} {\bibfnamefont {Igor~Sergeevich}\ \bibnamefont {Burmistrov}},\ }\bibfield  {title} {\enquote {\bibinfo {title} {Non-fermi liquid criticality and superuniversality in the quantum hall regime},}\ }\href@noop {} {\bibfield  {journal} {\bibinfo  {journal} {JETP letters}\ }\textbf {\bibinfo {volume} {87}},\ \bibinfo {pages} {220--224} (\bibinfo {year} {2008})}\BibitemShut {NoStop}%
\bibitem [{\citenamefont {Kumar}\ \emph {et~al.}(2022)\citenamefont {Kumar}, \citenamefont {Nosov},\ and\ \citenamefont {Raghu}}]{kumar2022interaction}%
  \BibitemOpen
  \bibfield  {author} {\bibinfo {author} {\bibfnamefont {Prashant}\ \bibnamefont {Kumar}}, \bibinfo {author} {\bibfnamefont {PA}~\bibnamefont {Nosov}}, \ and\ \bibinfo {author} {\bibfnamefont {S}~\bibnamefont {Raghu}},\ }\bibfield  {title} {\enquote {\bibinfo {title} {Interaction effects on quantum hall transitions: Dynamical scaling laws and superuniversality},}\ }\href@noop {} {\bibfield  {journal} {\bibinfo  {journal} {Physical Review Research}\ }\textbf {\bibinfo {volume} {4}},\ \bibinfo {pages} {033146} (\bibinfo {year} {2022})}\BibitemShut {NoStop}%
\bibitem [{\citenamefont {Witten}(1979)}]{witten1979dyons}%
  \BibitemOpen
  \bibfield  {author} {\bibinfo {author} {\bibfnamefont {Edward}\ \bibnamefont {Witten}},\ }\bibfield  {title} {\enquote {\bibinfo {title} {Dyons of charge e$\theta$/2$\pi$},}\ }\href@noop {} {\bibfield  {journal} {\bibinfo  {journal} {Physics Letters B}\ }\textbf {\bibinfo {volume} {86}},\ \bibinfo {pages} {283--287} (\bibinfo {year} {1979})}\BibitemShut {NoStop}%
\bibitem [{\citenamefont {Rosenberg}\ and\ \citenamefont {Franz}(2010)}]{PhysRevB.82.035105}%
  \BibitemOpen
  \bibfield  {author} {\bibinfo {author} {\bibfnamefont {G.}~\bibnamefont {Rosenberg}}\ and\ \bibinfo {author} {\bibfnamefont {M.}~\bibnamefont {Franz}},\ }\bibfield  {title} {\enquote {\bibinfo {title} {Witten effect in a crystalline topological insulator},}\ }\href {\doibase 10.1103/PhysRevB.82.035105} {\bibfield  {journal} {\bibinfo  {journal} {Phys. Rev. B}\ }\textbf {\bibinfo {volume} {82}},\ \bibinfo {pages} {035105} (\bibinfo {year} {2010})}\BibitemShut {NoStop}%
\bibitem [{\citenamefont {Aoki}\ \emph {et~al.}(2023)\citenamefont {Aoki}, \citenamefont {Fukaya}, \citenamefont {Kan}, \citenamefont {Koshino},\ and\ \citenamefont {Matsuki}}]{PhysRevB.108.155104}%
  \BibitemOpen
  \bibfield  {author} {\bibinfo {author} {\bibfnamefont {Shoto}\ \bibnamefont {Aoki}}, \bibinfo {author} {\bibfnamefont {Hidenori}\ \bibnamefont {Fukaya}}, \bibinfo {author} {\bibfnamefont {Naoto}\ \bibnamefont {Kan}}, \bibinfo {author} {\bibfnamefont {Mikito}\ \bibnamefont {Koshino}}, \ and\ \bibinfo {author} {\bibfnamefont {Yoshiyuki}\ \bibnamefont {Matsuki}},\ }\bibfield  {title} {\enquote {\bibinfo {title} {Magnetic monopole becomes dyon in topological insulators},}\ }\href {\doibase 10.1103/PhysRevB.108.155104} {\bibfield  {journal} {\bibinfo  {journal} {Phys. Rev. B}\ }\textbf {\bibinfo {volume} {108}},\ \bibinfo {pages} {155104} (\bibinfo {year} {2023})}\BibitemShut {NoStop}%
\bibitem [{\citenamefont {Guillemin}\ and\ \citenamefont {Pollack}(2010)}]{guillemin2010differential}%
  \BibitemOpen
  \bibfield  {author} {\bibinfo {author} {\bibfnamefont {Victor}\ \bibnamefont {Guillemin}}\ and\ \bibinfo {author} {\bibfnamefont {Alan}\ \bibnamefont {Pollack}},\ }\href@noop {} {\emph {\bibinfo {title} {Differential topology}}},\ Vol.\ \bibinfo {volume} {370}\ (\bibinfo  {publisher} {American Mathematical Soc.},\ \bibinfo {year} {2010})\BibitemShut {NoStop}%
\bibitem [{\citenamefont {Nielsen}\ and\ \citenamefont {Ninomiya}(1981{\natexlab{a}})}]{nielsen_absence_1981}%
  \BibitemOpen
  \bibfield  {author} {\bibinfo {author} {\bibfnamefont {H.~B.}\ \bibnamefont {Nielsen}}\ and\ \bibinfo {author} {\bibfnamefont {M.}~\bibnamefont {Ninomiya}},\ }\bibfield  {title} {\enquote {\bibinfo {title} {Absence of neutrinos on a lattice: ({II}). {Intuitive} topological proof},}\ }\href {\doibase 10.1016/0550-3213(81)90524-1} {\bibfield  {journal} {\bibinfo  {journal} {Nuclear Physics B}\ }\textbf {\bibinfo {volume} {193}},\ \bibinfo {pages} {173--194} (\bibinfo {year} {1981}{\natexlab{a}})}\BibitemShut {NoStop}%
\bibitem [{\citenamefont {Nielsen}\ and\ \citenamefont {Ninomiya}(1981{\natexlab{b}})}]{nielsen_absence_1981-1}%
  \BibitemOpen
  \bibfield  {author} {\bibinfo {author} {\bibfnamefont {H.~B.}\ \bibnamefont {Nielsen}}\ and\ \bibinfo {author} {\bibfnamefont {M.}~\bibnamefont {Ninomiya}},\ }\bibfield  {title} {\enquote {\bibinfo {title} {Absence of neutrinos on a lattice: ({I}). {Proof} by homotopy theory},}\ }\href {\doibase 10.1016/0550-3213(81)90361-8} {\bibfield  {journal} {\bibinfo  {journal} {Nuclear Physics B}\ }\textbf {\bibinfo {volume} {185}},\ \bibinfo {pages} {20--40} (\bibinfo {year} {1981}{\natexlab{b}})}\BibitemShut {NoStop}%
\bibitem [{\citenamefont {Gallego}\ \emph {et~al.}(2012)\citenamefont {Gallego}, \citenamefont {Tasci}, \citenamefont {Flor}, \citenamefont {Perez-Mato},\ and\ \citenamefont {Aroyo}}]{gallego_magnetic_2012}%
  \BibitemOpen
  \bibfield  {author} {\bibinfo {author} {\bibfnamefont {S.~V.}\ \bibnamefont {Gallego}}, \bibinfo {author} {\bibfnamefont {E.~S.}\ \bibnamefont {Tasci}}, \bibinfo {author} {\bibfnamefont {G.~de~la}\ \bibnamefont {Flor}}, \bibinfo {author} {\bibfnamefont {J.~M.}\ \bibnamefont {Perez-Mato}}, \ and\ \bibinfo {author} {\bibfnamefont {M.~I.}\ \bibnamefont {Aroyo}},\ }\bibfield  {title} {\enquote {\bibinfo {title} {Magnetic symmetry in the {Bilbao} {Crystallographic} {Server}: a computer program to provide systematic absences of magnetic neutron diffraction},}\ }\href {\doibase 10.1107/S0021889812042185} {\bibfield  {journal} {\bibinfo  {journal} {Journal of Applied Crystallography}\ }\textbf {\bibinfo {volume} {45}},\ \bibinfo {pages} {1236--1247} (\bibinfo {year} {2012})}\BibitemShut {NoStop}%
\bibitem [{\citenamefont {Perez-Mato}\ \emph {et~al.}(2015)\citenamefont {Perez-Mato}, \citenamefont {Gallego}, \citenamefont {Tasci}, \citenamefont {Elcoro}, \citenamefont {Flor},\ and\ \citenamefont {Aroyo}}]{perez-mato_symmetry-based_2015}%
  \BibitemOpen
  \bibfield  {author} {\bibinfo {author} {\bibfnamefont {J.~M.}\ \bibnamefont {Perez-Mato}}, \bibinfo {author} {\bibfnamefont {S.~V.}\ \bibnamefont {Gallego}}, \bibinfo {author} {\bibfnamefont {E.~S.}\ \bibnamefont {Tasci}}, \bibinfo {author} {\bibfnamefont {L.}~\bibnamefont {Elcoro}}, \bibinfo {author} {\bibfnamefont {G.~de~la}\ \bibnamefont {Flor}}, \ and\ \bibinfo {author} {\bibfnamefont {M.~I.}\ \bibnamefont {Aroyo}},\ }\bibfield  {title} {\enquote {\bibinfo {title} {Symmetry-{Based} {Computational} {Tools} for {Magnetic} {Crystallography}},}\ }\href {\doibase 10.1146/annurev-matsci-070214-021008} {\bibfield  {journal} {\bibinfo  {journal} {Annual Review of Materials Research}\ }\textbf {\bibinfo {volume} {45}},\ \bibinfo {pages} {217--248} (\bibinfo {year} {2015})}\BibitemShut {NoStop}%
\bibitem [{\citenamefont {Elcoro}\ \emph {et~al.}(2021)\citenamefont {Elcoro}, \citenamefont {Wieder}, \citenamefont {Song}, \citenamefont {Xu}, \citenamefont {Bradlyn},\ and\ \citenamefont {Bernevig}}]{elcoro_magnetic_2021}%
  \BibitemOpen
  \bibfield  {author} {\bibinfo {author} {\bibfnamefont {Luis}\ \bibnamefont {Elcoro}}, \bibinfo {author} {\bibfnamefont {Benjamin~J.}\ \bibnamefont {Wieder}}, \bibinfo {author} {\bibfnamefont {Zhida}\ \bibnamefont {Song}}, \bibinfo {author} {\bibfnamefont {Yuanfeng}\ \bibnamefont {Xu}}, \bibinfo {author} {\bibfnamefont {Barry}\ \bibnamefont {Bradlyn}}, \ and\ \bibinfo {author} {\bibfnamefont {B.~Andrei}\ \bibnamefont {Bernevig}},\ }\bibfield  {title} {\enquote {\bibinfo {title} {Magnetic topological quantum chemistry},}\ }\href {\doibase 10.1038/s41467-021-26241-8} {\bibfield  {journal} {\bibinfo  {journal} {Nature Communications}\ }\textbf {\bibinfo {volume} {12}},\ \bibinfo {pages} {5965} (\bibinfo {year} {2021})}\BibitemShut {NoStop}%
\bibitem [{\citenamefont {Oseledec}(1968)}]{oseledec1968multiplicative}%
  \BibitemOpen
  \bibfield  {author} {\bibinfo {author} {\bibfnamefont {Valery~Iustinovich}\ \bibnamefont {Oseledec}},\ }\bibfield  {title} {\enquote {\bibinfo {title} {A multiplicative ergodic theorem, lyapunov characteristic numbers for dynamical systems},}\ }\href@noop {} {\bibfield  {journal} {\bibinfo  {journal} {Transactions of the Moscow Mathematical Society}\ }\textbf {\bibinfo {volume} {19}},\ \bibinfo {pages} {197--231} (\bibinfo {year} {1968})}\BibitemShut {NoStop}%
\bibitem [{\citenamefont {Kainaris}\ \emph {et~al.}(2014)\citenamefont {Kainaris}, \citenamefont {Gornyi}, \citenamefont {Carr},\ and\ \citenamefont {Mirlin}}]{kainaris2014conductivity}%
  \BibitemOpen
  \bibfield  {author} {\bibinfo {author} {\bibfnamefont {Nikolaos}\ \bibnamefont {Kainaris}}, \bibinfo {author} {\bibfnamefont {Igor~V}\ \bibnamefont {Gornyi}}, \bibinfo {author} {\bibfnamefont {Sam~T}\ \bibnamefont {Carr}}, \ and\ \bibinfo {author} {\bibfnamefont {Alexander~D}\ \bibnamefont {Mirlin}},\ }\bibfield  {title} {\enquote {\bibinfo {title} {Conductivity of a generic helical liquid},}\ }\href@noop {} {\bibfield  {journal} {\bibinfo  {journal} {Physical Review B}\ }\textbf {\bibinfo {volume} {90}},\ \bibinfo {pages} {075118} (\bibinfo {year} {2014})}\BibitemShut {NoStop}%
\bibitem [{\citenamefont {S{\'e}n{\'e}chal}\ \emph {et~al.}(2004)\citenamefont {S{\'e}n{\'e}chal}, \citenamefont {Tremblay},\ and\ \citenamefont {Bourbonnais}}]{senechal2004theoretical}%
  \BibitemOpen
  \bibfield  {author} {\bibinfo {author} {\bibfnamefont {David}\ \bibnamefont {S{\'e}n{\'e}chal}}, \bibinfo {author} {\bibfnamefont {Andr{\'e}-Marie}\ \bibnamefont {Tremblay}}, \ and\ \bibinfo {author} {\bibfnamefont {Claude}\ \bibnamefont {Bourbonnais}},\ }\href@noop {} {\emph {\bibinfo {title} {Theoretical methods for strongly correlated electrons}}}\ (\bibinfo  {publisher} {Springer Science \& Business Media},\ \bibinfo {year} {2004})\BibitemShut {NoStop}%
\bibitem [{\citenamefont {Giamarchi}\ and\ \citenamefont {Millis}(1992)}]{giamarchi1992conductivity}%
  \BibitemOpen
  \bibfield  {author} {\bibinfo {author} {\bibfnamefont {T}~\bibnamefont {Giamarchi}}\ and\ \bibinfo {author} {\bibfnamefont {AJ}~\bibnamefont {Millis}},\ }\bibfield  {title} {\enquote {\bibinfo {title} {Conductivity of a luttinger liquid},}\ }\href@noop {} {\bibfield  {journal} {\bibinfo  {journal} {Physical Review B}\ }\textbf {\bibinfo {volume} {46}},\ \bibinfo {pages} {9325} (\bibinfo {year} {1992})}\BibitemShut {NoStop}%
\bibitem [{\citenamefont {Lee}\ and\ \citenamefont {Wang}(1996)}]{lee1996effects}%
  \BibitemOpen
  \bibfield  {author} {\bibinfo {author} {\bibfnamefont {Dung-Hai}\ \bibnamefont {Lee}}\ and\ \bibinfo {author} {\bibfnamefont {Ziqiang}\ \bibnamefont {Wang}},\ }\bibfield  {title} {\enquote {\bibinfo {title} {Effects of electron-electron interactions on the integer quantum hall transitions},}\ }\href@noop {} {\bibfield  {journal} {\bibinfo  {journal} {Physical review letters}\ }\textbf {\bibinfo {volume} {76}},\ \bibinfo {pages} {4014} (\bibinfo {year} {1996})}\BibitemShut {NoStop}%
\bibitem [{\citenamefont {Apalkov}\ and\ \citenamefont {Raikh}(2003)}]{apalkov2003interplay}%
  \BibitemOpen
  \bibfield  {author} {\bibinfo {author} {\bibfnamefont {VM}~\bibnamefont {Apalkov}}\ and\ \bibinfo {author} {\bibfnamefont {ME}~\bibnamefont {Raikh}},\ }\bibfield  {title} {\enquote {\bibinfo {title} {Interplay of short-range interactions and quantum interference near the integer quantum hall transition},}\ }\href@noop {} {\bibfield  {journal} {\bibinfo  {journal} {Physical Review B}\ }\textbf {\bibinfo {volume} {68}},\ \bibinfo {pages} {195312} (\bibinfo {year} {2003})}\BibitemShut {NoStop}%
\bibitem [{\citenamefont {Lee}(1996)}]{lee1996transitions}%
  \BibitemOpen
  \bibfield  {author} {\bibinfo {author} {\bibfnamefont {Dung-Hai}\ \bibnamefont {Lee}},\ }\bibfield  {title} {\enquote {\bibinfo {title} {Transitions between hall plateaux and the dimerization transition of a hubbard chain},}\ }\href@noop {} {\bibfield  {journal} {\bibinfo  {journal} {Philosophical magazine letters}\ }\textbf {\bibinfo {volume} {73}},\ \bibinfo {pages} {145--152} (\bibinfo {year} {1996})}\BibitemShut {NoStop}%
\bibitem [{\citenamefont {Wilczek}(1987)}]{wilczek1987two}%
  \BibitemOpen
  \bibfield  {author} {\bibinfo {author} {\bibfnamefont {Frank}\ \bibnamefont {Wilczek}},\ }\bibfield  {title} {\enquote {\bibinfo {title} {Two applications of axion electrodynamics},}\ }\href@noop {} {\bibfield  {journal} {\bibinfo  {journal} {Physical review letters}\ }\textbf {\bibinfo {volume} {58}},\ \bibinfo {pages} {1799} (\bibinfo {year} {1987})}\BibitemShut {NoStop}%
\bibitem [{\citenamefont {Castelnovo}\ \emph {et~al.}(2008)\citenamefont {Castelnovo}, \citenamefont {Moessner},\ and\ \citenamefont {Sondhi}}]{castelnovo2008magnetic}%
  \BibitemOpen
  \bibfield  {author} {\bibinfo {author} {\bibfnamefont {Claudio}\ \bibnamefont {Castelnovo}}, \bibinfo {author} {\bibfnamefont {Roderich}\ \bibnamefont {Moessner}}, \ and\ \bibinfo {author} {\bibfnamefont {Shivaji~L}\ \bibnamefont {Sondhi}},\ }\bibfield  {title} {\enquote {\bibinfo {title} {Magnetic monopoles in spin ice},}\ }\href@noop {} {\bibfield  {journal} {\bibinfo  {journal} {Nature}\ }\textbf {\bibinfo {volume} {451}},\ \bibinfo {pages} {42--45} (\bibinfo {year} {2008})}\BibitemShut {NoStop}%
\bibitem [{\citenamefont {Fennell}\ \emph {et~al.}(2009)\citenamefont {Fennell}, \citenamefont {Deen}, \citenamefont {Wildes}, \citenamefont {Schmalzl}, \citenamefont {Prabhakaran}, \citenamefont {Boothroyd}, \citenamefont {Aldus}, \citenamefont {McMorrow},\ and\ \citenamefont {Bramwell}}]{doi:10.1126/science.1177582}%
  \BibitemOpen
  \bibfield  {author} {\bibinfo {author} {\bibfnamefont {T.}~\bibnamefont {Fennell}}, \bibinfo {author} {\bibfnamefont {P.~P.}\ \bibnamefont {Deen}}, \bibinfo {author} {\bibfnamefont {A.~R.}\ \bibnamefont {Wildes}}, \bibinfo {author} {\bibfnamefont {K.}~\bibnamefont {Schmalzl}}, \bibinfo {author} {\bibfnamefont {D.}~\bibnamefont {Prabhakaran}}, \bibinfo {author} {\bibfnamefont {A.~T.}\ \bibnamefont {Boothroyd}}, \bibinfo {author} {\bibfnamefont {R.~J.}\ \bibnamefont {Aldus}}, \bibinfo {author} {\bibfnamefont {D.~F.}\ \bibnamefont {McMorrow}}, \ and\ \bibinfo {author} {\bibfnamefont {S.~T.}\ \bibnamefont {Bramwell}},\ }\bibfield  {title} {\enquote {\bibinfo {title} {Magnetic coulomb phase in the spin ice $\mathrm{Ho}_{2}\mathrm{Ti}_{2}\mathrm{O}_7$},}\ }\href {\doibase 10.1126/science.1177582} {\bibfield  {journal} {\bibinfo  {journal} {Science}\ }\textbf {\bibinfo {volume} {326}},\ \bibinfo {pages} {415--417} (\bibinfo {year} {2009})}\BibitemShut {NoStop}%
\bibitem [{\citenamefont {Morris}\ \emph {et~al.}(2009)\citenamefont {Morris}, \citenamefont {Tennant}, \citenamefont {Grigera}, \citenamefont {Klemke}, \citenamefont {Castelnovo}, \citenamefont {Moessner}, \citenamefont {Czternasty}, \citenamefont {Meissner}, \citenamefont {Rule}, \citenamefont {Hoffmann}, \citenamefont {Kiefer}, \citenamefont {Gerischer}, \citenamefont {Slobinsky},\ and\ \citenamefont {Perry}}]{doi:10.1126/science.1178868}%
  \BibitemOpen
  \bibfield  {author} {\bibinfo {author} {\bibfnamefont {D.~J.~P.}\ \bibnamefont {Morris}}, \bibinfo {author} {\bibfnamefont {D.~A.}\ \bibnamefont {Tennant}}, \bibinfo {author} {\bibfnamefont {S.~A.}\ \bibnamefont {Grigera}}, \bibinfo {author} {\bibfnamefont {B.}~\bibnamefont {Klemke}}, \bibinfo {author} {\bibfnamefont {C.}~\bibnamefont {Castelnovo}}, \bibinfo {author} {\bibfnamefont {R.}~\bibnamefont {Moessner}}, \bibinfo {author} {\bibfnamefont {C.}~\bibnamefont {Czternasty}}, \bibinfo {author} {\bibfnamefont {M.}~\bibnamefont {Meissner}}, \bibinfo {author} {\bibfnamefont {K.~C.}\ \bibnamefont {Rule}}, \bibinfo {author} {\bibfnamefont {J.-U.}\ \bibnamefont {Hoffmann}}, \bibinfo {author} {\bibfnamefont {K.}~\bibnamefont {Kiefer}}, \bibinfo {author} {\bibfnamefont {S.}~\bibnamefont {Gerischer}}, \bibinfo {author} {\bibfnamefont {D.}~\bibnamefont {Slobinsky}}, \ and\ \bibinfo {author} {\bibfnamefont {R.~S.}\ \bibnamefont {Perry}},\ }\bibfield  {title} {\enquote {\bibinfo {title} {Dirac strings and magnetic
  monopoles in the spin ice $\mathrm{Dy}_2 \mathrm{Ti}_2 \mathrm{O}_7$},}\ }\href {\doibase 10.1126/science.1178868} {\bibfield  {journal} {\bibinfo  {journal} {Science}\ }\textbf {\bibinfo {volume} {326}},\ \bibinfo {pages} {411--414} (\bibinfo {year} {2009})}\BibitemShut {NoStop}%
\bibitem [{\citenamefont {Sasaki}\ \emph {et~al.}(2014)\citenamefont {Sasaki}, \citenamefont {Imai},\ and\ \citenamefont {Kanazawa}}]{Sasaki_2014}%
  \BibitemOpen
  \bibfield  {author} {\bibinfo {author} {\bibfnamefont {T}~\bibnamefont {Sasaki}}, \bibinfo {author} {\bibfnamefont {E}~\bibnamefont {Imai}}, \ and\ \bibinfo {author} {\bibfnamefont {I}~\bibnamefont {Kanazawa}},\ }\bibfield  {title} {\enquote {\bibinfo {title} {Witten effect and fractional electric charge on the domain wall between topological insulators and spin ice compounds},}\ }\href {\doibase 10.1088/1742-6596/568/5/052029} {\bibfield  {journal} {\bibinfo  {journal} {Journal of Physics: Conference Series}\ }\textbf {\bibinfo {volume} {568}},\ \bibinfo {pages} {052029} (\bibinfo {year} {2014})}\BibitemShut {NoStop}%
\bibitem [{\citenamefont {Seradjeh}\ \emph {et~al.}(2009)\citenamefont {Seradjeh}, \citenamefont {Moore},\ and\ \citenamefont {Franz}}]{PhysRevLett.103.066402}%
  \BibitemOpen
  \bibfield  {author} {\bibinfo {author} {\bibfnamefont {B.}~\bibnamefont {Seradjeh}}, \bibinfo {author} {\bibfnamefont {J.~E.}\ \bibnamefont {Moore}}, \ and\ \bibinfo {author} {\bibfnamefont {M.}~\bibnamefont {Franz}},\ }\bibfield  {title} {\enquote {\bibinfo {title} {Exciton condensation and charge fractionalization in a topological insulator film},}\ }\href {\doibase 10.1103/PhysRevLett.103.066402} {\bibfield  {journal} {\bibinfo  {journal} {Phys. Rev. Lett.}\ }\textbf {\bibinfo {volume} {103}},\ \bibinfo {pages} {066402} (\bibinfo {year} {2009})}\BibitemShut {NoStop}%
\bibitem [{\citenamefont {Iwasawa}\ \emph {et~al.}(2019)\citenamefont {Iwasawa}, \citenamefont {Dudin}, \citenamefont {Inui}, \citenamefont {Masui}, \citenamefont {Kim}, \citenamefont {Cacho},\ and\ \citenamefont {Hoesch}}]{iwasawa2019buried}%
  \BibitemOpen
  \bibfield  {author} {\bibinfo {author} {\bibfnamefont {Hideaki}\ \bibnamefont {Iwasawa}}, \bibinfo {author} {\bibfnamefont {Pavel}\ \bibnamefont {Dudin}}, \bibinfo {author} {\bibfnamefont {Kyosuke}\ \bibnamefont {Inui}}, \bibinfo {author} {\bibfnamefont {Takahiko}\ \bibnamefont {Masui}}, \bibinfo {author} {\bibfnamefont {Timur~K}\ \bibnamefont {Kim}}, \bibinfo {author} {\bibfnamefont {Cephise}\ \bibnamefont {Cacho}}, \ and\ \bibinfo {author} {\bibfnamefont {Moritz}\ \bibnamefont {Hoesch}},\ }\bibfield  {title} {\enquote {\bibinfo {title} {Buried double cuo chains in $\mathrm{YBa}_2 \mathrm{Cu}_4 \mathrm{O}_8$ uncovered by nano-arpes},}\ }\href@noop {} {\bibfield  {journal} {\bibinfo  {journal} {Physical Review B}\ }\textbf {\bibinfo {volume} {99}},\ \bibinfo {pages} {140510} (\bibinfo {year} {2019})}\BibitemShut {NoStop}%
\bibitem [{\citenamefont {Wang}\ \emph {et~al.}(2012)\citenamefont {Wang}, \citenamefont {Sun}, \citenamefont {Chen}, \citenamefont {Franchini}, \citenamefont {Xu}, \citenamefont {Weng}, \citenamefont {Dai},\ and\ \citenamefont {Fang}}]{PhysRevB.85.195320}%
  \BibitemOpen
  \bibfield  {author} {\bibinfo {author} {\bibfnamefont {Zhijun}\ \bibnamefont {Wang}}, \bibinfo {author} {\bibfnamefont {Yan}\ \bibnamefont {Sun}}, \bibinfo {author} {\bibfnamefont {Xing-Qiu}\ \bibnamefont {Chen}}, \bibinfo {author} {\bibfnamefont {Cesare}\ \bibnamefont {Franchini}}, \bibinfo {author} {\bibfnamefont {Gang}\ \bibnamefont {Xu}}, \bibinfo {author} {\bibfnamefont {Hongming}\ \bibnamefont {Weng}}, \bibinfo {author} {\bibfnamefont {Xi}~\bibnamefont {Dai}}, \ and\ \bibinfo {author} {\bibfnamefont {Zhong}\ \bibnamefont {Fang}},\ }\bibfield  {title} {\enquote {\bibinfo {title} {Dirac semimetal and topological phase transitions in ${A}_{3}$bi ($a=\text{Na}$, k, rb)},}\ }\href {\doibase 10.1103/PhysRevB.85.195320} {\bibfield  {journal} {\bibinfo  {journal} {Phys. Rev. B}\ }\textbf {\bibinfo {volume} {85}},\ \bibinfo {pages} {195320} (\bibinfo {year} {2012})}\BibitemShut {NoStop}%
\bibitem [{\citenamefont {Liu}\ \emph {et~al.}(2014{\natexlab{b}})\citenamefont {Liu}, \citenamefont {Zhou}, \citenamefont {Zhang}, \citenamefont {Wang}, \citenamefont {Weng}, \citenamefont {Prabhakaran}, \citenamefont {Mo}, \citenamefont {Shen}, \citenamefont {Fang}, \citenamefont {Dai}, \citenamefont {Hussain},\ and\ \citenamefont {Chen}}]{doi:10.1126/science.1245085}%
  \BibitemOpen
  \bibfield  {author} {\bibinfo {author} {\bibfnamefont {Z.~K.}\ \bibnamefont {Liu}}, \bibinfo {author} {\bibfnamefont {B.}~\bibnamefont {Zhou}}, \bibinfo {author} {\bibfnamefont {Y.}~\bibnamefont {Zhang}}, \bibinfo {author} {\bibfnamefont {Z.~J.}\ \bibnamefont {Wang}}, \bibinfo {author} {\bibfnamefont {H.~M.}\ \bibnamefont {Weng}}, \bibinfo {author} {\bibfnamefont {D.}~\bibnamefont {Prabhakaran}}, \bibinfo {author} {\bibfnamefont {S.-K.}\ \bibnamefont {Mo}}, \bibinfo {author} {\bibfnamefont {Z.~X.}\ \bibnamefont {Shen}}, \bibinfo {author} {\bibfnamefont {Z.}~\bibnamefont {Fang}}, \bibinfo {author} {\bibfnamefont {X.}~\bibnamefont {Dai}}, \bibinfo {author} {\bibfnamefont {Z.}~\bibnamefont {Hussain}}, \ and\ \bibinfo {author} {\bibfnamefont {Y.~L.}\ \bibnamefont {Chen}},\ }\bibfield  {title} {\enquote {\bibinfo {title} {Discovery of a three-dimensional topological dirac semimetal, $\mathrm{Na}_3 \mathrm{Bi}$},}\ }\href {\doibase 10.1126/science.1245085} {\bibfield  {journal} {\bibinfo  {journal} {Science}\
  }\textbf {\bibinfo {volume} {343}},\ \bibinfo {pages} {864--867} (\bibinfo {year} {2014}{\natexlab{b}})}\BibitemShut {NoStop}%
\end{thebibliography}
\end{document}